\documentclass[12pt]{article}
\usepackage{bbm}
\usepackage{color}
\usepackage{authblk}
\usepackage{latexsym}
\usepackage{epsfig,amssymb,euscript, mathrsfs,cite}
\usepackage{amsmath}
\usepackage{slashed}
\usepackage{mathrsfs} 
\usepackage{enumerate}
\definecolor{MyDarkBlue}{rgb}{0.15,0.15,0.45}
\usepackage[linktocpage=true]{hyperref}
\hypersetup{
colorlinks=true,
citecolor=MyDarkBlue,
linkcolor=MyDarkBlue,
urlcolor=MyDarkBlue,
pdfauthor={},
pdftitle={},
pdfsubject={hep-th}
}
\newsavebox{\ns}
\newsavebox{\dbrane}
\newsavebox{\dbshort}

\def\be{\begin{equation}}
\def\ee{\end{equation}}
\def\bea{\begin{eqnarray}}
\def\eea{\end{eqnarray}}

\newcommand{\nn}{\nonumber\\}

\newcommand\R{\mathbb{R}}
\newcommand\Z{\mathbb{Z}}

\newcommand\C{\mathbb{C}}

\newcommand\diff{\mathrm{d}}

\newcommand{\ii}{\mathrm{i}}

\newcommand{\ex}{\mathrm{e}}

\newcommand{\q}{q}

\newcommand{\GGamma}{\hat\Gamma}

\newcommand{\cc}{\mathtt{k}}

\newcommand{\bn}{\mathtt{k}}

\newcommand{\z}{z}

\newcommand{\newepsilon}{\zeta}
\newcommand{\sa}{\mathtt{s}}

\newlength{\sswidth}

\numberwithin{equation}{section}       

\newcommand{\ff}{h}

\begin{document}

\begin{titlepage}

\begin{flushright}
Imperial/TP/2021/JG/05\\
\end{flushright}

\vskip 1cm

\begin{center}


{\Large \bf Supersymmetric spindles}

\vskip 1cm

\vskip 1cm
{Pietro Ferrero$^{\mathrm{a}}$, Jerome P. Gauntlett$^{\mathrm{b}}$,  
and James Sparks$^{\mathrm{a}}$}

\vskip 0.5cm

${}^{\,\mathrm{a}}$\textit{Mathematical Institute, University of Oxford,\\
Andrew Wiles Building, Radcliffe Observatory Quarter,\\
Woodstock Road, Oxford, OX2 6GG, U.K.\\}

\vskip 0.2 cm

${}^{\mathrm{b}}$\textit{Blackett Laboratory, Imperial College, \\
Prince Consort Rd., London, SW7 2AZ, U.K.\\}

\end{center}

\vskip 0.5 cm

\begin{abstract}
\noindent  
In the context of holography,
 we analyse aspects of  supersymmetric geometries based on two-dimensional orbifolds known as spindles. By analysing spin$^c$ spinors on a spindle with an azimuthal rotation symmetry we show that under rather general conditions there are just two possibilities, called the `twist' and the `anti-twist', which are determined by the quantized magnetic flux through the spindle. A special case of the twist is the standard topological twist which is associated with constant and chiral spinors. We construct solutions of $D=5$ and $D=4$ STU gauged supergravity theories that are dual to D3-branes and M2-branes wrapping spindles, respectively, which realize both the anti-twist, as seen before, but also the twist. 
 For the wrapped D3-brane solutions we reproduce the central charge of the gravity solution from the dual field theory by analysing the anomaly polynomial of $\mathcal{N}=4$ SYM theory. We also discuss M5-branes wrapped on spindles both from a gravity and a field theory point of view.
\end{abstract}

\end{titlepage}

\pagestyle{plain}
\setcounter{page}{1}
\newcounter{bean}
\baselineskip18pt

\tableofcontents

\section{Introduction}

A new facet of the AdS/CFT correspondence involves a class of two-dimensional orbifolds known as
spindles \cite{Ferrero:2020laf}. 
Recall that a spindle is topologically a two-sphere but with quantized conical deficit angles at the two poles. More precisely, we define
a spindle $\Sigma=\mathbb{WCP}^1_{[n_1,n_2]}$ to be a weighted projective space that is specified by two positive, co-prime integers $n_1$, $n_2$.
Various AdS$\times\Sigma$ supersymmetric solutions of gauged supergravity theories in $D=4,5$ and $7$ dimensions have now been constructed, which are naturally associated with\footnote{A recent interesting paper has constructed solutions
of massive type IIA that are associated with D4-branes wrapping spindles \cite{Faedo:2021nub}.}
  M2, D3 and M5-branes, respectively \cite{Ferrero:2020laf,Ferrero:2020twa,Hosseini:2021fge,Boido:2021szx,Ferrero:2021wvk,Ferrero:2021ovq,Couzens:2021rlk}. These can then be uplifted on certain compact manifolds to give supersymmetric solutions of $D=10$ and $D=11$ supergravity that are dual to new SCFTs. Remarkably, in the examples involving M2 and D3-branes, one finds uplifted solutions that are completely regular and free from any orbifold singularities. In other examples, involving M5-branes, orbifold singularities remain in the $D=11$ solutions,  but strong supporting evidence has been provided that they are in fact dual to \emph{bona fide} SCFTs.  
  
The regular uplifted solutions in $D=10$ and $D=11$ supergravity just mentioned were in fact first constructed some time ago
using a different approach \cite{Gauntlett:2006af,Gauntlett:2006ns}, but the dual SCFTs remained obscure. The new insight that they can be recast as uplifted solutions of a lower-dimensional gauged supergravity leads to the natural conjecture that the dual SCFTs are the low-energy limit of the field theories arising from branes wrapping a spindle. Evidence supporting  this picture has been provided in the examples involving D3 and M5-branes, by calculating the central charge of the proposed SCFT using anomaly polynomials, and finding exact agreement with the gravity result.

The picture that has emerged has a direct parallel with the well-known constructions involving branes wrapping Riemann surfaces $\Sigma_g$,
with genus $g$, but there are some crucial differences. In the standard constructions, starting with \cite{Maldacena:2000mw}, supersymmetry on the world-volume of the branes is realized with a partial ``topological twist", which involves switching on background R-symmetry gauge fields that are determined, locally, by the spin connection of the metric on the Riemann surface.  
An immediate consequence of the topological twist is that the integrated R-symmetry flux through the Riemann surface, suitably normalized, is equal to the Euler number of the Riemann surface. Another consequence of the topological twist is that the Killing spinors for the AdS$\times\Sigma_g$ solutions are  chiral and constant on $\Sigma_g$. As is also well-known, this standard set-up is realized by wrapping branes on a Riemann surface that is holomorphically embedded in a Calabi-Yau manifold.

By contrast, in the new AdS$\times \Sigma$ solutions involving spindles the Killing spinors are neither chiral nor constant
on $\Sigma$. In fact two distinct ways of realizing supersymmetry for spindles have now emerged, which we refer to 
as the \emph{twist} and \emph{anti-twist}, and they can be characterized via the R-symmetry flux through the spindle: 
\begin{align}\label{fluxchi}
\text{Twist:}\qquad \frac{1}{2\pi} \int_\Sigma F \, &= \, \chi \, \equiv\, \frac{n_1+ n_2}{n_1 n_2}~,\nn
\text{Anti-twist:}\qquad \frac{1}{2\pi} \int_\Sigma F \, &= \, \chi_{-} \, \equiv\, \frac{n_2- n_1}{n_1 n_2}~,
\end{align}
where $F$ is the field strength for the R-symmetry gauge field. 
Notice in the case of the twist $\chi$ is in fact the Euler number of the spindle, so this is a feature that is also shared 
with the standard topological twist.

All of the known AdS$\times \Sigma$ solutions have a spindle metric that is invariant under azimuthal rotations about the poles. 
One of the main results of this paper is to show, assuming such a spindle metric, that the twist and the anti-twist are the \emph{only} possible ways of
realizing supersymmetry on the spindle. The proof follows from analysing the regularity conditions for the two-component spin$^c$ spinors on the spindle $\Sigma$, which are charged with respect to the R-symmetry gauge field, near each of the two orbifold points. The only other assumption that goes into the proof is that the Killing spinor should be non-vanishing at these orbifold points, which holds on general grounds for supersymmetric AdS solutions, as we shall discuss. Our analysis also clarifies why the regularity arguments that have been used for supergravity solutions with spindles constructed in \cite{Ferrero:2020laf,Faedo:2021nub,Ferrero:2020twa,Hosseini:2021fge,Boido:2021szx,Ferrero:2021wvk,Ferrero:2021ovq,Couzens:2021rlk}
do in fact give the correct results. 

For the Killing spinors arising in the supersymmetric AdS$\times \Sigma$ solutions, we show that a vector bilinear which can be constructed from the Killing spinor is the Killing vector which generates the azimuthal rotations on the spindle. This is true for both the twist and the anti-twist cases. 
This aspect of the way in which the supersymmetry algebra is being realized on the spindle can be contrasted\footnote{From the dual field theory point of view, a closely related point is that in considering a SCFT on the spindle, one finds that the R-symmetry of the resulting lower-dimensional SCFT is a mixture of the spindle symmetry and the  R-symmetry of the higher-dimensional SCFT.}
 with the usual topological twist
for Riemann surfaces, where the corresponding vector bilinear vanishes identically. This further underscores the fact that while the twist case for the spindle and the topological twist share the feature that the R-symmetry flux is the Euler number as in \eqref{fluxchi}, they are in fact quite distinct. That being
said, if one formally sets $n_1=n_2=1$ in the central charge of the dual SCFTs for various AdS$\times \Sigma$ solutions, one gets
the same result that one gets for some known AdS$\times S^2$ solutions with a topological twist. 

In this paper we will also re-examine the known constructions of AdS$\times \Sigma$ solutions of various gauged supergravity theories, and find some new
solutions that were missed in the original analyses. In particular, previous constructions 
of AdS$_3\times \Sigma$ D3-brane solutions and AdS$_2\times \Sigma$ M2-brane solutions 
only realized the \emph{anti-twist} in \eqref{fluxchi}, but we shall see that for both D3-branes and M2-branes 
there also exist new \emph{twist} solutions.\footnote{With the caveat that these require 
multiple magnetic charges, and thus in particular do not exist in minimal gauged supergravity, as shown in \cite{Ferrero:2020laf, Ferrero:2020twa}.}
Moreover, these twist solutions have a natural physical interpretation as the IR limit of 
D3-branes and M2-branes wrapped on a spindle that is holomorphically embedded inside 
a Calabi-Yau four-fold and five-fold, respectively. Interestingly, 
for AdS$_5\times \Sigma$ M5-brane solutions the construction 
in \cite{Ferrero:2021wvk} only leads to twist solutions, 
while for AdS$_3\times \Sigma\times\Sigma_g$ M5-brane 
solutions we find new twist solutions, as well as the previously known anti-twist solutions. 
It is an interesting open problem to identify a UV geometry for the many anti-twist 
solutions that have now been constructed, involving wrapped branes, a point we discuss further in
section \ref{fincoms}.

The outline of the paper is as follows. In section \ref{sectwists} we 
give a general proof that~\eqref{fluxchi} are the only possibilities 
for the R-symmetry flux through a spindle, under some rather general assumptions. 
Our discussion includes a global description of $U(1)$ orbibundles, the associated Killing spinors and some properties of the uplifted solutions. 
We will explain how the standard topological twist arises 
as a special case of the twist, but we emphasize that in actual solutions the Killing spinors have distinct features. 
In section \ref{sec:D3} we present 
a fairly comprehensive analysis of AdS$_3\times\Sigma$ D3-brane solutions, 
in particular constructing new twist solutions. 
We interpret the new twist solutions 
as near-horizon limits of D3-branes wrapped on $\Sigma$ embedded holomorphically inside a Calabi-Yau four-fold, 
and reproduce the supergravity central charge by computing the anomaly polynomial of $\mathcal{N}=4$ 
SYM theory wrapped on $\Sigma$. Some of the new $D=5$ gauged supergravity solutions also uplift to 
new AdS$_3\times\Sigma\times\Sigma_g$ twist M5-brane solutions, or more generally the product of $\Sigma$ with a punctured Riemann surface. Section \ref{sec:M2} 
presents a similar analysis for AdS$_2\times\Sigma$ M2-brane solutions, 
although here the regularity conditions are more difficult to solve in closed 
form, and we content ourselves with showing that new twist solutions exist,
without classifying them. In section \ref{sec:M5} we briefly discuss
AdS$_5\times \Sigma$ M5-brane solutions, in particular constructing the 
explicit Killing spinors for the solutions found in \cite{Ferrero:2021wvk}. 
We conclude with some discussion in section \ref{fincoms}. Three appendices 
contain some additional results. Appendix 
\ref{appA} shows how various previously 
known AdS$\times S^2$ solutions 
arise from limits of the (local) AdS$\times \Sigma$ solutions 
presented here,  while appendix \ref{AdSspinors} discusses Killing spinors in AdS.
In appendix \ref{app:c} we present some details on our derivation of the explicit Killing spinors for the case
of the D3-brane solutions; the derivation for the other cases is somewhat similar.

\section{Twists and anti-twists}\label{sectwists}

As already mentioned, 
several different classes of supersymmetric supergravity solutions involving 
spindles have recently been constructed, including \cite{Ferrero:2020laf,Ferrero:2020twa,Hosseini:2021fge,Boido:2021szx,Ferrero:2021wvk,Ferrero:2021ovq,Couzens:2021rlk}. Typically these are realized as solutions to a lower-dimensional gauged supergravity theory, 
and then uplifted to $D=10$ string theory or $D=11$ M-theory solutions, where the solutions describe 
D3-brane, M2-brane or M5-brane backgrounds. In 
every case the total magnetic flux for the R-symmetry gauge field through the spindle  $\Sigma$
has been found to be one of the two possibilities in \eqref{fluxchi}. 
Here $A$ is the Abelian R-symmetry gauge field, with curvature $F=\diff A$, normalized so that the Killing spinor has charge $\tfrac{1}{2}$. 

In this section we give a general proof that 
these are the only two possibilities, under some fairly general assumptions:
 (i) azimuthal rotations of the spindle are a symmetry of the solution, 
(ii) 
the Killing spinor is non-zero at the poles of the spindle. We shall discuss the assumption (ii)  
further in section \ref{sec:nowhere}, where we argue that the Killing spinor is in fact nowhere zero
for supersymmetric AdS solutions in string and M-theory.
The precise definition of assumption (i) will be made clear in the next subsections. 
The proof follows from global regularity conditions on the Killing spinor, independently
of precisely which equation it solves.

\subsection{Set-up}\label{sec:setup}

We begin in this subsection by summarizing the basic set-up.
We focus on a gauged supergravity perspective, as this is the framework where we construct solutions in later sections, but we also discuss the implications for the uplifted supergravity solutions in section \ref{sec:uplift}.

The precise details of the Killing spinor equation will not be relevant for what follows. 
In general we can take the Killing spinor to carry a possible ``flavour index" and write 
$\newepsilon^i$, which for each $i=1,\ldots,n$ is a two-component spin$^c$ spinor on $\Sigma$. We   
assume it satisfies\footnote{The Killing spinor might also satisfy additional algebraic equations, but
 these play no role in the following arguments.}
 a Killing spinor equation (KSE) of the schematic form
\begin{align}\label{kseqnschem}
(\nabla-\tfrac{\ii}{2} A)\newepsilon^i+\mathcal{M}^i{}_j\newepsilon^j \, = \, 0\,.
\end{align}
Here $\nabla=\diff +\frac{1}{4}\omega^{ab}\gamma_{ab}$ is the Levi-Civita spin connection and, for definiteness, we have taken the spinor
to have charge $\tfrac{1}{2}$ with respect to the R-symmetry gauge field $A$.
 The matrix $\mathcal{M}^i{}_j$ could depend on scalar fields, as well as other gauge fields that act on the flavour indices but, as we will see, this matrix will not play any role in the following discussion; indeed we can suppress the flavour index on $\newepsilon^i$ and simply write $\newepsilon$. 

We will be interested in studying such Killing spinors $\zeta$ on a spindle.  
By definition, a spindle $\Sigma$ is topologically a two-sphere with conical deficit angles 
 $2\pi(1-1/n_S)$ at the ``south pole'' and $2\pi(1-1/n_N)$ at the ``north pole'', where
 $n_S, n_N\in \mathbb{N}$ are relatively prime with hcf$(n_S,n_N)=1$.
 This may equivalently be described as the weighted projective space $\Sigma=\mathbb{WCP}^1_{[n_S,n_N]}$, 
 which is an orbifold Riemann surface. 
We will exclude the special case $n_S=n_N=1$ from the definition of a spindle,
 but point out that these values give rise to a smooth two-sphere. 
 We assume that the metric on the spindle is invariant under azimuthal rotations around the poles. 
 That is, there is a Killing vector $k\equiv \partial_\varphi$, with  coordinates $(\rho,\varphi)$ in which the 
metric on the spindle takes the form
\begin{align}\label{spindlemetric}
\diff s^2_\Sigma\, = \, \diff\rho^2+f^2(\rho)\diff\varphi^2\, .
\end{align}
We normalize $\varphi$ to have period $\Delta\varphi=2\pi$, and take $\rho\in [\rho_1,\rho_2]$.
The metric \eqref{spindlemetric} describes a spindle with the above deficit angles if $f(\rho)$ has the boundary behaviour
$f(\rho)=(\rho-\rho_1)/n_N+O((\rho-\rho_1)^2)$ and $f(\rho)=(\rho_2-\rho)/n_S+O((\rho_2-\rho)^2)$
near the north and south poles, respectively, with $f$ strictly positive on the open interval $(\rho_1,\rho_2)$. 

The Killing vector $k=\partial_\varphi$ is assumed to generate a symmetry of the full solution. In particular we 
assume that the matrix $\mathcal{M}$ in the KSE \eqref{kseqnschem} invariant, i.e. $\mathcal{L}_k\mathcal{M}=0$,  where $\mathcal{L}_k$ is the Lie derivative.  
Similarly we assume that $\mathcal{L}_kF=0$, and will typically work in local gauges which 
 satisfy $\mathcal{L}_kA=0$, 
 and hence we can write
 \begin{align}\label{Agauge}
A\, = \, a(\rho)\diff \varphi\, .
\end{align}
Notice this leaves a freedom to make \emph{local}
gauge transformations 
\begin{align}\label{residualgauge}
A \, \rightarrow \, A + \gamma\, \diff\varphi\, ,
\end{align}
 with $\gamma$  a constant, which equivalently shifts $a(\rho)\rightarrow a(\rho)+\gamma$. 
 We shall present a careful global discussion of gauge fields on spindles in section \ref{sec:orbs} below. 

With the above assumptions, the KSE \eqref{kseqnschem} is invariant under $k$. A consequence is that
the Killing spinor solution $\newepsilon$ has definite charge $Q$ under $k=\partial_\varphi$. 
That is, $\mathcal{L}_k \newepsilon = \ii Q\, \newepsilon$, where 
$\mathcal{L}_k= \nabla_k + \tfrac{1}{8}(\diff k^\flat)^{ab}\gamma_{ab}$ denotes the spinor Lie derivative along $k$, where
$k^\flat$ is the one-form  metrically dual to $k$. Indeed, since the KSE is invariant 
under $k$, its space of solutions will form a representation of the 
$U(1)$ symmetry generated by $k$. But the irreducible representations 
are then one-dimensional, classified by their charge 
 $Q$, and it follows that without loss of generality we may take a solution $\newepsilon$ with definite charge.\footnote{Equivalently, 
 notice that if $\newepsilon$ solves the KSE then so does $\mathcal{L}_k \newepsilon$, which is then also a Killing spinor.}
Notice that the charge $Q$ of the Killing spinor depends on the residual local gauge freedom 
\eqref{residualgauge}, which acts on $\newepsilon$ as
\begin{align}
\newepsilon\, \rightarrow \, \ex^{\ii\gamma\varphi/2}\, \newepsilon~,
\end{align}
and thus shifts $Q\rightarrow Q+\gamma/2$. 

\subsection{Orbibundles on spindles}\label{sec:orbs}

Since a spindle is a singular space, with orbifold singularities at its poles, we need to be 
careful about what we mean by a ``regular gauge field'' and ``regular spinor'' on a spindle. We begin in this 
subsection with a global description of an Abelian gauge field on an orbifold Riemann surface, 
in terms of a connection on a principal $U(1)$-orbibundle, and then move 
on to spinors in section \ref{sec:spinors}.

In appendix A of \cite{Ferrero:2020twa}, $U(1)$-orbibundles over spindles were described
in terms of the quotient  of $S^3$ by a weighted $U(1)$ action on $\C^2\supset S^3$.
 This makes use
of the fact that a locally free\footnote{Meaning all isotropy groups are finite.} action of 
a compact Lie group $G$ on a manifold $M$ has orbit space $M/G$ which is naturally an 
orbifold, with the total space $M$ then being a principal $G$-orbibundle over $M/G$. 
Starting with $M=S^3$ gives a particular $U(1)$-orbibundle over 
the spindle $\Sigma=S^3/U(1)$, and different bundles may then be constructed by factoring through the 
Lens space quotient $S^3/\Z_\lambda$, where $\Z_\lambda\subset U(1)$. 
Here, we  instead present the more usual local 
description of orbibundles, and associated connections, in terms of covering spaces and gluing, specializing to the 
case of interest where $G=U(1)$ is fibred over a spindle. 

We begin with a local model near to one of the poles of the spindle (or a more general orbifold Riemann surface). 
Thus let $\C\times S^1$ be a trivial principal $U(1)=S^1$-bundle 
over $\C\cong \R^2$, and introduce complex coordinates 
$(z,\xi)$, where $|\xi|=1$ is a unit norm complex number parametrizing 
the circle fibre. The copy of $\C$ will be the covering space for 
$U\equiv \C/\Z_n$, where the generator of $\Z_n$ acts on 
$z\in \C$ via $z\mapsto \omega z$, with $\omega\equiv \ex^{2\pi \ii/n}$ 
being a primitive $n^{\mathrm{th}}$ root of unity. 
In general, to specify a principal $G$-orbibundle over $\R^k/\Gamma$, one 
 picks a 
homomorphism $h$ from the local orbifold group $\Gamma$ into the fibre group $G$. 
In the current setting this amounts to picking a ``charge'' $m\in\Z$:
\begin{align}
h: \Z_n\, \rightarrow\, U(1)\, , \quad \mbox{where}\quad h(\omega)\, = \, \omega^m\, . 
\end{align}
More precisely, since $\omega^n=1$, we have $m\in\Z_n$. 
A principal $G$-orbibundle over $\R^k/\Gamma$ is then 
defined as the quotient $B\equiv (\R^k\times G)/\Gamma$, where the action on 
$G$ is given by the homomorphism $h$. In our case this is 
\begin{align}\label{actiondiscappc}
B \, \equiv \, (\C\times S^1)/\Z_n\, , \quad \mbox{where}\quad \omega\cdot (z,\xi) \, = \, (\omega z, \omega^m \xi)\, .
\end{align}
The bundle projection map $\pi:B\rightarrow U=\C/\Z_n$ projects onto the first factor. Since 
the action of $h(\Gamma)$ on the fibre $G$ is free, it follows that 
the total space of $B$ will be smooth ({\it i.e.} the lifted action of $\Gamma$ using $h$ has no fixed points) 
if $h$ is injective. In our case, this is so precisely\footnote{If $\mathrm{hcf}(m,n)\ne1$ we can write $m=b m'$, $n=b n'$ for some $b\in\mathbb{Z}$. Then if we carry out the action of \eqref{actiondiscappc} $n'$ times, it takes $(z,\xi) \to (\omega^{n/b}z,\xi)$, which defines
a $\mathbb{Z}_b$ sub-action that leaves the fibre fixed.} when $\mathrm{hcf}(m,n)=1$. 

We may further analyse this local model by writing $\xi=\ex^{\ii\chi}$, 
$z=\rho\, \ex^{\ii \hat\phi}$, where $\rho\geq 0$ is a radial coordinate, 
and on the covering space $\C\times S^1$ we have $\Delta\chi=\Delta\hat\phi=2\pi$. 
We may geometrize a connection one-form for the $U(1)$ bundle by writing the metric
\begin{align}\label{CS1metric}
\diff s^2_{\C\times S^1} \, = \, (\diff \chi + A_{(0)})^2+ \diff \rho^2 + \ff(\rho)^2\diff\hat\phi^2\, ,
\end{align}
where $\ff(\rho)=\rho+O(\rho^2)$ near $\rho=0$ in order for this rotationally-invariant metric on $\C$ to be smooth, 
and importantly $A_{(0)}$ is an arbitrary  global one-form on $\C$ that is invariant 
under the $\Z_n$ action. For the case of interest in the paper we 
impose the stronger condition that 
\begin{align}\label{liederiv}
\mathcal{L}_{\partial_{\hat\phi}} A_{(0)}\, = \, 0\,,
\end{align}
so that 
rotations of $\C$ generated by $\partial_{\hat\phi}$ are a symmetry.  
The generator of $\Z_n$  acts on the angular coordinates as
\begin{align}\label{shiftangles}
(\chi,\hat\phi) \, \mapsto\, \left(\chi+\frac{2\pi m}{n}, \hat\phi+\frac{2\pi}{n}\right)\, .
\end{align}

We may then make the $SL(2,\Z)$ transformation of the torus parametrized by 
$(\chi,\hat\phi)$ given by
\begin{align}\label{SL2}
\begin{pmatrix}\psi \\ \phi \end{pmatrix} \, \equiv \, \begin{pmatrix} 1 & -m \\ 0 & 1\end{pmatrix}\begin{pmatrix}\chi \\ \hat\phi \end{pmatrix}\,,
\end{align}
with  $\Delta \psi =\Delta \phi=2\pi$ on the covering space. 
Notice that this  transformation 
is only defined for $\rho>0$, as the azimuthal 
coordinate $\hat\phi$ is not defined at the origin $\rho=0$ of $\C$.
Moreover, although so far $m\in\Z_n$, the transformation 
\eqref{SL2} requires picking a particular choice  of integer 
lift $m\in\Z$. This then introduces some redundancy/gauge freedom 
in the subsequent description, that we comment on further below.
The new coordinate $\psi=\chi-m \, \hat\phi$ is invariant under \eqref{shiftangles}, while the metric \eqref{CS1metric} 
becomes
\begin{align}\label{CS1metricbetter}
\diff s^2_{\C\times S^1} \, = \, (\diff \psi + A)^2+ \diff \rho^2 + \ff(\rho)^2\diff\phi^2\, .
\end{align}
Notice that now $\Z_n$ acts only on $\phi$:
\begin{align}
(\psi,\phi)\, \mapsto \, \left(\psi,\phi+\frac{2\pi}{n} \right)\,,
\end{align}
while the new connection one-form  is
\begin{align}\label{gcplusflat}
A \, = \, A_{(0)} + m \, \diff\phi\, .
\end{align}

The metric on the quotient $B=(\C\times S^1)/\Z_n$ is then similarly given by \eqref{CS1metricbetter}, but where 
$\psi$ has period $2\pi$ and $\phi$ has period $2\pi/n$. Because of the latter period, the connection term 
$m\, \diff \phi$ is a non-trivial flat connection on $(\C\setminus\{0\})/\Z_n$, {\it i.e.} it cannot be removed by any
non-singular gauge transformation. 
To match with the conventions in \eqref{spindlemetric}, on the quotient space we write 
$\phi=\varphi/n$, with $\varphi$ having period $2\pi$, so that $A=A_{(0)}+ \frac{m}{n}\diff\varphi$. 
This gauge field is not defined at the origin in $\C$, precisely because 
the azimuthal coordinates $\hat\phi$, $\varphi$ are not defined 
there. 

The above construction shows that we may equivalently view $U(1)$-orbibundles 
over $U=\C/\Z_n$ as specifying a flat connection that is pulled back from the unit circle in $(\C\setminus\{0\})/\Z_n$, with 
holonomy precisely $\exp\left(\ii \int_{S^1_\varphi} A_{\mathrm{flat}}\right)=\omega^m=h(\omega)\in U(1)$. Also notice that 
if $m=pn$ is an integer multiple $p$ of $n$, then $A=A_{(0)}-\ii g^{-1}\diff  g$ is gauge equivalent to the global connection form $A_{(0)}$ on\footnote{While $A=A_{(0)}$ is globally defined on $\C/\Z_n$, the large gauge
 transformation $g=\ex^{\ii p \varphi}$ is singular at the origin of $\C$.}
   $(\C\setminus\{0\})/\Z_n$, where $g=\ex^{\ii p \varphi}$ 
is a large gauge transformation on this circle ({\it i.e.} not homotopic to the identity if $p\ne 1$). This is another manifestation of the fact that $m\in\Z_n$.

With this local model to hand, it is now straightforward to glue two\footnote{We can similarly glue an arbitrary number of such local models together using
$U(1)$ gauge transformations; the spindle is special in that we can consider metrics with azimuthal symmetry as we are interested in here. We also note that the total space of the $U(1)$-orbibundles that we are constructing, when
$\mathrm{hcf}(m,n)=1$ for all local models, give smooth 3-dimensional manifolds known as Seifert manifolds.}  
copies to obtain a spindle. Associated with the north and south poles, we take local models $(\C \times S^1)/\Z_{n_N}$ and bundle data specified by $m_N \in \Z_{n_N}$, along  
with $(\C \times S^1)/\Z_{n_S}$ and bundle data specified by $m_S \in \Z_{n_S}$, respectively. The metrics in each patch on the total space of the bundle are as in \eqref{CS1metricbetter} and
given by
\begin{align}
\diff s^2_{(\C\times S^1)/\Z_{n_N}} \, &= \, \left(\diff \psi_N+ A_{(0)}^N + \frac{m_N}{n_N}\diff\varphi\right)^2+ \diff \rho_N^2 + \ff(\rho_N)^2\frac{1}{n_N^2}\diff\varphi^2\, ,\nn
\diff s^2_{(\C\times S^1)/\Z_{n_S}} \, &= \, \left(\diff \psi_S+ A_{(0)}^S + \frac{m_S}{n_S}\diff\varphi\right)^2+ \diff \rho_S^2 + \ff(\rho_S)^2\frac{1}{n_S^2}\diff\varphi^2\, ,
\end{align}
where we may use the same name for the coordinate $\varphi$ in each patch, with $\Delta\varphi=2\pi$, as this is identified in the gluing.  Here $\psi_N$, $\psi_S$ each have period $2\pi$, and 
$A^N_{(0)}$, $A^S_{(0)}$ are global one-forms in each of their respective covering space patches. 
The total gauge fields in each patch, $A^N  = A^N_{(0)} + \frac{m_N}{n_N}\, \diff\varphi$ and $A^S = A^S_{(0)} + \frac{m_S}{n_S}\, \diff\varphi$,
are defined only on an open disc in $\C\setminus\{0\}$, with $\rho_N,\rho_S>0$,  as discussed above, with the 
flat connection capturing the orbibundle data. 
We glue 
on the circles at some fixed values of $\rho_N,\rho_S>0$, reversing the orientation 
of the south patch before gluing, where here 
$A^N$ and $A^S$ may in general differ by a $U(1)$ gauge transformation $g:S^1_{\mathrm{glue}}\rightarrow U(1)$, 
of the form $g=\ex^{\ii p \varphi}$ with $p\in \Z$. 
That is, on the overlap
\begin{align}\label{A2A1p}
A^N \, = \, A^S -\ii g^{-1}\diff g \, = \, A^S + p\, \diff\varphi \, .
\end{align}

The total gauge field flux $F=\diff A$ through the spindle may then be computed using Stokes' theorem, being 
careful around the poles where the flat gauge fields are singular. This leads to an integral 
over two small circles around each pole, $S^1_{N, \epsilon}$, $S^1_{S, \epsilon}$,  
where $A_N^{(0)}$,  $A_S^{(0)}$ are non-singular and hence 
give zero as $\epsilon\rightarrow 0$, but the flat gauge fields contribute. We must also be 
careful 
 where we glue over $S^1_{\mathrm{glue}}$ and \eqref{A2A1} holds. 
Thus we compute
\begin{align}\label{fluxlambdap}
\frac{1}{2\pi}\int_\Sigma F & \, = \, -\frac{m_N}{n_N}+\frac{1}{2\pi}\int_{S^1_{\mathrm{glue}}}A^N - \left[-\frac{m_S}{n_S}+\frac{1}{2\pi}\int_{S^1_{\mathrm{glue}}}A^S\right]\nn
& \, = \,  p - \frac{m_N}{n_N}+\frac{m_S}{n_S} \, \equiv\ \frac{\lambda}{n_Nn_S}\, ,
\end{align}
where we have defined the integer $\lambda = -m_Nn_S+m_Sn_N+ p\, n_Nn_S\in\Z$. This 
shows that $U(1)$-orbibundles over a spindle have 
gauge field flux given by $\lambda/n_Nn_S$ with $\lambda \in \Z$.  We refer to the associated complex line orbibundles as $\mathcal{O}(\lambda)$. 

In fact $\lambda\in\Z$ specifies uniquely the orbibundle, up to isomorphism, and in particular it determines the local models near the orbifold points. 
To see this, note that using Bezout's Lemma it is immediate that for given 
$n_N$, $n_S$ it is possible to choose the $m_N$, $m_S$ with $p=0$ so that $\lambda=-m_Nn_S +m_Sn_N=1$. 
These $m_N$, $m_S$ are unique up to simultaneously shifting $m_N\mapsto m_N + kn_N$, $m_S\mapsto m_S + kn_S$, with $k\in\Z$, 
but this means that the mod reductions $m_N\in\Z_{n_N}$, $m_S\in \Z_{n_S}$  are unique, specifying 
the local model uniquely at each pole. Notice here that different choices of integer lifts $m_N\in\Z$, $m_S\in\Z$ 
in the $SL(2,\Z)$ transformation \eqref{SL2} may also then be absorbed into the choice of $p\in\Z$ in \eqref{A2A1p}. Indeed, the gluing \eqref{A2A1p}
may also be regarded as an $SL(2,\Z)$ transformation (see \eqref{psip} below). 
Having discussed the case $\lambda=1$, other integer values for $\lambda$
may be obtained simply by multiplying $m_N$, $m_S$ by $\lambda$. 

Notice from \eqref{A2A1p} that the non-singular gauge fields 
in each patch are related on the intersection by the gluing
\begin{align}\label{fracgt}
A^N_{(0)}\, = \, A^S_{(0)}+\frac{\lambda}{n_N n_S}\diff \varphi\, .
\end{align}
While this characterization of the orbibundle on the spindle is in some sense equivalent to what we have just discussed, it
is potentially confusing as one is gluing the gauge fields in each patch using fractional gauge transformations. 
By contrast,
we reiterate that the gauge fields $A_N$  and $A_S$, with their flat connection pieces, are glued using $U(1)$ gauge transformations on the overlap as 
in \eqref{A2A1p}. Nevertheless, in previous papers discussing spindles we note 
that smooth gauges for the gauge fields have been used at the poles, with 
the corresponding fractional transition function in \eqref{fracgt}. While \eqref{fracgt} is perfectly correct for a spindle, 
the discussion we have presented in this subsection is a precise global 
description of the construction and, moreover, also extends straightforwardly to more general orbifold Riemann surfaces.

 \subsection{Spinors on spindles}\label{sec:spinors}
 
In the previous section we have ``geometrized'' the R-symmetry gauge field, introducing the $U(1)$ fibre 
direction for which it is a corresponding connection. In the cases where the spindles arise in
gauged supergravity, with an uplift to $D=10$ or $D=11$ dimensions, 
 this is in any case part of the construction.  We elaborate on this 
 further in section \ref{sec:uplift}
 
 The total space of the $U(1)$-orbibundle is a Seifert fibred space, which is a smooth three-manifold $M_3$
 when $\mathrm{hcf}(m,n)=1$ for all the local models. The one-form 
 \begin{align}
 \eta \, \equiv \, \diff \psi + A\, ,
 \end{align}
  is a smooth global one-form on this three-manifold. Focusing on the case of the spindle with two orbifold points, in the above construction we defined this in patches, 
  and in particular on the overlap we have
  \begin{align}\label{psip}
  \psi_N \, = \, \psi_S - p\, \varphi\, ,
  \end{align} 
  which from \eqref{A2A1p} then ensures that $\diff\psi_N + A^N=\diff\psi_S + A^S$ on the overlap. 
  This is precisely the $U(1)$ gauge transformation $\ex^{\ii\psi_N}=\ex^{\ii\psi_S}\cdot \ex^{-\ii p\varphi}$. 
Near to the poles one should invert the $SL(2,\Z)$ transformation \eqref{SL2} to
  obtain the connection one-forms $\diff\chi_N+A^N_{(0)}$, $\diff\chi_S+A^S_{(0)}$, which are smooth one-forms on the covering 
  copies of~$\C$. Whenever one checks smoothness  of local expressions in the orbifold setting, one necessarily has to 
 resort to using patches and covering spaces in this way. We note in passing that 
 the total space $M_3$ is a Lens space, when the orbifold base $\Sigma$ is a spindle.
  
 By definition the Killing spinor $\zeta$ has charge $\tfrac{1}{2}$ under $A$. One can then introduce a complex 
 line bundle $L$, on which $A$ is a connection, so that the spinor is a section of 
 $S\otimes L^{1/2}$, where $S$ is the spin bundle on $\Sigma$.
 The vector field  $\partial_\psi$  on $M_3$ generates the $U(1)$ action, such 
that $\Sigma=M_3/U(1)$ is the quotient. Moreover, for any function 
or more generally tensor field on $M_3$ we may Fourier expand 
in eigenmodes of $\mathcal{L}_{\partial_\psi}$. A function of charge $r$ under 
$\partial_\psi$, with corresponding phase $\ex^{\ii r\psi}$, 
is then, in its dependence on $\Sigma$, precisely a 
section of $L^r$. 
One thus obtains a spinor on $\Sigma$ of charge 
$\tfrac{1}{2}$ under $A$ by starting with a spinor 
field on $M_3$ that has charge $\tfrac{1}{2}$ under $\mathcal{L}_{\partial_\psi}$. 
Indeed, notice that the tangent bundle of $M_3$ naturally splits as 
$T_{\partial_\psi}\oplus T_\Sigma$, and the spinor bundle 
on $M_3$ then likewise splits as a tensor product of the spinor bundle $S$ 
on $\Sigma$ with a spinor on the circle fibres of $M_3$. A
charge $\tfrac{1}{2}$ spinor under $\mathcal{L}_{\partial_\psi}$ then has phase 
$\ex^{\ii \psi/2}$, which is the anti-periodic spin structure 
around the circle fibres. The resulting spinor field on $M_3$ is then 
a genuine spinor, rather than a spinor charged under a gauge field. 
This is precisely how supersymmetry is realized in consistent truncations to gauged supergravity. 
The comments above apply whether $\Sigma$ is smooth or an orbifold, as long as 
one uses the appropriate patches and covering spaces in the orbifold setting when
gluing local descriptions together. 

With these preliminary global comments to hand, as in the previous 
subsection we may now introduce local coordinate patches and covering spaces in order to describe 
the Killing spinor $\zeta$. We  thus introduce coordinate 
patches $\mathcal{U}_{N}\cong \C/\Z_{n_N}$ and
$\mathcal{U}_{S}\cong \C/\Z_{n_S}$, containing the north and south poles, respectively, and analyse how the spinor $\newepsilon$ behaves near to the poles, as well as how it patches together at the equator, where  $\mathcal{U}_{N}$ and $\mathcal{U}_{S}$ overlap. 
We begin by considering $\mathcal{U}_N$, equipped with polar coordinates $(\rho_N\ge 0,\varphi)$ such that 
$\rho_N=0$ is the north pole.  We may also introduce corresponding
Euclidean coordinates $x_N + \ii y_N \equiv \rho_N\,  \ex^{\ii \varphi/n_N}$, 
identifying $\mathcal{U}_{N}\cong \C/\Z_{n_N}$. 
Recalling \eqref{gcplusflat} and \eqref{Agauge} we write the gauge field in this patch as 
a sum of a regular one-form plus a flat connection:
\begin{align}\label{gfieldnorth}
A^N\, = \, a^N_{(0)}(\rho_N)\diff\varphi+\frac{m_N}{n_N}\diff \varphi\,,
\end{align}
where  $m_N\in \mathbb{Z}$. We demand that $a^N_{(0)}(0)=0$ to ensure that first term on the right hand side is a well-defined one-form 
at the origin of polar coordinates (where $\diff\varphi$ is a singular one-form) and notice this condition fixes the residual gauge freedom in~\eqref{residualgauge}. 

Since we wish to describe spinors we introduce
an orthonormal frame $e^1_N, e^2_N$, such that 
$e^1_N(0)=\diff x_N$, $e^2_N(0)=\diff y_N$ agrees with the usual Euclidean frame at the north pole. 
We use gamma matrices  $\gamma_1=\sigma^1$, $\gamma_2=\sigma^2$, where $\sigma^a$ are Pauli matrices, and take the chirality
operator to be $\gamma_3\equiv-\ii\gamma_1\gamma_2=\sigma^3$. We may then write our spinor $\newepsilon$ in this patch and in the given frame in the form
\bea\label{epsilonN}
\newepsilon^N \, = \,  \begin{pmatrix} \newepsilon^N_+ \\ \newepsilon^N_-\end{pmatrix}~.
\eea
We will discuss the smoothness of $\newepsilon^N_\pm$ at the north pole momentarily.

The frame $e_N^1$, $e_N^2$ is not invariant under $k=\partial_\varphi$, so the charge of the spinor \eqref{epsilonN} under $k$ is not immediately manifest in the Euclidean frame. 
We can make it manifest by 
changing to a $k$-invariant frame, defined by 
\begin{align}\label{kframe}
\hat{e}^1 \, \equiv \, \diff\rho~, \qquad \hat{e}^2\, \equiv\, f(\rho)\, \diff\varphi~.
\end{align}
This is an orthonormal frame for the metric \eqref{spindlemetric}, where note that near the north pole this 
is a polar coordinate frame with $\hat{e}^1 = \diff\rho_N$, $\hat{e}^2 = \rho_N\, \diff\varphi/n_N + O(\rho_N^2)$. 
The two frames are related by the $U(1)\cong SO(2)$ rotation
\bea\label{rotmatnorth}
\begin{pmatrix} \hat{e}^1 \\ \hat{e}^2 \end{pmatrix}=
\begin{pmatrix} \cos \frac{\varphi}{n_N} & \sin \frac{\varphi}{n_N} \\ -\sin \frac{\varphi}{n_N} & \cos\frac{\varphi}{n_N}\end{pmatrix} 
\begin{pmatrix} e^1_N \\ e^2_N \end{pmatrix}
~.
\eea
When vectors are rotated by an orthogonal matrix $L^a{}_b=\exp(s^a{}_b)$, spinors  transform in the spin representation with $\tilde L=\exp(\tfrac{1}{4}s^{ab}\gamma_{ab})$. Hence
in the new  frame \eqref{kframe} the spinor $\newepsilon^N$ reads
\bea\label{spinorphases}
\hat\newepsilon^N \, = \,  \begin{pmatrix} \ex^{\ii \varphi/2n_N}\newepsilon_+^N \\  \ex^{-\ii \varphi/2n_N}\newepsilon_-^N\end{pmatrix}~.
\eea
Since the frame \eqref{kframe} is $k$-invariant, $\mathcal{L}_k \hat{e}^a=0$, we have $\mathcal{L}_k\hat\newepsilon^N
= \partial_\varphi\hat\newepsilon^N$.  Recall that in a fixed gauge of the form \eqref{Agauge} 
we take the spinor to have definite charge under $k$. 
More precisely,
in the patch $\mathcal{U}_N$ 
covering the north pole with gauge field \eqref{gfieldnorth} we take the charge of the spinor to be $Q_N$. It then follows that the 
spinor in the two frames takes the form
\begin{align}\label{hatnothatspin}
\newepsilon^N \, = \,  \begin{pmatrix}  \ex^{\ii \left(Q_N-\frac{1}{2n_N}\right) \varphi} f^N_+(\rho_N) \\  \ex^{\ii \left(Q_N+\frac{1}{2n_N}\right)\varphi}f_-^N(\rho_N)\end{pmatrix}~,
\qquad
\hat\newepsilon^N \, = \, \ex^{\ii Q_N\varphi}\begin{pmatrix} f^N_+(\rho_N)\\ f_-^N(\rho_N)\end{pmatrix}\,.
\end{align}

We now return to the issue of smoothness at the north pole. 
If the spinor $\zeta$ were \emph{uncharged} under the gauge field $A$, 
then by definition it is smooth at the north pole of the spindle precisely 
if its lift to the covering space $\C$, where $\varphi$ has 
period $2\pi n_N$, is smooth at the origin $\rho_N=0$. The expression for $\zeta^N$ in \eqref{hatnothatspin} 
is with respect to a smooth frame at the origin, and thus the spinor 
will be regular precisely if its components are smooth functions 
at the origin. Notice now that if the positive chirality component $\zeta^N_+$
is \emph{non-zero} at the origin, so $f^N_+(0)\neq 0$, then 
we must have $Q_N=1/2n_N$ for $\zeta^N_+$ to be smooth, as $\varphi$ is singular at the origin. 
Similarly, if instead $\zeta^N_-$ is non-zero at the origin we would instead deduce that 
$Q_N=-1/2n_N$.

However, the spinor $\zeta$ instead has charge $\tfrac{1}{2}$ under $A$ in the KSE as in \eqref{kseqnschem}, 
where in the north pole patch $A=A^N$ takes the form in \eqref{gfieldnorth}. As discussed earlier in this subsection, 
this charged spinor $\zeta$ arises via reduction of a spinor on $M_3$ that has charge $\tfrac{1}{2}$ 
under $\mathcal{L}_{\partial_\psi}$. Moreover, to examine regularity of this spinor 
at the origin of $\C$ we must invert the $SL(2,\mathbb{Z})$ transformation \eqref{SL2}, which 
removes the singular flat connection term $\frac{m_N}{n_N}\diff\varphi$ in the gauge field $A^N$. 
This leads to coordinates $(\rho_N,\hat\phi_N,\chi_N)$, with $\hat\phi_N$ and $\chi_N$ having period $2\pi$ in the covering space, 
with 
the twisted $\mathbb{Z}_{n_N}$ action:
\begin{align}\label{shiftanglesns}
(\chi_N,\hat\phi_N) \, \mapsto\, \left(\chi_N+\frac{2\pi m_N}{n_N}, \hat\phi_N+\frac{2\pi}{n_N}\right)\, .
\end{align}
After uplifting the spinor $\zeta^N$ to $M_3$ we thus obtain
\begin{align}\label{upliftreg}
\zeta^N\, \to \,  
\ex^{\ii\frac{\chi_N}{2}}\begin{pmatrix}  \ex^{\ii \left(Q_N-\frac{1}{2n_N}-\frac{m_N}{2n_N}\right) n_N\hat\phi_N} f^N_+(\rho_N) \\  \ex^{\ii \left(Q_N+\frac{1}{2n_N}-\frac{m_N}{2n_N}\right)n_N\hat\phi_N}f_-^N(\rho_N)\end{pmatrix}~.
\end{align}
This is then smooth precisely if the components 
are smooth functions at $\rho_N\to 0$. As in the paragraph above, we then immediately deduce that 
if the positive chirality (upper) component is 
non-vanishing at the north pole then we must have $Q_N-\frac{1}{2n_N}-\frac{m_N}{2n_N}=0$, and
similarly if the 
 negative  chirality (lower)
component is non-vanishing at the north pole then we must have $Q_N+\frac{1}{2n_N}-\frac{m_N}{2n_N}=0$.

A similar analysis may be carried out in the south pole patch $\mathcal{U}_S$, equipped with polar coordinates 
$(\rho_S \geq 0,\varphi)$ such that $\rho_S=0$ is the south pole. 
The corresponding Euclidean coordinates are $x_S+\ii y_S \equiv -\rho_S\, \ex^{-\ii \varphi/n_S}$, 
and we take
\begin{align}\label{gfieldsouth}
A^S\, = \, a^S_{(0)}(\rho_S)\diff\varphi+\frac{m_S}{n_S}\diff\varphi\,,
\end{align}
where $m_S\in \mathbb{Z}$ and we demand that $a^S_{(0)}(0)=0$ to ensure that first term on the right hand side is 
smooth one-form at the origin of polar coordinates 
The only difference now is that the $k$-invariant frame \eqref{kframe} 
has $\hat{e}^1=-\diff\rho_S$, $\hat{e}^2=\rho_S\, \diff\varphi/n_S + O(\rho_S^2)$. The analogous $U(1)\cong SO(2)$ 
rotation to \eqref{rotmatnorth} reads
\bea\label{rotmatsouth}
\begin{pmatrix} \hat{e}^1 \\ \hat{e}^2 \end{pmatrix}=
\begin{pmatrix} \cos \frac{\varphi}{n_S} & -\sin \frac{\varphi}{n_S} \\ \sin \frac{\varphi}{n_S} & \cos\frac{\varphi}{n_S}\end{pmatrix} 
\begin{pmatrix} e^1_S\\ e^2_S \end{pmatrix}
~,
\eea
where at the south pole we take $e^1_S(0)=\diff x_S$, $e^2_S(0)=\diff y_S$. With respect to the 
non-singular frame $e^a_S$, and the $k$-invariant frame \eqref{kframe}, 
a similar analysis to that at the north pole then leads to the respective spinors
\begin{align}\label{hatnothatspinsouth}
\newepsilon^S \, = \,  \begin{pmatrix}   \ex^{\ii \left(Q_S+\frac{1}{2n_S}\right) \varphi}f^S_+(\rho_S) \\  \ex^{\ii \left(Q_S-\frac{1}{2n_S}\right)\varphi} f_-^S(\rho_S)\end{pmatrix}~,
\qquad
\hat\newepsilon^S \, = \, \ex^{\ii Q_S\varphi}\begin{pmatrix} f^S_+(\rho_S)\\ f_-^S(\rho_S)\end{pmatrix}\,.
\end{align}
Also, after uplifting the spinor $\zeta^S$ to $M_3$ we obtain
\begin{align}\label{upliftregs}
\zeta^S\, \to \,  
\ex^{\ii\frac{\chi_S}{2}}\begin{pmatrix}  \ex^{\ii \left(Q_S+\frac{1}{2n_S}-\frac{m_S}{2n_S}\right) n_S\hat\phi_S} f^S_+(\rho_S) \\  \ex^{\ii \left(Q_S-\frac{1}{2n_S}-\frac{m_S}{2n_S}\right)n_S\hat\phi_S}f_-^S(\rho_S)\end{pmatrix}~,
\end{align}
with the components smooth functions at $\rho_S\to 0$.

Although 
the hatted spinors $\hat{\newepsilon}^N$, $\hat{\newepsilon}^S$ are defined 
using the same frame \eqref{kframe}, note that the former uses the gauge field $A^N$, while the latter uses the gauge field $A^S$.
As in \eqref{A2A1} these gauge fields are related by a $U(1)$ gauge transformation: 
\begin{align}\label{A2A1}
A^N \, = \, A^S -\ii g^{-1}\diff g \, = \, A^S + p\, \diff\varphi \, ,
\end{align}
with $p\in\mathbb{Z}$. 
Since the Killing spinor has charge $\tfrac{1}{2}$ under $A$, we correspondingly have the relation
\begin{align}
\hat\newepsilon^N \, = \, \ex^{\ii (p/2) \varphi}\, \hat\newepsilon^S\,.
\end{align}
This is satisfied by the expressions in \eqref{hatnothatspin}, \eqref{hatnothatspinsouth}  provided that $f_\pm^N(\rho_N)=f_\pm^S(\rho_S)$ on the overlap,
as well as
\bea\label{charges}
Q_N\, =\, Q_S + \frac{p}{2}~.
\eea

With these results in hand, we may now investigate the consequences of our assumption (ii), namely that the 
spinor is non-zero at the poles. At the north pole at least one of the chiral components 
must be non-zero. Let us assume for the moment that $\newepsilon^N_+(0)\neq 0$, i.e. $f_+^N(0)\ne0$. Then  from \eqref{upliftreg}, as already discussed regularity of the spinor $\newepsilon^N$ at the north pole  implies 
\begin{align}\label{qnqsp}
Q_N\, = \, \frac{1}{2n_N}+\frac{m_N}{2n_N}\,.
\end{align}
The negative chirality component of the uplifted spinor is then $\ex^{\ii\chi_N/2}\ex^{\ii \varphi/n_N}f_-^N(\rho_N)$, 
and for this to be regular at the north pole we must have $f_-^N(\rho_N)=0$ so that the 
spinor is necessarily \emph{chiral} at the pole. In fact we shall see in specific examples that 
$f_-^N(\rho_N)$ vanishes linearly in $\rho_N$, near to $\rho_N=0$, which implies that to leading order near the pole
$\ex^{-\ii\chi_N/2} \newepsilon_-^N\sim z_N\equiv \rho_N\,  \ex^{\ii\varphi/n_N}$, which is smooth.

Looking next at the south pole, there are two possibilities that follow from regularity of $\newepsilon^S$:
\begin{align}\label{qscases}
\mbox{Twist}: & \quad \ \newepsilon^S_+(0) \, \neq \, 0 \quad \Rightarrow \quad Q_S \, = \, -\frac{1}{2n_S}+\frac{m_S}{2n_S}~, \nonumber\\
\mbox{Anti-twist}: & \quad \ \newepsilon^S_-(0) \, \neq \, 0 \quad \Rightarrow \quad Q_S \, = \, +\frac{1}{2n_S}+\frac{m_S}{2n_S}~.
\end{align}
Notice that similarly to the north pole, if 
$\newepsilon^S_{\pm}(0)  \neq 0$ then correspondingly $\newepsilon^S_{\mp}(0)  = 0$. 

We now recall the formula for the total flux \eqref{fluxlambdap}. Using \eqref{charges} we
can rewrite this as
$\frac{\lambda}{n_N n_S} \, = \, 2(Q_N -Q_S)- \frac{m_N}{n_N}+\frac{m_S}{n_S}$
and hence using \eqref{qnqsp}, \eqref{qscases} we deduce that
\begin{align}\label{twistcases}
\mbox{Twist}: &\quad  \ \lambda \, = \, n_S + n_N~, \nonumber\\
\mbox{Anti-twist}: & \quad \  \lambda \, = \, n_S - n_N~.
\end{align}
These are precisely the two possibilities \eqref{fluxchi}, and imply that the gauge field 
$A$ is a connection on $\mathcal{O}(n_S\pm n_N)$, respectively. 
In the twist case, with the upper sign, this is the tangent bundle of 
the spindle $\Sigma=\mathbb{WCP}^1_{[n_S,n_N]}$. 

In reaching this conclusion we assumed that $\newepsilon^N_+(0)\neq 0$. 
If instead $\newepsilon^N_-(0)\neq 0$, a similar analysis leads to the two cases
\begin{align}\label{otherlambdas}
\lambda \, = \, -n_S \mp n_N~.
\end{align}
These are the twist and anti-twist cases \eqref{twistcases}, but with the opposite overall sign 
for the gauge field. Such solutions will be related to twist and anti-twist solutions
by charge conjugation, in the following sense. 
Recall that we took the gamma matrices to be $\gamma_1=\sigma^1$, 
$\gamma_2=\sigma^2$. The charge conjugation matrix\footnote{We can also run the same argument that follows
by replacing $B$ with $\tilde B\equiv B\gamma_3$, which satisfies $\tilde B^{-1}\gamma_a \tilde B = -\gamma_a^*$.}
is then $B\equiv\sigma^1$, which 
satisfies $B^{-1}\gamma_a B = \gamma_a^*$, where we define the charge 
conjugate spinor as 
\begin{align}
\newepsilon^c\, \equiv\, B \newepsilon^* \, = \, \begin{pmatrix}\newepsilon_-^* \\ \newepsilon_+^*\end{pmatrix}~.
\end{align}
If $\newepsilon$ satisfies \eqref{kseqnschem}, then it is straightforward to check 
that $\newepsilon^c$ satisfies the same equation with $A$ replaced by 
$A^c\equiv -A$ and $\mathcal{M}$ replaced by $\mathcal{M}^c$, where 
$\mathcal{M}^*=B^{-1} \mathcal{M}^c B$. 
Since our analysis did not depend on the precise form of $\mathcal{M}$, 
the solutions with $\lambda$ given by \eqref{otherlambdas} 
may equivalently be viewed as twist and anti-twist solutions 
\eqref{twistcases}, but solving the Killing spinor equation with 
$\mathcal{M}$ replaced by $\mathcal{M}^c$. It might be the case that there is a $\Z_2$ action on 
the fields such that $\mathcal{M}^c$ is the same function of the transformed fields as $\mathcal{M}$, 
i.e. charge conjugation leaves the Killing spinor equation \eqref{kseqnschem} invariant. 
However, either way one can effectively reduce to the two cases in 
\eqref{twistcases}.

The above analysis
showed that 
 the spinor is necessarily \emph{chiral}
at each of the two poles. The twist case is then compatible with having $\newepsilon_-\equiv 0$, 
so that the spinor is positive chirality. This is the case for the topological 
twist, where the spinor is chiral and moreover constant. 
On the other hand, for the anti-twist the spinor $\newepsilon$ is necessarily non-chiral, 
 with opposite chirality components vanishing at the two poles.  

It is illuminating to consider the global structure of spin$^c$ bundles we have constructed above. 
Recall that on any oriented two-manifold or orbifold, chiral spinors
are sections of the bundles
\bea
S^+\, = \, K^{1/2}~, \qquad S^- \, = \, K^{-1/2} \, = \, \Lambda^{0,1}\otimes K^{1/2}~.
\eea
Here $K=\Lambda^{1,0}$ is the canonical bundle, namely the cotangent bundle, 
and $\Lambda^{0,1}$ denote $(0,1)$-forms with respect to the canonical complex 
structure, specified by the orientation. For the spindle 
$K=\mathcal{O}(-n_S-n_N)$, which notice has a genuine square root 
only when $n_S+n_N$ is even. On the other hand, our Killing spinor 
is also charged under $A$, which is a connection on the line bundle 
$L\equiv \mathcal{O}(\lambda)$. Since the spinor has charge $\tfrac{1}{2}$, the chiral 
components $\newepsilon_\pm$ of $\newepsilon$ are globally sections of the bundles
\begin{align}
\newepsilon_+  : \ S^+\otimes L^{1/2} & \, = \, \mathcal{O}(-\tfrac{1}{2}(n_S+n_N))\otimes \mathcal{O}(\tfrac{1}{2}(n_S\pm n_N)) \nn
& \, = \, \begin{cases} 
\ \mathcal{O}(0) & \quad \quad \ \mbox{Twist}\\ \ \mathcal{O}(-n_N)&  \quad \quad \ \mbox{Anti-twist}\end{cases}~,\nn
\newepsilon_-  : \ S^-\otimes L^{1/2} & \, = \, \mathcal{O}(\tfrac{1}{2}(n_S+n_N))\otimes \mathcal{O}(\tfrac{1}{2}(n_S\pm n_N)) \nn & \, = \, \begin{cases} 
\ \mathcal{O}(n_S+n_N)& \quad  \mbox{Twist} \\ \ \mathcal{O}(n_S)& \quad \mbox{Anti-twist}\end{cases}~.
\end{align}
Notice that these are well-defined line bundles over the spindle, even though 
in general neither $S^\pm$ nor $L^{1/2}$ is. This is a consequence of the fact that 
$\lambda=n_S\pm n_N \equiv n_S + n_N$ mod 2, meaning that $A$ is a 
spin$^c$ gauge field. 
In the twist case  notice that $\newepsilon_+$ is a section of the trivial line bundle $\mathcal{O}(0)$. 

Before continuing, we briefly pause to contrast the detailed global analysis for the regularity of the
spin$^c$ spinors that we have just carried out with what has appeared in the literature in the context of specific solutions
\cite{Ferrero:2020laf,Faedo:2021nub,Ferrero:2020twa,Hosseini:2021fge,Boido:2021szx,Ferrero:2021wvk,Ferrero:2021ovq,Couzens:2021rlk}. In essence, these papers used patches with regular gauge fields $A^N_{(0)}$, $A^S_{(0)}$ that are related by
fractional gauge transformations as in \eqref{fracgt}. This is then combined with demanding smoothness of the unhatted spinors
given in \eqref{hatnothatspin}, rather than \eqref{upliftreg}. While this approach does in fact give the correct results for
a spindle, our new discussion here provides a precise global description.

\subsection{Topological twist}\label{sec:toptwist}

The ``topological twist'' is a particular case of the twist solutions. 
Specifically, one  solves the original Killing spinor equation 
\eqref{kseqnschem} by taking $\newepsilon_-\equiv 0$, so that 
$\newepsilon=\newepsilon_+$ is a chiral spinor. Since then
$\gamma_{12}\newepsilon = \ii\sigma^3\newepsilon = \ii \newepsilon$, 
equation \eqref{kseqnschem} reads\footnote{If instead $\newepsilon=\newepsilon_-$ has negative 
chirality, one instead takes $A=-\omega_{12}$, which is the upper sign 
solution in \eqref{otherlambdas} for which $A$ is a connection on the cotangent bundle 
of $\Sigma$, rather than tangent bundle.}
\bea
\diff \newepsilon + \tfrac{\ii}{2}(\omega_{12}-A)\newepsilon + \mathcal{M}\newepsilon \, = \, 0~.
\eea
Assuming that $\mathcal{M}\newepsilon=0$, this is solved by taking
\bea
A \, = \, \omega_{12}~, \qquad \newepsilon \, = \, \mathrm{constant}~.
\eea
Taking $A$ to be the spin connection $\omega_{12}$ immediately implies that this 
is a connection on the tangent bundle of  $\Sigma$, and hence 
\eqref{fluxchi} holds with the upper sign. Moreover, 
$\newepsilon=\newepsilon_+$ being nowhere zero is then necessarily a section of a trivial 
line bundle $\mathcal{O}(0)$. This is  clearly a special case 
of what we have called ``twist'' solutions in the previous subsection. 

However, more generally there are twist solutions that are not 
``topological twists'' in the above strict sense, as seen in \cite{Ferrero:2021wvk}. 
We will later see new examples of twist solutions with 
$\newepsilon_-$ not identically zero (but necessarily zero at the poles of the 
spindle), and similarly $\mathcal{M}\newepsilon\neq 0$ and $\newepsilon_+$ non-constant. 
We thus find it convenient to distinguish between topological twist 
and twist, although this distinction is not always made in the literature.

Finally, we note here that for the topological twist case we immediately
deduce that the vector bilinears vanish: $\newepsilon^\dagger\gamma_a\newepsilon=\newepsilon^\dagger\gamma_a\gamma_3\newepsilon=0$.
This is true not just for the two-sphere but also for a topological twist on an arbitrary Riemann surface.
By contrast, as we will see later in explicit examples, for the twist case on a spindle there is a vector bilinear constructed from the Killing spinors which gives the Killing vector of the spindle. This demonstrates, at the level of gauged supergravity, 
that supersymmetry for the topological twist and the twist on the spindle are realized in quite distinct ways. 

\subsection{Nowhere zero spinors}\label{sec:nowhere}

The analysis so far was quite general, and apart from 
$k=\partial_\varphi$ generating a symmetry of the solution, 
the only other assumption we made was that the Killing spinor $\newepsilon$ 
is non-zero at the poles of the spindle. In fact in almost all cases of interest
$\newepsilon$ is \emph{nowhere zero}, as 
we now briefly discuss.

Starting with a non-trivial solution to the Killing spinor equation \eqref{kseqnschem}, we can follow an argument given in \cite{Closset:2012ru}.
Assume that $\zeta(x_0)=0$ for some point $x_0$. Assuming the manifold is path-connected we can connect to any other point by a smooth curve $x(s)$. Along this curve \eqref{kseqnschem} has the schematic form $\frac{\diff}{\diff s}\zeta(x(s))=\tilde M(s)\zeta(x(s))$
for a smooth matrix $\tilde M$. Invoking uniqueness of first order linear ODEs, with $\zeta(x_0)=0$  we deduce that
$\zeta(x(s))=0$ and hence $\zeta\equiv 0$ identically, which contradicts our assumption of non-triviality. Thus, we deduce that
$\zeta$ is nowhere vanishing.

This result is immediately applicable to AdS$\times \Sigma$ solutions\footnote{The AdS$\times D$ solutions of e.g.
 \cite{Bah:2021mzw,Bah:2021hei,Couzens:2021tnv,Suh:2021ifj,Couzens:2021rlk}, where $D$ is topologically a disc, are constructed from the same family
of local solutions that we are considering in this paper. In these solutions the Killing spinors vanish as one approaches the boundary of the disc. However, since this is a boundary, in fact a singular boundary (since
the warp factor vanishes),
our argument does not apply. }
of gauged supergravity. After uplifting to
string theory and M-theory the supergravity metric takes the general form
\begin{align}\label{adsans}
\diff s^2 \, = \, \ex^{2\Delta}\diff s^2_{\mathrm{AdS}} + \diff s^2_{\mathrm{internal}}~,
\end{align}
where the warp factor  $\Delta$ is a function of the internal space 
coordinates, so as to preserve the symmetries of AdS. The argument given in the preceding paragraph can in fact be
immediately adapted to arbitrary supersymmetric solutions of the form \eqref{adsans}, utilizing the Killing spinor
equations for $D=10, 11$ supergravity. In particular one can conclude that the internal space Killing spinor $\eta$ 
is nowhere vanishing. In particular, for solutions where $\diff s^2_{\mathrm{internal}}$ contains a spindle, the restriction of $\eta$ to the spindle is also nowhere zero.

As somewhat of an aside it is interesting to note that in many cases the internal space Killing spinor $\eta$ 
for such supersymmetric solutions satisfies 
\begin{align}\label{etanorm}
\bar\eta\eta\, = \, \ex^{\Delta}~,
\end{align}
which implies that $\eta$ is nowhere zero. 
 For example, the equation \eqref{etanorm} holds for general supersymmetric AdS$_3$ D3-brane solutions and  M2-brane solutions
i.e. $\mathcal{N}=(2,0)$ AdS$_3$ solutions of type IIB with five-form flux and $\mathcal{N}=2$ AdS$_2$ solutions of $D=11$ supergravity
 with electric four-form flux, respectively. 
The same formula holds for the most general supersymmetric AdS$_5$ solutions of $D=11$ 
 supergravity \cite{Gauntlett:2004zh},  and for the most general $\mathcal{N}=2$ AdS$_4$ solutions of $D=11$ \cite{Gabella:2012rc}.  
A similar conclusion is reached for the $\mathcal{N}=(1,0)$ AdS$_3$ solutions of type IIB studied in \cite{Passias:2019rga}.

The analysis in the previous subsections can also be applied to rigid supersymmetric backgrounds that contain 
a spindle. For supersymmetric backgrounds of so-called new minimal 
supergravity in $d=3$ and $d=4$ dimensions, the Killing spinors 
are necessarily nowhere zero -- see,  e.g.  \cite{Closset:2012ru} and \cite{Dumitrescu:2012ha}, respectively, using the above
argument. Localization results for supersymmetric field theories with four supercharges in three and four dimensions have mainly been derived in this setting. However, we note that if one uses conformal supergravity 
the spinor can have zeroes\footnote{To see why the above argument can fail, consider the conformal Killing spinor equation
$(\delta_\mu^\nu-\frac{1}{d}\gamma_\mu\gamma^\nu)\nabla_\nu\zeta=\ldots$ , where the right hand side could contain other fields, and notice that $(\delta_\mu^\nu-\frac{1}{d}\gamma_\mu\gamma^\nu)$ is a projection.}; it would be interesting to know if supersymmetry on a spindle can be realised in this specific context
in a manner that is different from the twist and the anti-twist. Furthermore, it would also be of interest to know if such an alternative does exist, then could it ever arise with the rigid background arising as the conformal boundary of an AdS solution.

\subsection{Uplifted solutions}\label{sec:uplift}

Let us return to the general formula \eqref{fluxlambdap} for the flux of the gauge field strength through the spindle. 
For $\lambda=n_S\pm n_N$ in the twist and anti-twist cases, respectively, we may solve this by taking 
$p=0$ and $m_N=-1$, $m_S=\pm 1$. As discussed after equation \eqref{fluxlambdap} 
these choices are not unique, but $m_N\in\Z_{n_N}$, $m_S\in\Z_{n_S}$ \emph{are} unique, 
and this is the invariant data for the local models at each pole. Moreover, 
notice that $m=\mp 1$ precisely give the tangent and cotangent bundle actions 
at a pole: if the generator $\omega$ of $\Z_n$ acts as $z\mapsto \omega z$ on 
$\C$, then the one-form $\diff z\mapsto \omega\,  \diff z$, while the 
tangent vector $\partial_z\mapsto \omega^{-1}\partial_z$. 
With the above choices we then have that
\begin{align}
\tilde{A} \ \equiv\  A^N \, = \, A^S
\end{align}
is a \emph{global} one-form on the spindle minus both of its poles. We may also then write this near each pole as
\begin{align}\label{Agauges}
\tilde{A} \, = \, A^N_{(0)} -\frac{1}{n_N}\diff\varphi \, = \, A^S_{(0)} \pm \frac{1}{n_S}\diff\varphi\, ,
\end{align}
where recall that $A^N_{(0)}$ and $A^S_{(0)}$ are smooth one-forms in their respective patches, and 
in particular smooth at each pole, respectively. 
It then follows that $\tilde{A}$ is singular at \emph{both} poles, where from \eqref{Agauges} we deduce
\bea\label{Aatpoles}
\left.\tilde{A}\, \right|_{N}\, = \, -\frac{1}{n_N}\diff\varphi~, \qquad \left.\tilde{A}\, \right|_{S}\, = \, 
\begin{cases} \  
\displaystyle
+\frac{1}{n_S}\diff\varphi & \quad \mbox{Twist}\vspace{0.15cm}\\  \ 
\displaystyle
-\frac{1}{n_S}\diff\varphi & \quad \mbox{Anti-twist}\end{cases}~.
\eea
We shall make use of this gauge for the field theory analysis later in the paper. 
In particular, from \eqref{qnqsp} and \eqref{qscases} we see that 
$Q_N=Q_S=0$, and the spinor $\newepsilon$ is \emph{uncharged} 
in this gauge.

Recall that in sections \ref{sec:orbs} and \ref{sec:spinors} we 
``geometrized'' the R-symmetry gauge field by introducing 
a corresponding circle direction $\psi$, where after appropriate gluing 
one verifies that $\eta=\diff\psi+A$ 
is a global one-form on the total space $M_3$. This precisely 
happens when the Killing spinor equation \eqref{kseqnschem} arises by dimensionally reducing a 
$D=10$ or $D=11$ Killing spinor equation, associated with a solution of the form
\eqref{adsans}.  On uplifting, the Abelian R-symmetry gauge field $A$ becomes a connection in the internal 
space metric, with the gauge transformations we have discussed becoming coordinate transformations. 
The Killing vector $\partial_\psi$ then generates an isometry of the internal space. 
For example, to be concrete we may consider the original D3-brane spindle 
solutions in \cite{Ferrero:2020laf}. These are AdS$_3\times Y_7$ solutions of type IIB 
supergravity, where $Y_7$ is the total space of a regular Sasaki-Einstein 5-manifold $SE_5$
fibred over a spindle $\Sigma$. Here $SE_5$ by definition itself fibres over a K\"ahler-Einstein 
4-manifold $KE_4$, and fixing a point in the later space inside $Y_7$ is precisely our (Lens space) $M_3$, with $\psi$
being fibred both over the $KE_4$ and also over the spindle. The fibration 
over the $KE_4$ doesn't play any role in the regularity analysis we have been discussing thus far.

On the other hand, in  the lifted geometry one can phrase things more globally and invariantly. 
The $D=10$ or $D=11$ Killing spinor will form a representation of the isometry group, 
and 
if there is a single complex spinor then this statement means it will have definite 
charge $Q_i$ under the $i$th $U(1)$ isometry $\partial_{\varphi_i}$. Generically, in the examples discussed
in this paper, there are at least two such isometries. However, we can always choose a 
basis in which the spinor has charge $\tfrac{1}{2}$ under $\partial_{\varphi_1}\equiv \partial_\psi$, 
and is uncharged under $\partial_{\varphi_2}$. The latter is by definition then a flavour 
isometry. 
The gauge $\tilde{A}$ for $A$ in \eqref{Agauges}
is then precisely 
a choice of gauge for which $\partial_\varphi = \partial_{\varphi_2}$ is a flavour symmetry.  

In the known examples of the anti-twist and twist on the spindle,
as highlighted at the end of section \ref{sec:toptwist}, there is a vector bilinear for the Killing spinors of the
gauged supergravity solution that involves the Killing direction on the spindle. After uplifting, one finds this implies that there
is a vector bilinear for the higher-dimensional Killing spinors which is a linear combination of this Killing vector with the Killing vector $\partial_\psi$. The dual interpretation of this is that the lower-dimensional SCFT has an R-symmetry that is a linear combination
of the rotational symmetry on the spindle with the R-symmetry of the higher-dimensional SCFT upon which the SCFT is wrapped.


\section{D3-branes}\label{sec:D3}

We now re-examine the case of D3-branes wrapping spindles that was considered in \cite{Ferrero:2020laf,Hosseini:2021fge,Boido:2021szx}. For the 
cases associated with solutions of minimal $D=5$ gauged supergravity we recover exactly the same results as \cite{Ferrero:2020laf} where it was  shown, in particular,
that only the anti-twist case is allowed. For the cases associated with the $D=5$ STU gauged supergravity we will recover the results of 
\cite{Hosseini:2021fge,Boido:2021szx} for the anti-twist case, but we will show here, as a new result, that the twist case is also possible. 
In contrast to the anti-twist solutions, these new twist solutions are naturally interpreted as the near-horizon limits 
of D3-branes wrapping a spindle embedded as a holomorphic curve inside a Calabi-Yau four-fold. 
Moreover, taking a (formal) limit of the central charge for these twist solutions, in which the spindle becomes a two-sphere, 
we reproduce the central charge of the \emph{topologically} twisted solutions of \cite{Benini:2013cda}.
We will also discuss how a special sub-class of solutions are associated with M5-branes wrapped on
the product of a spindle with a Riemann surface. We also make some comparisons with field theory.

\subsection{The $D=5$ gauged supergravity model}\label{sec:D3gaugedSUGRA}

The $D=5$ STU, $U(1)^3$ gauged supergravity model that we shall consider has a Lagrangian given by (see  {\it e.g.} \cite{Behrndt:1998ns}) 
\begin{align}\label{d3lagoverall}
\mathcal{L}\, = \, \, &
\sqrt{-g}\Big[R-\mathcal{V}-\frac{1}{2} \sum_{I=1}^{3}\left(X^{(I)}\right)^{-2}\left(\partial X^{(I)}\right)^{2} 
-\frac{1}{4} \sum_{I=1}^{3}\left(X^{(I)}\right)^{-2}(F^{(I)})^{2}\Big]
\nn
&\qquad\ \  -F^{(1)} \wedge F^{(2)} \wedge A^{(3)}\,,
\end{align}
where $A^{(I)}$ are three $U(1)$ gauge fields, $I=1,2,3$, with field strengths $F^{(I)}=\diff A^{(I)}$. 
The three scalar fields $X^{(I)}>0$ satisfy the constraint $X^{(1)}\,X^{(2)}\,X^{(3)}=1$ and can be written
in terms
of two canonically normalized scalars $\varphi_{1}$, $\varphi_2$ via
\begin{align}
X^{(1)}\, = \, \ex^{-\frac{\varphi_1}{\sqrt{6}}-\frac{\varphi_2}{\sqrt{2}}}\,, \qquad
X^{(2)}\, = \, \ex^{-\frac{\varphi_1}{\sqrt{6}}+\frac{\varphi_2}{\sqrt{2}}}\,, \qquad
X^{(3)}\, = \, \ex^{\frac{2\varphi_1}{\sqrt{6}}}\,.
\end{align}
The potential is given by 
\begin{align}
\mathcal{V}\, = \, -4 \sum_{I=1}^{3}\left(X^{(I)}\right)^{-1}\,.
\end{align}

The model is a consistent truncation of $D=5$, $\mathcal{N}=2$ gauged supergravity coupled to two vector multiplets.
For a bosonic solution to preserve supersymmetry, we require that the variation of the gravitino and the two gauginos
both vanish
\begin{align}\label{vargrav}
0 \, = & \, \, \Big[\nabla_{\mu}-\frac{\ii}{2}\sum_{I=1}^3A^{(I)}_{\mu}+\frac{1}{6}\sum_{I=1}^3 X^{(I)}\,\Gamma_{\mu}+\frac{\ii}{24}\sum_{I=1}^3\,\left(X^{(I)}\right)^{-1}\,\left(\Gamma_{\mu}^{\,\,\,\nu\rho}-4\,\delta_{\mu}^{\nu}\,\Gamma^{\rho}\right)\,F^{(I)}_{\nu\rho}\Big]\,\epsilon\,,\nn
0 \, = & \, \, \Big[\slashed{\partial}\varphi_i-2\sum_{I=1}^3\partial_{\varphi_i} X^{(I)}+\frac{\ii}{2}\sum_{I=1}^3\partial_{\varphi_i} \left(X^{(I)}\right)^{-1}\,{F}^{(I)}_{\mu\nu}\Gamma^{\mu\nu}\Big]\,\epsilon\, \,,
\end{align}
where $\Gamma^\mu$ are $D=5$ Gamma matrices, $\nabla=\diff+\frac{1}{4}\omega^{ab}\Gamma_{ab}$, with $\omega^{ab}$ the 
Levi-Civita connection, 
and $\epsilon$ is a Dirac spinor. 
We can define the R-symmetry gauge field via
\begin{align}\label{rgfdef}
A^R\, \equiv \, \sum_{I=1}^3 A^{(I)}\,,
\end{align}
which is then normalized as in section \ref{sectwists}.

Any supersymmetric solution of this model can be uplifted on $S^5$ to give a supersymmetric solution of type IIB supergravity using the results of
\cite{Cvetic:1999xp}.\footnote{As pointed out in \cite{Gauntlett:2006ns} there is a typo in the five-form flux in \cite{Cvetic:1999xp}. The corrected 
expression may be found in \cite{Boido:2021szx}.}
In particular, the vacuum AdS$_5$ solution, which has unit radius, uplifts to
the AdS$_5\times S^5$ solution, dual to $d=4$ $\mathcal{N}=4$ SYM theory.
In the special case that two gauge fields are equal, and also reducing to one scalar field, for example $A^{(1)}=A^{(2)}$ and $X^{(1)}=X^{(2)}$ ({\it i.e.} $\varphi_2=0$), we can then uplift
on a class of $M_6$ geometries that arise in the 
most general AdS$_5\times M_6$ solutions of $D=11$ supergravity dual to $\mathcal{N}=2$ SCFTs \cite{Lin:2004nb},
to give supersymmetric solutions of $D=11$ supergravity \cite{Gauntlett:2007sm}. 
In particular, this includes uplifting on $M_6$ which are an $S^4$ fibration over a Riemann surface $\Sigma_g$ with genus $g>1$, that arise
in the Maldacena-N\'u\~nez solution describing M5-branes wrapped on a Riemann surface embedded in a Calabi-Yau two-fold \cite{Maldacena:2000mw}, or more generally a punctured Riemann surface as in \cite{Gaiotto:2009gz}. 
Setting three gauge fields equal and the scalars to zero leads to minimal gauged supergravity, and the solutions can be uplifted to type IIB on a Sasaki-Einstein manifold \cite{Buchel:2006gb}; this case is discussed in \cite{Ferrero:2020laf}. 

\subsection{Supergravity solution and global analysis}\label{sec:D3global}

We consider the supersymmetric solutions of $D=5$ STU gauged supergravity that were presented in \cite{Kunduri:2007qy}, and then recently analysed in
\cite{Hosseini:2021fge,Boido:2021szx}. Following\footnote{We have slightly changed the parametrization of the solutions as compared
to \cite{Boido:2021szx} in order to be more uniform with later sections.}
 \cite{Boido:2021szx}, 
the local form of the solutions is given by 
\begin{align}\label{D3sugrasol}
\diff s^2_5&\, = \, H^{1/3}\,\left[\diff s^2_{\text{AdS}_3}+\frac{1}{4P}\diff y^2+\frac{P}{H}\diff z^2\right]\, = \, H^{1/3}\,\left[\diff s^2_{\text{AdS}_3}+\diff s^2_{\Sigma}\right]\,,\nonumber\\
A^{(I)}&\, = \, \frac{y}{h_I}\diff z\,,
\qquad X^{(I)}\, = \, \frac{H^{1/3}}{h_I}\,,
\end{align}
where $\diff s^2_{\text{AdS}_3}$ is a unit radius metric and 
$h_I, H$ and $P$ are all functions of $y$ given by
\begin{align}\label{D3functions}
h_I\, = \, y+q_I\,,\qquad
H&\, = \, \prod_{I=1}^3 h_I\,,\qquad
P\, = \, H-y^2\,,
\end{align}
and $q_I$,  $I=1,2,3$ are constants.
Note that the field strength can be written as
\begin{align}\label{fluxD3branes}
\left(X^{(I)}\right)^{-2} F^{(I)}\, = \, \frac{2 q_I}{H^{1/2}}\widetilde{\text{vol}}_{\Sigma}\,,
\end{align}
where $\widetilde{\text{vol}}(\Sigma)$ is the volume form associated to the conformally rescaled spindle metric 
$\widetilde{\diff s^2}_{\Sigma}=H^{1/3}\diff s^2_{\Sigma}$.

We want to choose the parameters so that $y,z$ parametrize a compact spindle and with a positive warp factor, $H>0$.  Since $X^{(I)}>0$, we must have $h_I>0$. Also note that $P>0$ implies that $H>0$.
We assume that $P$ has three real roots,  
which we take to be $y_1<y_2<y_3$,  and we take $y\in[y_1,y_2]$ with $P\ge 0$.   Expanding near the boundaries of this interval, after setting 
$\varrho_i=2|y-y_i|^{1/2}$ for $i=1,2$,  we find
\begin{align}\label{kdefs}
\diff s^2_{\Sigma}\, \simeq \, \frac{1}{4\, | P'(y_i)|}\left(\diff \varrho_i^2+\kappa^2_i\,\varrho^2_i\,\diff z^2\right)\,, \quad \text{where} \quad
\kappa_i \, = \, \left|\frac{P'(y_i)}{y_i}\right|\,, \quad i =  1,2\,.
\end{align}
We would now like to remove the absolute value in the definition of $\kappa_i$.  
From the expression for $P$ it is clear that $P'(y_1)>0$ and $P'(y_2)<0$ and so, with $y_1<y_2$, we need to distinguish three cases corresponding to the three possible signs for $y_i$ ($i=1,2$). This is conveniently achieved by writing
\begin{align}\label{casesabc}
\eta_1\,y_1\, <\, 0\, ,\qquad
\eta_2\,y_2\, > \, 0\, ,
\end{align}
with the three cases determined by the signs 
\begin{align}\label{etasD3}
\text{Case A}:\quad (\eta_1, \eta_2)&\, = \, (-1,+1)\,,\nn
\text{Case B}: \quad(\eta_1, \eta_2)&\, = \, (+1,+1)\,,\nn
\text{Case C}:\quad (\eta_1, \eta_2)&\, = \, (+1,-1)\,.
\end{align}
Then, for all three cases we can write
\begin{align}\label{kapexp}
 \kappa_i\, = \, -\frac{\eta_i P'(y_i)}{y_i}\,,
\end{align}
and $\Sigma$ will be a spindle provided we impose
\begin{align}\label{D3periodicity}
\Delta z\, = \, \frac{2\pi}{\kappa_1 n_1}\, = \, \frac{2\pi}{\kappa_2 n_2}\,,
\end{align}
where $n_1,n_2\in \mathbb{N}$ with hcf$(n_1,n_2)=1$.  
As we will see, only cases A and B will in fact be realized.

We next demand that the three gauge field fluxes on the spindle are suitably quantized. Specifically, 
as in \cite{Boido:2021szx} with the $D=5$ spinors carrying charge $\tfrac{1}{2}$ with respect to each of the gauge fields, we demand
\begin{align}\label{D3fluxes}
Q_I&\, \equiv\, \frac{1}{2\pi}\int_{\Sigma}F^{(I)}\, = \, \left(\frac{y_2}{h_I(y_2)}-\frac{y_1}{h_I(y_1)}\right)\,\frac{\Delta z}{2\pi}\,,\nn
&\, \equiv \, \frac{p_I}{n_1\,n_2}\,,\qquad p_I\in \mathbb{Z}\,.
\end{align}
After writing
\begin{align}\label{prootsexp}
P\, = \, (y-y_1)\,(y-y_2)\,(y-y_3)\,,
\end{align} 
and comparing with the expression for $P$ in \eqref{D3functions}, one
can now solve \eqref{D3periodicity} and \eqref{D3fluxes} for $y_{1,2}$,  $q_{1,2,3}$ and $\Delta z$ in terms of 
the spindle data $n_1, n_2$ and the quantized fluxes $p_{1,2,3}$. We will give the explicit solution below.

However, without knowing the explicit solution 
one can compute the Euler number of the spindle, $\chi(\Sigma)$.  
Using the metric
\begin{align}
\diff s^2_{\Sigma}\, = \, \frac{1}{4P}\diff y^2+\frac{P}{H}\diff z^2\,,
\end{align}
one finds
$\sqrt{g}_{\Sigma}\,R_{\Sigma}=2\, [H^{-3/2}(PH'-HP')]'$
from which we calculate
\begin{align}
\chi(\Sigma)\, = \, \frac{1}{4\pi}\int_{\Sigma}\sqrt{g}_{\Sigma}\,R_{\Sigma}\,\diff y\,\diff z \, = \, \frac{\Delta z}{2\pi}\left(\frac{P'(y_1)}{|y_1|}-\frac{P'(y_2)}{|y_2|}\right)\, = \, \frac{n_1+n_2}{n_1\,n_2}\,,
\end{align}
where we have used $H(y_i)=y_i^2$, \eqref{kdefs}--\eqref{D3periodicity}, and the result is independent of which case is considered.  
Equivalently, using $\ex^1=\diff y/(2P^{1/2})$, $\ex^2=(P^{1/2}/H^{1/2}) \diff z$ as an orthonormal frame, the spin connection
for $\diff s^2_\Sigma$ is given by $\omega^{12}_{\Sigma}= [H^{-3/2}(PH'-HP')] \diff z$. When evaluated at the roots we have
\begin{align}\label{spinatroots}
\omega_{\Sigma}^{12}(y_1)=-\frac{\diff\varphi}{n_1},\qquad
\omega_{\Sigma}^{12}(y_2)=+\frac{\diff\varphi}{n_2},
\end{align}
and hence $\chi(\Sigma)=(2\pi)^{-1}\int_\Sigma \diff\omega^{12}_\Sigma$ gives the same result. Now $\omega^{12}_{\Sigma}$
is an $SO(2)\cong U(1)$ connection on the tangent bundle, and it is interesting to highlight that the behaviour at the poles
given in \eqref{spinatroots} is in alignment with the comments in section \ref{sec:uplift}.

In addition,  we can also compute the total R-symmetry flux $Q^R\equiv (Q_1+Q_2+Q_3)$.  To that end,  one can use the identities 
\begin{align}
P'&\, = \, h_1\,h_2+h_2\,h_3+h_1\,h_3-2 y\,,\nonumber \\
y_i^2&\, = \, h_1(y_i)\,h_2(y_i)\,h_3(y_i)\, , \qquad (i=1,2)\,,
\end{align}
in \eqref{kapexp},  which allows us to rewrite \eqref{D3periodicity} as
\begin{align}\label{deltaphiD3}
\frac{\Delta z}{2\pi}\,\Big[2-y_i\sum_{I=1}^3\frac{1}{h_I(y_i)}\Big]&\, = \, \frac{\eta_i}{n_i}\,,\quad
\text{(no sum on $i$)}\,.
\end{align}
One can now immediately prove that the R-symmetry flux is given by 
\begin{align}\label{D3sumcharges}
Q^{R}\, = \, \frac{\eta_1\,n_2-\eta_2\,n_1}{n_1 n_2}
\qquad
\Leftrightarrow
\qquad
p_1+p_2+p_3\, = \, {\eta_1\,n_2-\eta_2\,n_1}\,.
\end{align}
Case B is the anti-twist that was considered in \cite{Hosseini:2021fge,Boido:2021szx}. 
We now see that the same local solutions can also potentially realize the twist in cases A and C, although 
we will ultimately find that  case C is not possible, while case A is.

The central charge for the $(0,2)$ SCFT dual to the AdS$_3\times \Sigma$ solution 
can be expressed in the form
\begin{align}
c\, = \, \frac{3}{2\pi}\,\Delta z\,(y_2-y_1)\,N^2\,,
\end{align}
as in \cite{Boido:2021szx}.
To express this in terms of spindle data we now need to solve for 
$y_{1,2}$ and $\Delta z$ in terms of $n_i$ and $p_1, p_2$ (recall that $p_3$ is given by the constraint \eqref{D3sumcharges}), as outlined below \eqref{prootsexp}.
The explicit solution given in \cite{Boido:2021szx} is valid for case B.
With a little effort one can rewrite the solution of \cite{Boido:2021szx} by reintroducing $p_3$ and then using the signs $\eta_{1,2}$ introduced in \eqref{etasD3} to deduce expressions that are valid for all three cases. Specifically,  for the constants $q_I$ we find
\begin{align}\label{qexpsex}
q_1&=\frac{8}{s^3}\,p_2\,p_3\,(n_1+\eta_2\,p_2)\,(n_1+\eta_2\,p_3)\,(n_2-\eta_1\,p_2)\,(n_2-\eta_1\,p_3)\,, \nonumber\\
q_2&=\left. q_1\right|_{p_1\leftrightarrow p_2}\,, \nonumber\\ 
q_3&=\left. q_1\right|_{p_1\leftrightarrow p_3}\,, 
\end{align}
where we have introduced 
\begin{align}\label{sdefD3}
 s \, \equiv \, n_1^2+n_2^2-( p_1^2+p_2^2+p_3^2)\,.
\end{align}
The roots $y_{1,2}$ are given by
\begin{align}\label{exxsol}
y_1\, = \, &-\frac{8\eta_2}{s^3}p_1p_2p_3 (n_1+\eta_2\,p_1)(n_1+\eta_2\,p_2)(n_1+\eta_2\,p_3)\,,\nonumber\\
y_2\, = \, &\left. y_1\right|_{n_1\leftrightarrow n_2,\,\,\eta_2\to -\eta_1}\,,
\end{align}
in terms of which $y_3$ is fixed using $\sum_{I=1}^3(y_I+q_I)=1$, and explicitly we have 
\begin{align}
y_3\, = \, &\frac{8\eta_1 \eta_2}{s^3} \prod_{I=1}^3 \, (n_1 + \eta_2 p_I)(n_2 - \eta_1 p_I)\, .
\end{align}
 Finally, we have
\begin{align}\label{delphi}
\frac{\Delta z}{2\pi}\, = \, \frac{s}{2n_1 n_2(\eta_2\, n_1+\eta_1\, n_2)}\,.
\end{align}
With these results we can now express the central charge, for all three cases, in the simple form
\begin{align}\label{centchge}
c\, = \, \frac{6\,p_1p_2p_3}{n_1 n_2\,s} N^2\,.
\end{align}

We now examine whether it is possible to choose the spindle data $n_{1,2}$ and quantized fluxes $p_I$, satisfying \eqref{D3sumcharges},  such that cases A, B and C are in fact realized as \emph{bona fide} solutions. Recall we have assumed from the start that
$n_{1,2}>0$. We also need to check that  
\begin{align}\label{D3positivity}
h_I>0\, , \qquad 
y_1<y_2<y_3\, ,\qquad 
\eta_1\,y_1<0\, ,\qquad
\eta_2\,y_2>0\, ,
\end{align}
where the first condition comes from the fact that the scalar fields $X^{(I)}>0$, and
the last two come from \eqref{casesabc}. To analyse these constraints one can utilize the expressions given in 
\eqref{qexpsex}, \eqref{exxsol}. Doing so, we find that case C is not possible, leaving just cases A and B.

For case B, which realizes the anti-twist with $\eta_1=\eta_2=1$, we find that the above conditions \eqref{D3positivity} are all satisfied provided that 
\begin{align}
\text{Anti-twist}: \qquad p_1,p_2>0\, ,\qquad p_3 \, = \,  n_2-n_1-p_1-p_2>0\, .
\end{align}
We can then show that, in particular, the central charge given in \eqref{centchge} is positive, $c>0$.
This is the same result that was obtained in \cite{Boido:2021szx}.

We also find that case A, which realizes the twist with $\eta_1=-1$, $\eta_2=1$, is possible provided that 
\begin{align}
\text{Twist}: \qquad n_2>n_1\quad \text{and two $p_i$ are positive}\, .
\end{align}
Notice here that since $p_1+p_2+p_3=-n_1-n_2$, given that two of the $p_i$ are positive, the third must then be negative. 
Moreover, these conditions imply that the central charge is positive, $c>0$. We also note that
after \emph{formally} setting $n_1=n_2$ in the expression for the central charge in case A
with $\eta_1=-1$, $\eta_2=1$, we recover the expression for the central charge obtained in \cite{Benini:2013cda} for D3-branes wrapping a two-sphere with a topological twist. However, we emphasize that
 this is a formal statement, since {\it e.g.} from \eqref{exxsol} we see that $y_1=y_2$ in this limit, giving a double root  for the polynomial $P$, which
 is incompatible with having a compact spindle. Instead the solutions of \cite{Benini:2013cda}
 may be recovered from our local solutions by utilizing a scaling limit, as shown in appendix~\ref{appA_D3}.

It is interesting to ask whether case A or B can be realized in minimal gauged supergravity. We will obtain solutions of minimal gauged supergravity after setting $p_1=p_2=p_3=1/3(\eta_1 n_2-\eta_2 n_1)$, and we immediately find that only the anti-twist, case B, is allowed, confirming what was already shown in
\cite{Ferrero:2020laf}. It is also worth noting that if we formally set one of the fluxes to zero, say $p_1=0$, we will obtain a local solution that preserves twice as much supersymmetry. However, we can immediately see from the expression for the central charge, which vanishes in this limit, that this cannot make sense. Indeed, from \eqref{exxsol} we see that we obtain a double root for the polynomial $P$ in this limit, and
hence we will not obtain a compact spindle. 

Another interesting limit to consider is when two charges are equal, for example $p_1=p_2\neq 0$,  
with  $p_3=-p_1-p_2+(\eta_1\,  n_2-\eta_2\,  n_1)$. For case B, the anti-twist, we just need $p_1>0$ and $n_2-n_1>2p_1$, as shown in 
\cite{Boido:2021szx}. We also find that case A, the twist, is possible when we have $n_2>n_1$ and $p_1>0$.
As mentioned earlier, these solutions can be uplifted on $M_6$ to give solutions of $D=11$ supergravity,
as we discuss further
in section \ref{sec:M5anom}.

\subsection{Killing spinors}\label{sec:D3Kill}
In this subsection we explicitly present the solutions to the Killing spinor equations~\eqref{vargrav}.
Some details of the derivation are presented in appendix~\ref{app:c}.

We begin by choosing the following orthonormal frame for the metric \eqref{D3sugrasol}:
\begin{align}
\ex^a&\, = \, H^{1/6}\,\bar \ex^a,\quad\text{$a=0,1,2$}\, ,\qquad \ex^3\, =\, \frac{H^{1/6}}{2P^{1/2}}\,\diff y\,, \qquad
\ex^4\, = \, \frac{P^{1/2}}{H^{1/3}}\,\diff z\,,
\end{align}
where $\bar\ex^a$ is an orthonormal frame for the unit radius metric on
 AdS$_3$. 
We use five-dimensional gamma matrices
\begin{align}\label{gammas5d}
\Gamma_a&\, = \, \beta_a\otimes \gamma_3\,, \qquad 
\Gamma_3 \, = \, 1\otimes \gamma_1\,, \qquad
\Gamma_4\, = \, 1\otimes \gamma_2\,,
\end{align}
where $\beta_0=\ii \sigma^2$, $\beta_1=\sigma^1$, $\beta_2=\sigma^3$
are three-dimensional gamma matrices for AdS$_3$ directions,  while $\gamma_1=\sigma^1$,  $\gamma_2=\sigma^2$ are two-dimensional gamma matrices for the spindle directions,  with chirality matrix $\gamma_3=-\ii\,\gamma_1\gamma_2=\sigma^3$.  

We next introduce two-component AdS$_3$ Killing spinors $\vartheta$ satisfying 
\begin{align}\label{ads3ks}
\nabla_a\vartheta\, = \, \frac{1}{2}\,\beta_a \vartheta\, ,
\end{align}
whose two solutions $\vartheta^{(A)}_+$,  $A=1,2$,  are described explicitly in appendix~\ref{AdSspinors}.
We then find that the two $D=5$ Killing spinors $\epsilon^{(A)}$,  $A=1,2$,  solving \eqref{vargrav} are
\begin{align}\label{epdec}
\epsilon^{(A)} \, =\, \vartheta^{(A)}_+ \otimes \newepsilon\,,
\end{align} 
where the two-component spinor $\newepsilon$ on the spindle is given by
\begin{align}\label{d3kses}
\newepsilon\, \equiv \, \begin{pmatrix}
\newepsilon_+\\
\newepsilon_-
\end{pmatrix}
\, = \, \frac{\ex^{\ii z}}{\sqrt{2}H^{1/6}}\,
\begin{pmatrix}
[\sqrt{H}-y]^{1/2}\\
-[\sqrt{H}+y]^{1/2}
\end{pmatrix} \, = \, H^{1/12}\, \ex^{\ii z}\begin{pmatrix} \sin\frac{\alpha}{2} \\ -\cos\frac{\alpha}{2}\end{pmatrix}\, ,
\end{align}
and the function $\alpha=\alpha(y)$ satisfies
\begin{align}\label{alphaD3branes}
\sin\alpha \, = \, \frac{P^{1/2}}{H^{1/2}}\, , \qquad \cos\alpha \, = \, \frac{y}{H^{1/2}}\, .
\end{align} 
In particular, the Killing spinors given by \eqref{epdec} satisfy the single projection condition
\begin{align}\label{spinorprojectionD3}
M\epsilon^{(A)}\, = \, -\ii \, \epsilon^{(A)}~, \quad \mbox{where}\quad M\, \equiv\, \cos\alpha\, \Gamma_{34}+\ii \sin\alpha\, \Gamma_3\, ,
\end{align}
and one can verify that $M^2=-1$. We thus preserve half of the supersymmetries, corresponding to 
$d=2$ $\mathcal{N}=(0,2)$ supersymmetry. 
One also sees from \eqref{d3kses} that the squared norm of the Killing spinor is
\begin{align}
\newepsilon^\dagger\newepsilon\, = \, H^{1/6}\, , 
\end{align}
which is nowhere vanishing, and proportional to the square root of the warp factor, as expected from the general discussion in 
section \ref{sec:nowhere}. Moreover, on the spindle we have the Killing vector bilinear
\begin{align}
\frac{\ii}{\zeta^\dagger\zeta}(\zeta^\dagger\gamma^m\gamma_3\zeta)\partial_m\, = \, \partial_z\, ,
\end{align}
where here $x^m=(y,z)$ are coordinates on the spindle. 
On the $D=5$ spacetime we then deduce
\begin{align}\label{susyalg}
(\bar\epsilon^{(A)}\Gamma^\mu\epsilon^{(B)})\partial_\mu \, = \, (\bar\vartheta^{(A)}_+\beta^a\vartheta^{(B)}_+)\partial_a+\ii(\bar\vartheta^{(A)}_+\vartheta^{(B)}_+)\partial_z\,,
\end{align}
which reveals the way in which supersymmetry is being realized for the AdS$_3\times\Sigma$ solutions. In particular, recalling
the explicit Killing spinors on AdS$_3$ (see appendix~\ref{AdSspinors}) we conclude that, schematically, the last term in \eqref{susyalg} 
implies that the azimuthal rotations on the spindle are acting as an R-symmetry for the 
$d=2$, $\mathcal{N}=(0,2)$ supersymmetry algebra. By contrast this term is absent in the supersymmetry algebra for AdS$_3\times\Sigma_g$ solutions associated with a topological twist on the Riemann surface $\Sigma_g$. 

Using the expression \eqref{d3kses} for the Killing spinors, together with \eqref{casesabc}, one can immediately 
deduce 
\begin{itemize}
\item Case A, Twist: $\newepsilon_+(y_1)=0$ and $\newepsilon_+(y_2)=0$,  with $\newepsilon_-\neq 0$ everywhere.
\item Case B, Anti-Twist: $\newepsilon_-(y_1)=0$ but $\newepsilon_-$ is non-vanishing everywhere else, while $\newepsilon_+(y_2)=0$ but $\newepsilon_+$ is non-vanishing everywhere else.
\end{itemize}
The corresponding R-symmetry flux in the two cases is
\begin{align}\label{Rfluxsummary}
Q^R\,  \equiv\, \frac{1}{2\pi}\int_\Sigma F^R \, = \,\frac{p_1+p_2+p_3}{n_1n_2} \,  = \, \begin{cases} 
\ -\displaystyle\frac{n_2+n_1}{n_1n_2} & \mbox{Case A, Twist}\, , \\ 
\quad \displaystyle\frac{n_2-n_1}{n_1n_2} & \mbox{Case B, Anti-Twist}\, .\end{cases}
\end{align}
Comparing to the general analysis in section \ref{sec:spinors}, recall that 
to obtain the R-symmetry flux in \eqref{twistcases} we assumed 
$\zeta_+^N(0)$, $\zeta_+^S(0)\neq 0$, while 
if instead $\zeta_-^N(0)$, $\zeta_-^S(0)\neq 0$ one obtains 
the upper sign in \eqref{otherlambdas}. The latter then agrees with the 
R-symmetry flux in \eqref{Rfluxsummary}, where $n_1$ 
may be identified with either of $n_S$ or $n_N$ (and then respectively $n_2$ with either $n_N$ or $n_S$). On the other hand, for case B since 
$\zeta_+$ is non-vanishing at $y_1$, we identify $y_1=y_N$, 
$y_2=y_S$, and the R-symmetry flux in \eqref{Rfluxsummary} agrees 
with the anti-twist in equation \eqref{twistcases}.

Note that in the gauge we are using for $A^{(I)}$, the R-symmetry gauge field 
is not regular at either pole located at $y=y_1$ or $y=y_2$.
However, using \eqref{deltaphiD3} we may obtain a smooth gauge field at $y=y_1$ via the gauge transformation $A^R\to A^R_{(0)}=A^R-2\, \diff z+\frac{\eta_1}{n_1} \diff\varphi$. Here we have introduced $\varphi=\frac{2\pi}{\Delta z} z$, with $\Delta\varphi=2\pi$, and 
the subscript on $A^R_{(0)}$ indicates that this gauge field is smooth at the pole, 
as in the notation and discussion in section~\ref{sectwists}. In particular the shift 
by $\frac{\eta_1}{n_1}\diff\varphi$ is equivalent to an $SL(2,\Z)$ transformation in the covering space.
This gauge transformation  will act on the spinor, which is only charged under the R-symmetry, and replace the $\ex^{\ii z}$ phase in
\eqref{d3kses}
with $\ex^{\tfrac{\ii \eta_1\varphi}{2n_1}}$. Noting that the frame we are using is the analogue of the hatted frame in section 2 
(which is the frame that is not regular at $y=y_1$), we see that this behaviour is
in agreement with \eqref{hatnothatspin}. Recall here that for the twist case A, $\zeta_+(y_1)=0$ and $\eta_1=-1$, while 
for the anti-twist case B, $\zeta_-(y_1)=0$ and $\eta_1=+1$. It then follows from 
 \eqref{d3kses} that in the locally Euclidean frame and in the new gauge, for the twist case A we have 
 $\zeta_+\sim \ex^{-\tfrac{\ii \varphi}{n_1}}\rho_1$, where $\rho_1$ is the radial coordinate at $y=y_1$, while 
 for the anti-twist case B we instead have 
 $\zeta_-\sim \ex^{+\tfrac{\ii \varphi}{n_1}}\rho_1$.  In both cases we have a simple holomorphic zero of the spinor to leading order 
 at the pole $\rho_1=0$. Similar comments apply, \emph{mutatis mutandis}, to the behaviour of the spinors at the other pole located
at $y=y_2$.

Finally, as already noted, the spinor \eqref{d3kses} is charged under azimuthal 
rotations of the spindle, due to the phase $\ex^{\ii z}$. 
We consider the following two-parameter family of gauge transformations on the gauge fields
\begin{align}\label{D3tildeAR}
\tilde{A}^{(I)}\, \equiv\, A^{(I)}+\sa^I\diff z \, = \, \rho_I(y)\,\diff\varphi\, ,
\end{align}
where the $\sa^I$ are constants satisfying the constraint
\begin{align}\label{aIconD3}
\sa^1+\sa^2+\sa^3\, = \, -2\,.
\end{align}
The latter ensures that the R-symmetry gauge field transforms to 
$\tilde{A}^R = A^R - 2\, \diff z$, which cancels the
phase $\ex^{\ii z}$ in the spinor as the spinor has charge $\tfrac{1}{2}$ under $A^R$. 
In \eqref{D3tildeAR} we have  introduced the functions
\begin{align}\label{rhoIdef}
\rho_I(y)\, \equiv\, \left(\frac{y}{h_I}+\sa^I\right)\frac{\Delta z}{2\pi}\, ,
\end{align}
which will play a role in the field theory analysis in the next subsection (where we will also see that
the gauge choice of all $\sa^I$ equal, namely $\sa^I=-2/3$, has a notable property).
From \eqref{D3tildeAR} one can check that
\begin{align}\label{D3tildeARpoles}
\left.\tilde A^R\, \right|_{y=y_1}&\, = \, -\frac{\eta_1}{n_1}\diff\varphi\,, \qquad
\left.\tilde A^R\, \right|_{y=y_2}\, = \, -\frac{\eta_2}{n_2}\diff\varphi\,.
\end{align}
This agrees with the general conditions for a local gauge
in which the spinor is uncharged with respect to azimuthal rotations that we derived earlier \eqref{Aatpoles},
where in particular recall that our twist case A is a charge conjugate 
of the twist considered in section~\ref{sectwists}.

\subsection{Comparison with field theory}\label{sec:D3anom}

As in \cite{Boido:2021szx}, there is a natural field theory interpretation\footnote{See \cite{Ferrero:2020laf} for a more general interpretation for the solutions
of minimal gauged supergravity and also section \ref{sec:M5anom} for an M5-brane interpretation of a sub-class.}
  of the solutions we have constructed. 
One begins with $d=4$, $\mathcal{N}=4$ SYM theory, which is dual to the AdS$_5\times S^5$ vacuum of the $D=5$, 
STU $U(1)^3$ gauged supergravity theory. One then compactifies this field theory on the spindle 
$\Sigma$, together with magnetic fluxes $Q_I$ \eqref{D3fluxes} for the three Abelian gauge fields. The existence of the
 AdS$_3$ supergravity solutions suggests that this compactified theory flows to a $d=2$, $(0,2)$ SCFT in the IR. 
There is an important distinction here between the twist and anti-twist solutions, that we discuss further 
in section \ref{fincoms}.

Following \cite{Ferrero:2020laf,Boido:2021szx}, we consider the anomaly polynomial of $\mathcal{N}=4$ SYM, with background gauge fields fluxes $F^{(I)}$ for the $U(1)^3\subset SO(6)$ Abelian global symmetry group.  In the large $N$ limit this reads
\begin{align}
\mathcal{A}_{4\diff}\, = \, c_1(F^{(1)})\,c_1(F^{(2)})\,c_1(F^{(3)})\,\frac{N^2}{2}\,.
\end{align}
Here $c_1(F^{(I)})$ denote the first Chern classes of the $U(1)$ bundles with gauge field curvatures $F^{(I)}$. 
Recalling the expression \eqref{D3tildeAR} for the gauge field, we next introduce connection one-forms 
\begin{align}
\mathscr{A}^{(I)}\, = \, \rho_I(y)\left(\diff \varphi+\mathcal{A}_{\mathcal{J}}\right)\,,
\end{align} 
with curvature
\begin{align}
\mathscr{F}^{(I)}\, = \, \diff \mathscr{A}^{(I)}\, = \, \rho'_I(y)\,\diff y \wedge \left(\diff \varphi+\mathcal{A}_{\mathcal{J}}\right)+\rho_I(y)\,\mathcal{F}_{\mathcal{J}}\,,
\end{align}
where $\mathcal{F}_{\mathcal{J}}=\diff \mathcal{A}_{\mathcal{J}}$. Here $\mathcal{A}_\mathcal{J}$ gauges the 
azimuthal symmetry of the spindle. In particular, this means that setting $\mathcal{A}_\mathcal{J}=0$ 
the  $\mathscr{A}^{(I)}$ reduce to the gauge fields $\tilde{A}^{(I)}$ in \eqref{D3tildeAR} {\it i.e.} in a gauge where the Killing spinor is 
uncharged under azimuthal rotations of the spindle generated by $\partial_\varphi$.  This implies the flux conditions
\begin{align}\label{rhodiffs}
\frac{1}{2\pi}\int_\Sigma \mathscr{F}^{(I)}\,  = \, \rho_I(y_2)-\rho_I(y_1)\, = \, \frac{p_I}{n_2 n_1}\qquad (I=1,2,3)\,,
\end{align}
together with the constraints \eqref{D3tildeARpoles} at the poles:
\bea\label{rhoatpolesx1x2}
\sum_{I=1}^3\rho_I(y_1)\, = \, -\frac{\eta_1}{n_1}~, \qquad \sum_{I=1}^3\rho_I(y_2)\, = \, -\frac{\eta_2}{n_2}~.
\eea
The general solution to the constraints \eqref{rhodiffs}, \eqref{rhoatpolesx1x2} for $\rho_I(y_i)$, with $p_3$ determined by \eqref{D3sumcharges}, 
 is
\begin{align}
\rho_1(y_1)&\, = \, \alpha_1\,, \quad
\rho_2(y_1) \, = \, \alpha_2\,, \quad
\rho_3(y_1) \, = \, -\frac{\eta_1}{n_1}-\alpha_1-\alpha_2\,,\nonumber\\
\rho_1(y_2)& \, = \, \frac{p_1}{n_1\,n_2}+\alpha_1\, , \quad
\rho_2(y_2)\, = \, \frac{p_2}{n_1\,n_2}+\alpha_2\, , \nonumber\\
\rho_3(y_2)& \, = \,  -\frac{\eta_2}{n_2}-\frac{p_1}{n_1\, n_2}-\frac{p_2}{n_2 n_1}-\alpha_1-\alpha_2\, ,
\end{align}
where $\alpha_1$, $\alpha_2$ are {\it a priori} arbitrary. Notice that the AdS$_3$ supergravity solution 
gave explicit functions \eqref{rhoIdef} for $\rho_I(y)$, although these are determined up to a choice of $\sa^I$ satisfying the constraint \eqref{aIconD3}. 
This is equivalent to the freedom of choosing $\alpha_1$, $\alpha_2$ above. 
However, we shall see momentarily that the central charge of the theory 
is independent of $\alpha_1$, $\alpha_2$. 

The curvature form $\mathscr{F}^{(I)}$ defines a corresponding first Chern class 
$c_1(\mathcal{L}_I) = [\mathscr{F}^{(I)}/2\pi]$, and similarly we define 
$c_1(\mathcal{J})=[\mathcal{F}_\mathcal{J}/2\pi]$. We then write
\begin{align}
c_1(F^{(I)})\, = \, \Delta_I\,c_1(R_{2\diff})+c_1(\mathcal{L}_I)\,,
\end{align}
with the trial R-charges $\Delta_I$ constrained to satisfy 
 $\Delta_1+\Delta_2+\Delta_3=2$, so that the superpotential of $\mathcal{N}=4$ SYM has R-charge 2, in $\mathcal{N}=1$ language. The 
$d=2$ anomaly polynomial is then obtained by integrating $\mathcal{A}_{4\diff}$ over $\Sigma$:
\begin{align}\label{anomaly2d}
\mathcal{A}_{2\diff}\, = \, \int_{\Sigma}\mathcal{A}_{4\diff}\, =  & \, \left[\left(\Delta_1\Delta_2 I_3+\Delta_2\Delta_3I_1+\Delta_3\Delta_1I_2\right)c_1(R_{2\diff})^2\right.\nonumber \\
&\left.\quad +\left(\Delta_1I_4+\Delta_2I_5+\Delta_3I_6\right)c_1(R_{2\diff})c_1(\mathcal{J})+I_7c_1(\mathcal{J})^2\right]\frac{N^2}{2}\, .
\end{align}
Here 
\begin{align}
I_I & \, = \, \rho_I(y_2)-\rho_I(y_1) \, \equiv\, [\rho_I]^{y_2}_{y_1}~, \quad I_4\, = \, [\rho_2\rho_3]^{y_2}_{y_1}~, \quad I_5\, = \, [\rho_3\rho_1]^{y_2}_{y_1}\,, \nonumber\\
 I_6 & \, = \, [\rho_1\rho_2]^{y_2}_{y_1}\,, \quad I_7 \, = \, [\rho_1\rho_2\rho_3]^{y_2}_{y_1}\, .
\end{align}

The $d=2$ superconformal R-symmetry extremizes the trial function
\begin{align}\label{centralfromtrace}
c_{\text{trial}}\, = \, 3\,\text{tr}\,\gamma_3 R^2_{\text{trial}}\,,
\end{align}
where
\begin{align}
R_{\text{trial}}\, = \, R_{2\diff}+\varepsilon\,\mathcal{J}\,.
\end{align}
From \eqref{anomaly2d} we find 
\begin{align}
c_{\text{trial}} & \, = \, 3\, \Big[\left(\Delta_1\Delta_2I_3+\Delta_2\Delta_3 I_1+\Delta_3\Delta_1I_2\right)+\left(\Delta_1I_4+\Delta_2I_5+\Delta_3I_6\right)\varepsilon \nn 
& \qquad \ \ +I_7\varepsilon^2\Big]N^2\,.
\end{align}
Extremizing over $\Delta_1$, $\Delta_2$ and $\varepsilon$, we obtain the extremal values
\begin{align}
\Delta_1^{\star}&\, = \, -2\, \frac{p_1}{s}(\eta_2\,n_1+p_1) -\varepsilon^*\, \alpha_1 
 \,,  && \Delta_2^{\star}\, = \, -2\, \frac{p_2}{s}(\eta_2\,n_1+p_2) -\varepsilon^*\, \alpha_2  \,, \nonumber\\
\varepsilon^{\star}&\, = \, 2\, \frac{n_2n_1}{s}(\eta_1\,n_2+\eta_2\,n_1)\,,
 &&\,\,c^{\star}\, = \, \frac{6\, p_1p_2p_3}{n_2n_1\,s}N^2\,,\end{align}
where note that $c^{\star}$ agrees with the general supergravity result \eqref{centchge}. 
Notice that neither of the critical values $\varepsilon^\star$ nor $c^\star$ depend 
on the parameters $\alpha_1$, $\alpha_2$, but the critical 
R-charges $\Delta_I^\star$ do. This is a gauge freedom, and in fact 
the parameters $\alpha_1$, $\alpha_2$ may 
be removed from $c_{\text{trial}}$ via the change of variable 
$\Delta_I\to \Delta_I-\varepsilon\,\alpha_I$, $I=1,2$ (where $\Delta_3=2-\Delta_1-\Delta_2$). 
We also note that by choosing $\alpha_1,\alpha_2$ with $\sa^I=-2/3$ in \eqref{rhoIdef} we find the
extremal values $\Delta_1^{\star}=\Delta_2^{\star}=\Delta_3^{\star}=2/3$, the notable property we alluded to earlier
(just below \eqref{rhoIdef}).

In the limit $n_1=n_2=1$,  when the spindle reduces to a two-sphere,  we should be able to 
\emph{formally}
compare our results with (3.12) of reference \cite{Benini:2013cda}, where their $\eta_{\Sigma}=2$, $d_G=N^2$,  $\mathfrak{g}=0$ and $\kappa=1$.  We find
\begin{align}\label{c_BB}
c_{BB}\,= \, \frac{24\,a_1a_2 (1+a_1+a_2)}{1+4(a_1^2+a_2^2+a_1a_2+a_1+a_2)}N^2\, .
\end{align}
This indeed agrees with the $n_1=n_2=1$ limit of $c^\star$ in the twist case A, identifying 
$a_i=p_i/2$. 
Notice the configurations of \cite{Benini:2013cda} are a topological twist for the two-sphere, and 
arise here as a limit of our more general twist case A spindle configurations. Note, however, that the supergravity solutions
of \cite{Benini:2013cda} do not arise as a limit of the solutions for the spindle, but instead can be obtained from the 
local solutions \eqref{D3sugrasol} after taking a certain scaling limit, as discussed in appendix \ref{appA_D3}. 

\subsection{M5-branes}\label{sec:M5anom}
As noted above there is an alternative field theory interpretation of the solutions of $D=5$ gauged supergravity that
we have constructed in the special case that two of the magnetic fluxes are equal. 
Start with a $d=4$, $\mathcal{N}=2$ field theory that
is dual to an AdS$_5\times M_6$ solution of $D=11$ supergravity of the type discussed in \cite{Lin:2004nb,Gaiotto:2009gz}. One can then compactify on the spindle
$\Sigma$, together with magnetic fluxes $Q_I$ \eqref{D3fluxes} for the three Abelian gauge fields, with two equal. The existence of our
corresponding 
AdS$_3$ supergravity solutions, uplifted to $D=11$ using the results of \cite{Gauntlett:2007sm},
indicates that this compactified theory flows to a $d=2$, $\mathcal{N}=(0,2)$ SCFT in the IR. 
  
Focusing, for simplicity,
on the explicit case of the AdS$_5\times M_6$ solution given in \cite{Maldacena:2000mw}, associated with $N$ M5-branes wrapping a Riemann
 surface of genus $g$, and setting $p_1=p_2$ for definiteness, it was shown in \cite{Boido:2021szx} for the anti-twist solutions that the central charge of the 
 $d=2$ SCFT can be written in the form  
\begin{align}\label{centchgem5}
c\, = \, \frac{8\,p_1^2p_3}{n_1 n_2\,s}\, (g-1)N^3\,,
\end{align}
where 
\begin{align}
\text{Anti-Twist}:\quad p_1\, = \, p_2>0\,,\qquad p_3\, = \, n_2-n_1-2p_1>0\,.
\end{align}
In fact the analysis in \cite{Boido:2021szx} can immediately be adapted to the twist case and we find a central charge 
as in \eqref{centchgem5}, but now with 
$p_1=p_2>0$, $n_2>n_1$ and 
\begin{align}
\text{Twist}:\quad  p_1\, = \, p_2>0,\quad n_2>n_1\quad\Rightarrow\quad p_3\, = \, -n_1-n_2-2p_1<0\,.
\end{align}
Furthermore, the field theory analysis of \cite{Boido:2021szx} using the large $N$ limit of the M5-brane anomaly polynomial
which was used to recover the gravity central charge for the anti-twist case, can also be adapted for the case of the twist and again we find exact 
agreement.

All of these configurations preserve $\mathcal{N}=(0,2)$ supersymmetry. Configurations with 
$\mathcal{N}=(0,4)$ or $\mathcal{N}=(2,2)$ supersymmetry are linked with setting $p_1=p_2=0$ or $p_3=0$,
associated, in the twist class, with M5-branes wrapping a product of $\Sigma\times \Sigma_g$ in a Calabi-Yau three fold or
a product of two Calabi-Yau two-folds, respectively. However, we do not have supergravity solutions with compact spindles
for either of these values as one can conclude from the vanishing of the central charge.

 It is also interesting to point out that if we \emph{formally} set $n_1=n_2=1$ in the twist class we find
\begin{align}\label{centchgem52}
c\, = \, \frac{8\,p_1^2}{1+3p_1} (g-1)N^3\,,
\end{align}
and this should be compared with the central charge of the $S^2\times \Sigma_g$ $D=11$ supergravity solutions, first found in
section 3.4 of \cite{Gauntlett:2007sm} and subsequently discussed in \cite{Benini:2013cda}. In the notation of the latter reference one should set $\kappa_1=1$, $\kappa_2=-1$, 
$\eta_1=2$, $\eta_2=2(g-1)$, $z_1=2a_1+1$ and $z_2=2 a_2-1$, and then set $a_2=0$ ({\it i.e.} $z_2=-1$). One then finds the central charge
\begin{align}
c_{BB}\,  = \, \frac{8a_1^2}{1+3a_1}(g-1)N^3\, .
\end{align}
This agrees with \eqref{centchgem52}, identifying $a_1=p_1$. 

\section{M2-branes}\label{sec:M2}

In this section we re-examine M2-branes wrapping spindles, generalizing earlier work in \cite{Ferrero:2020twa, Ferrero:2021ovq, Couzens:2021rlk}. 
The analysis is more involved than for the D3-brane case studied in section~\ref{sec:D3}, and our treatment will 
correspondingly be less comprehensive. However, a key point is that there again exist new twist solutions in addition
to the previously known anti-twist solutions, with the twist solutions 
naturally interpreted as near-horizon limits of M2-branes wrapping 
a holomorphically embedded spindle inside a Calabi-Yau five-fold. 

\subsection{The $D=4$ gauged supergravity model}\label{sec:M2gaugedSUGRA}

The $D=4$, $U(1)^4$ gauged supergravity model that we shall consider has a Lagrangian given by (see  {\it e.g.} \cite{Donos:2011pn})
\begin{align}\label{m2lagoverall}
\mathcal{L}\, = \, \sqrt{- g}  \Big[R-\mathcal{V}-\frac{1}{2} \sum_{I=1}^{4}\left(X^{(I)}\right)^{-2}\left(\partial X^{(I)}\right)^{2}
-\frac{1}{4} \sum_{I=1}^{4}\left(X^{(I)}\right)^{-2}\left(F^{(I)}\right)^{2}\Big]\,.
\end{align}
Here $A^{(I)}$,  $I=1,...,4$,  are four $U(1)$ gauge fields, with field strengths $F^{(I)}=\diff A^{(I)}$. The four scalar fields $X^{(I)}>0$, which satisfy the constraint $X^{(1)}\,X^{(2)}\,X^{(3)}\,X^{(4)}=1$, can be written in terms of 
three canonically normalized scalars $\varphi_{1,2,3}$ via
\begin{align}
X^{(1)}&\, = \,  \ex^{+\frac{\varphi_1}{2}+\frac{\varphi_2}{2}+\frac{\varphi_3}{2}}\,,\quad 
X^{(2)}\, = \, \ex^{+\frac{\varphi_1}{2}-\frac{\varphi_2}{2}-\frac{\varphi_3}{2}}\,,\nn 
X^{(3)}&\, = \, \ex^{-\frac{\varphi_1}{2}+\frac{\varphi_2}{2}-\frac{\varphi_3}{2}}\,,\quad 
X^{(4)}\, = \, \ex^{-\frac{\varphi_1}{2}-\frac{\varphi_2}{2}+\frac{\varphi_3}{2}}\,,
\end{align}
in terms of which the potential can be written
\begin{align}
\mathcal{V}\, = \, -2\,\sum_{i=1}^3\cosh(\varphi_i)\, = \, -\sum_{I<J}X^{(I)} X^{(J)}\,.
\end{align}

This model is a truncation of an $\mathcal{N}=2$ gauged supergravity theory coupled to three vector multiplets, with three axion fields set to zero.
A solution of our model with $F^{(I)}\wedge F^{(J)}=0$ will be a solution of the $\mathcal{N}=2$ theory, and can be uplifted to $D=11$ supergravity on an $S^7$ using the results of \cite{Cvetic:1999xp}. 
In particular, the vacuum AdS$_4$ solution, which has unit radius, uplifts to
the AdS$_4\times S^7$ solution, dual to $d=3$ ABJM theory.

A solution will preserve supersymmetry provided that we can find solutions to the Killing spinor 
equations
\begin{align}\label{susym2}
0 \, = & \, \, \Big[\nabla_{\mu}-\frac{\ii}{4}\sum_{I=1}^4 A^{(I)}+\frac{1}{8}\sum_{I=1}^4 X^{(I)} \,\Gamma_{\mu}+\frac{\ii}{16}\sum_{I=1}^4 (X^{(I)})^{-1}{F}^{(I)}_{\nu\rho}\Gamma^{\nu\rho}\Gamma_{\mu}\Big]\,\epsilon\,,\nn
0\, = & \, \, \Big[\slashed{\partial}\varphi_i-\,\sum_{I=1}^4\partial_{\varphi_i}X^{(I)}+\frac{\ii}{2}\,\sum_{I=1}^4 \partial_{\varphi_i}(X^{(I)})^{-1}{F}^{(I)}_{\mu\nu}\Gamma^{\mu\nu}\Big]\,\epsilon\, ,
\end{align}
where $\Gamma^\mu$ are $D=4$ Gamma matrices, $\nabla=\diff+\frac{1}{4}\omega^{ab}\Gamma_{ab}$, with $\omega^{ab}$ the 
Levi-Civita connection, 
and $\epsilon$ is a Dirac spinor. 
We can define the R-symmetry gauge field via
\begin{align}\label{rgfdefm2}
A^R\, \equiv \, \frac{1}{2}\sum_{I=1}^4 A^{(I)}\,,
\end{align}
which is then normalized as in section \ref{sectwists}.

\subsection{Supergravity solution and global analysis}\label{sec:M2global}
We consider an AdS$_2\times \Sigma$ solution of this theory with four independent magnetic charges, which was introduced in \cite{Gauntlett:2006ns} as a solution of $D=11$ supergravity (after uplifting on $S^7$), 
and revisited in \cite{Ferrero:2021ovq,Couzens:2021rlk} from a purely four-dimensional point of view.  
The metric,  gauge fields and scalars of this solution read 
\begin{align}\label{4chargemagneticAdS22}
\diff s^2_4&\, = \,H^{1/2}\Big[ \frac{1}{4}\diff s^2_{\text{AdS}_2}+\frac{1}{P}\diff y^2+\frac{P}{4H}\diff \z^2\Big]
\, = \, H^{1/2}\,\Big[\frac{1}{4}\diff s^2_{\text{AdS}_2}+\diff s^2_{\Sigma}\Big]\,,\nn
A^{(I)}&\, =\, \frac{y}{h_I}\diff z\,, \qquad
X^{(I)}\, = \, \left(\frac{h_1h_2 h_3 h_4}{h_I^4}\right)^{1/4}\,,
\end{align}
where {$\diff s^2_{\text{AdS}_2}$ is a unit radius metric and $h_I$, $H$ and $P$ are all functions of $y$ given by 
\begin{align}\label{functions4chargeAdS2}
h_I\, = \, y+q_I\, , \qquad
H\, = \, h_1h_2h_3h_4\,, \qquad
P\, = \,H-4y^2\,,
\end{align}
and $q_I$, $I=1,2,3,4$ are constants. Note that \eqref{4chargemagneticAdS22} is a solution of the equations of motion only when all $h_I$ have the same sign,  either positive or negative. 
Note that we can write the field strength as
\begin{align}
\left(X^{(I)}\right)^{-2} F^{(I)}\, = \, \frac{2 q_I}{H^{1/2}}\widetilde{\text{vol}}_{\Sigma}\,,
\end{align}
where $\widetilde{\text{vol}}(\Sigma)$ is the volume form associated to the conformally rescaled spindle metric 
$\widetilde{\diff s^2}_{\Sigma}=H^{1/2}\diff s^2_{\Sigma}$.

We want to choose the parameters so that $(y,z)$ parametrize a compact spindle, with positive warp factor $H>0$ as well
as $X^{(I)}>0$. It is technically difficult to be as explicit as we were for the case of the solutions of $D=5$ gauged supergravity but
we can still make significant progress. We assume that the quartic polynomial $P$ has four distinct real roots 
$y_0<y_1<y_2<y_3$,  and we take $y\in [y_1,y_2]$ with $P\ge 0$ in this range. Expanding near the boundaries of this
interval and setting $\varrho_i=2|y-y_i|^{1/2}$ for $i=1,2$ we find
\begin{align}
\diff s^2_{\Sigma}\, = \, \frac{1}{|P'(y_i)|}\,\left(\diff \varrho_i^2+\kappa_i^2\varrho_i^2\diff z^2\right)\,,
\quad  \text{where} \quad
\kappa_i\, = \, \left|\frac{P'(y_i)}{8y_i}\right|\,,\qquad i=1,2\, .
\end{align}
With $P'(y_1)>0$ and $P'(y_2)<0$, we again distinguish three cases corresponding to the three possible signs for $y_i$ ($i=1,2$), by
setting
\begin{align}\label{casesabc2}
\eta_1\,y_1\, <\, 0\, ,\qquad
\eta_2\,y_2\, > \, 0\, ,
\end{align}
with 
\begin{align}\label{etasm2}
\text{Case A}:\quad (\eta_1, \eta_2)&\, = \, (-1,+1)\,,\nn
\text{Case B}: \quad(\eta_1, \eta_2)&\, = \, (+1,+1)\,,\nn
\text{Case C}:\quad (\eta_1, \eta_2)&\, = \, (+1,-1)\,.
\end{align}
Then, for all three cases we can write
\begin{align}\label{kapexpm2}
 \kappa_i\, = \, -\frac{\eta_i P'(y_i)}{8y_i}\,,
\end{align}
and $\Sigma$ will be a spindle provided we impose
\begin{align}\label{M2periodicity}
\Delta z\, = \, \frac{2\pi}{\kappa_1 n_1}\, = \, \frac{2\pi}{\kappa_2 n_2}\,,
\end{align}
where $n_1,n_2\in \mathbb{N}$ with hcf$(n_1,n_2)=1$.  As in the previous section we then can immediately
determine the Euler number of the spindle, finding $\chi(\Sigma)=(n_1+n_2)/(n_1 n_2)$ as expected.

We now demand that the four gauge field fluxes on the spindle are suitably quantized. Observing from \eqref{susym2}
that the $D=4$ spinors carry charge $\tfrac{1}{4}$ with respect to each of the gauge fields, we demand 
\begin{align}
Q_I&\, = \, \frac{1}{2\pi}\int_{\Sigma} F^{(I)}\, = \, \left(\frac{y_2}{h_I(y_2)}-\frac{y_1}{h_I(y_1)}\right)\frac{\Delta z}{2\pi}\nn
&\equiv\frac{2p_I}{n_1 n_2}\,, \qquad p_I\in\mathbb{Z}\,.
\end{align}
We can now compute the total R-symmetry flux defined as
$Q^R\equiv \frac{1}{2}(Q_1+Q_2+Q_3+Q_4)$, which is normalized as in section \ref{sectwists}.  
To that end,  one can use the identities 
\begin{align}
P'&\, = \, h_1h_2 h_3+h_1h_2h_4+h_2h_3h_4+h_1 h_3 h_4 -8y\,,\nonumber \\
y_i^2&\, = \, \frac{1}{4}h_1(y_i)h_2(y_i)h_3(y_i)h_4(y_i)\, ,\qquad (i=1,2)\,,
\end{align}
in \eqref{kapexpm2},  which allows us to rewrite \eqref{M2periodicity} as
\begin{align}\label{deltaphiM2}
\frac{\Delta z}{2\pi}\,\Big[1-\frac{y_i}{2}\sum_{I=1}^4\frac{1}{h_I(y_i)}\Big]&\, = \, \frac{\eta_i}{n_i}\,,\quad
\text{(no sum on $i$)}\,.
\end{align}
One can now immediately prove that the R-symmetry flux is given by 
\begin{align}\label{M2sumcharges}
Q^{R}\, = \, \frac{\eta_1\,n_2-\eta_2\,n_1}{n_1 n_2}
\qquad
\Leftrightarrow
\qquad
p_1+p_2+p_3+p_4 \, = \, {\eta_1\,n_2-\eta_2\,n_1}\,.
\end{align}
Case B is the anti-twist; this is the only case realized when all four charges are equal, hence giving a solution of minimal gauged supergravity 
\cite{Ferrero:2020twa}.
Cases A and C correspond to twist solutions,  which in fact can only be realized in the model with four distinct charges, as discussed below.  

As mentioned earlier,  we could not solve explicitly the conditions for regularity in general,  so one might wonder whether there exist choices of the parameters $q_I$ for which the three cases A, B, C are realized in practice. To end this section,  in Table \ref{tab:valuesM2} we  give some possible values of the $q_I$ which lead to regular spindle solutions in the three cases.
\begin{table}
\begin{center}
\begin{tabular}{c||c|c|c|c|c|c}
Case & $q_1$ & $q_2$ & $q_3$ & $q_4$ & $y_1$ & $y_2$\\
\hline
\hline
A & $-0.01$& $0.44$& $0.72$& $0.97$& $0.01$& $0.11$\\
\hline
B & $0.03$& $0.11$& $0.67$& $0.95$& $-0.01$& $0.05$\\
\hline
C & $0.01$& $-0.44$& $-0.72$& $-0.97$& $-0.11$& $-0.01$
\end{tabular}
\end{center}
\caption{Values of the parameters $q_I$,  and corresponding roots $y_{1,2}$,  for which the three cases are realized.}
\label{tab:valuesM2}
\end{table}
In particular,  let us observe that analysing the roots numerically it seems that case A can only be realized when all $h_I>0$,  case B admits both all $h_I>0$ and all $h_I<0$,  while case C is only possible if all $h_I<0$.  Note that the map 
\begin{align}
y\to -y\,, \qquad q_I \to -\,q_I\,,
\end{align}
is a symmetry of the solution,  and it maps case A solutions to case C solutions,  while case B is mapped to itself.

Let us make a comment on the case with pairwise equal charges.  For concreteness, 
set $q_1=q_2$ and $q_3=q_4$,  so that $h_1=h_2$ and $h_3=h_4$.  One can wonder whether twist solutions (for either case A or case C) exist in this case,  but it turns out that the two middle roots of $P$ have the same sign only when ${h}_1 {h}_3<0$,  which violates the requirement that all $h_I$ have the same sign.  This rules out twist solutions,  leading to the anti-twist solutions discussed in \cite{Ferrero:2021ovq,Couzens:2021rlk}.

As conjectured in appendix C of \cite{Ferrero:2021ovq},  the AdS$_2\times \Sigma$ solutions with four magnetic charges discussed in this section should arise as the near-horizon limit of certain accelerating black holes in AdS$_4$,  which are the electro-magnetic dual of the black holes found in \cite{Lu:2014sza}.  Their entropy can be computed directly from the near-horizon solution,  and reads
\begin{align}
S_{BH}\, = \, \frac{A_{\text{horizon}}}{4G_{(4)}}\, = \, \frac{y_2-y_1}{8G_{(4)}}\,\Delta z\,,
\end{align}
where $G_{(4)}$ is the four-dimensional Newton constant.  Following the conventions of \cite{Ferrero:2021ovq},  this can be expressed in terms of the number $N$ of M2-branes as
\begin{align}
\frac{1}{G_{(4)}}\, = \, \frac{2\sqrt{2}}{3}N^{3/2}\,.
\end{align}

\subsection{Killing spinors}\label{sec:M2Kill}
We begin by choosing the following orthonormal frame for the metric \eqref{4chargemagneticAdS22}:
\begin{align}
\ex^a&\, = \, \frac{1}{2}H^{1/4}\,\bar \ex^a,\quad\text{$a=0,1$}\, ,\qquad \ex^2\, =\, \frac{H^{1/4}}{P^{1/2}}\,\diff y\,, \qquad
\ex^3\, = \, \frac{P^{1/2}}{2H^{1/4}}\,\diff z\,,
\end{align}
where $\bar\ex^a$ is an orthonormal frame for the unit radius metric on
 AdS$_2$. 
We use $D=4$ gamma matrices
\begin{align}\label{gammas4d}
\Gamma_a&\, = \, \beta_a\otimes \gamma_3\,, \qquad 
\Gamma_2 \, = \, 1\otimes \gamma_1\,, \qquad
\Gamma_3\, = \, 1\otimes \gamma_2\,,
\end{align}
where $\beta_0=\ii \sigma^2$, $\beta_1=\sigma^1$
are two-dimensional gamma matrices for the AdS$_2$ directions,  while $\gamma_1=\sigma^1$,  $\gamma_2=\sigma^2$ are two-dimensional gamma matrices for the spindle directions,  with chirality matrix $\gamma_3=-\ii\,\gamma_1\gamma_2=\sigma^3$.  

We next introduce two-component AdS$_2$ Killing spinors $\vartheta$ satisfying  
\begin{align}
\nabla_a\,\vartheta\, = \, \frac{\bn}{2}\gamma_a\vartheta\,.
\end{align}
As in appendix \ref{AdSspinors},  we have two solutions $\vartheta_+^{1,2}$ when $\bn=+1$ and two solutions $\vartheta_-^{1,2}$ when $\bn=-1$,  which in particular are Majorana spinors.  We then find that the $D=4$ Killing spinors $\epsilon^{(A)}$,  $A=1,2$,  solving \eqref{susym2} are different according to whether the functions $h_I$ are all positive or all negative.  In particular, we find 
\begin{align}\label{AdS2spindlespinors}
\epsilon^{(A)} \, = \, 
\begin{cases}\ 
\vartheta_+^{(A)}\otimes \zeta \qquad \text{all $h_I>0$\, ,}\\ \ 
\vartheta_-^{(A)}\otimes \tilde{\zeta} \qquad \text{all $h_I<0$\, ,}\\
\end{cases}
\end{align}
where the spinors $\zeta$ and $\tilde{\zeta}$ on the spindle are given by
\begin{align}\label{M2spinors1}
\zeta\, \equiv \, 
\begin{pmatrix}
\zeta_+\\
\zeta_-
\end{pmatrix} \, = \, \frac{\ex^{\tfrac{\ii z}{2}}}{\sqrt{2}H^{1/8}}\,
\begin{pmatrix}
[\sqrt{H}-2y]^{1/2}\\
- [\sqrt{H}+2y]^{1/2}
\end{pmatrix} \, = \, \ex^{\frac{\ii z}{2}}H^{1/8}
\begin{pmatrix}
\sin \frac{\alpha}{2}\\
-\cos \frac{\alpha}{2}
\end{pmatrix}\,,
\end{align}
and
\begin{align}\label{M2spinors2}
\tilde{\zeta}\, \equiv \, 
\begin{pmatrix}
\tilde{\zeta}_+\\
\tilde{\zeta}_-
\end{pmatrix}\, = \, \frac{\ex^{\tfrac{\ii z}{2}}}{\sqrt{2}H^{1/8}}\,
\begin{pmatrix}
[\sqrt{H}-2y]^{1/2}\\
[\sqrt{H}+2y]^{1/2}
\end{pmatrix} \, = \, \ex^{\frac{\ii z}{2}}H^{1/8}
\begin{pmatrix}
\sin \frac{\alpha}{2}\\
\cos \frac{\alpha}{2}
\end{pmatrix} \,,
\end{align}
respectively, where the function $\alpha=\alpha(y)$ satisfies
\begin{align}
\sin\alpha\, = \, \frac{P^{1/2}}{H^{1/2}}\,, \qquad
\cos\alpha \, = \, \frac{2y}{H^{1/2}}\,.
\end{align}
In particular the spinors $\epsilon^{(A)}$ given by \eqref{AdS2spindlespinors} satisfy
\begin{align}
M\,\epsilon^{(A)} \, = \, -\ii\, \epsilon^{(A)}\,, \quad \text{where} \quad
M\,  \equiv  \, \begin{cases}\ 
\cos\alpha \,\Gamma_{23}+\ii \sin \alpha\, \Gamma_2\,,\quad \text{all $h_I>0$\, ,}\\
\ \cos\alpha \,\Gamma_{23}-\ii \sin \alpha\, \Gamma_2\, \quad \text{all $h_I<0$\, ,}
\end{cases}
\end{align}
and one can verify that in both cases $M^2=-1$.  We thus preserve half of the supersymmetries, corresponding to the superalgebra of a $d=1$, $\mathcal{N}=2$ SCFT.

 Notice that the squared norms of these Killing spinors are given by
\begin{align}
\zeta^\dagger\zeta \, = \, \tilde{\zeta}^\dagger\tilde{\zeta} \, = \, H^{1/4}\,,
\end{align}
which are proportional to the square root of the warp factor, as expected from the general discussion in section \ref{sec:nowhere}.
We then calculate that on the spindle we have the Killing vector bilinear
\begin{align}
\frac{\ii}{\zeta^\dagger\zeta}(\zeta^\dagger\gamma^m\gamma_3\zeta)\partial_m\, =\, -\, \frac{\ii}{\tilde{\zeta}^\dagger\tilde{\zeta}}(\tilde{\zeta}^\dagger\gamma^m\gamma_3\tilde{\zeta})\partial_m\, = \, 2 \partial_z\, ,
\end{align}
where here $x^m=(y,z)$ are coordinates on the spindle.

Using the solutions \eqref{M2spinors1}, \eqref{M2spinors2} one can immediately deduce the following behaviour at the poles $y_1$, $y_2$ in the three cases, where the statements below 
apply to both $\zeta$ and~$\tilde\zeta$: 
\begin{itemize}
\item Case A, Twist: $\zeta^+(y_1)=0$, $\zeta^+(y_2)=0$,  with $\zeta^-\neq 0$ everywhere.
\item Case B, Anti-Twist: $\zeta^-(y_1)=0$ and non-vanishing everywhere else, $\zeta^+(y_2)=0$ and non-vanishing everywhere else.
\item Case C, Twist: $\zeta^-(y_1)=0$, $\zeta^-(y_2)=0$,  with $\zeta^+\neq 0$ everywhere.
\end{itemize}

Note that in the gauge we are using for $A^{(I)}$, the R-symmetry gauge field 
is not regular at either pole located at $y=y_1$ or $y=y_2$.
However, we may obtain  a smooth gauge field at $y=y_1$ via the gauge transformation $A^R\to A^R_{(0)}=A^R- \diff z+\frac{\eta_1}{n_1} \diff\varphi$. Here we have introduced $\varphi=\frac{2\pi}{\Delta z} z$, with $\Delta\varphi=2\pi$. 
This gauge transformation will act on the spinor, which is only charged under the R-symmetry, and replace the $\ex^{\tfrac{\ii z}{2}}$ phase in
\eqref{M2spinors1} or \eqref{M2spinors2}
with $\ex^{\tfrac{\ii \eta_1\varphi}{2n_1}}$. Noting that the frame we are using is the analogue of the hatted frame in section 2 
(which is the frame that is not regular at $y=y_1$), we see that this behaviour is
in agreement with \eqref{hatnothatspin}. Recall here that for the twist case A, $\zeta_+(y_1)=0$ and $\eta_1=-1$, while 
for the anti-twist case B and twist case C, $\zeta_-(y_1)=0$ and $\eta_1=+1$ (where here and in the following the same formulas hold also for $\tilde\zeta$). It then follows from 
 \eqref{M2spinors1}, \eqref{M2spinors2} that in the locally Euclidean frame and in the new gauge, for the twist case A we have 
 $\zeta_+\sim \ex^{-\tfrac{\ii \varphi}{n_1}}\rho_1$, where $\rho_1$ is the radial coordinate at $y=y_1$, while 
 for the anti-twist case B and twist case C we instead have 
 $\zeta_-\sim \ex^{+\tfrac{\ii \varphi}{n_1}}\rho_1$.  In both cases we have a simple holomorphic zero of the spinor to leading order 
 at the pole $\rho_1=0$. Similar comments apply, \emph{mutatis mutandis}, to the behaviour of the spinors at the other pole located
at $y=y_2$.

Finally, notice that the spinors \eqref{M2spinors1}, \eqref{M2spinors2} are charged under azimuthal 
rotations of the spindle, due to the phase $\ex^{\ii z/2}$. 
We consider the following three-parameter family of local gauge transformations on the gauge fields
\begin{align}\label{M2tildeAR}
\tilde{A}^{(I)}\, \equiv\, A^{(I)}+\sa^I\diff z \, = \, \rho_I(y)\,\diff\varphi\, ,
\qquad
\rho_I(y)\, \equiv\, \left(\frac{y}{h_I}+\sa^I\right)\frac{\Delta z}{2\pi}\, ,
\end{align}
where the $\sa^I$ are constants satisfying the constraint 
\begin{align}\label{aIconm2}
\sa^1+\sa^2+\sa^3+\sa^4\, = \, -2\,.
\end{align}
The latter ensures that the R-symmetry gauge field transforms to 
$\tilde{A}^R = A^R -  \diff z$, which cancels the
phase $\ex^{\ii z/2}$ in the spinor as the spinor has charge $\tfrac{1}{2}$ under $A^R$. 
Using \eqref{deltaphiM2} one can check that
\begin{align}\label{M2tildeARpoles}
\left.\tilde A^R\, \right|_{y=y_1}&\, = \, -\frac{\eta_1}{n_1}\diff\varphi\,, \qquad
\left.\tilde A^R\, \right|_{y=y_2}\, = \, -\frac{\eta_2}{n_2}\diff\varphi\,.
\end{align}
This again agrees with the general conditions in section \ref{sectwists} we derived earlier, 
where in particular recall that our twist case A is a charge conjugate 
of the twist considered in section~\ref{sectwists}.

\section{M5-branes}\label{sec:M5}
In this section we will consider the case of M5-branes wrapping a spindle. The relevant solutions were constructed in a $U(1)^2$ gauged
supergravity in $D=7$ and then uplifted to $D=11$ supergravity on an $S^4$ in \cite{Ferrero:2021wvk}. There it was shown that only the twist case is allowed. 
Here we explicitly construct the $D=7$ Killing spinors and confirm that they indeed have the structure that we elucidated in section 2. 

\subsection{The $D=7$ gauged supergravity model}

As in \cite{Ferrero:2021wvk} the $U(1)^2$ gauged supergravity model that we shall consider has a Lagrangian 
given by 
\begin{align}\label{dsevenlag}
\mathcal{L}\, = \, \sqrt{-g}\Big[R-\mathcal{V}-\frac{1}{2}\partial_{\mu}\vec{\varphi}\cdot\partial^{\mu}\vec{\varphi}-\frac{1}{4}\sum_{i=1}^2
(X^{(I)})^{-2}\,(F^{(I)})^2\Big]\,,
\end{align}
where $A^{(I)}$, $I=1,2$, are two $U(1)$ gauge fields with field strengths $F^{(I)}=\diff A^{(I)}$. There are two
real and canonically normalized scalar fields, $\vec{\varphi}=(\varphi_1,\varphi_2)$ in terms of which 
\begin{align}
X^{(1)}\, = \, \ex^{-\frac{1}{\sqrt{2}}\varphi_1-\frac{1}{\sqrt{10}}\varphi_2}\, \equiv \, \ex^{2\lambda_1}\,,
\qquad
X^{(2)}\, = \, \ex^{+\frac{1}{\sqrt{2}}\varphi_1-\frac{1}{\sqrt{10}}\varphi_2}\, \equiv \, \ex^{2\lambda_2}\, ,
\end{align}
where we have also introduced the scalars $\lambda_{1,2}$ for convenience. The scalar potential is given by
\begin{align}
\mathcal{V}\, = \, \frac{1}{2}\ex^{-8\lambda_1-8\lambda_2}-2\ex^{-2\lambda_1-4\lambda_2}-2\ex^{-4\lambda_1-2\lambda_2}-4\ex^{2\lambda_1+2\lambda_2}\, .
\end{align}
Using the results of \cite{Liu:1999ai}, any solution to the equations of motion with $F^{(I)}\wedge F^{(J)}=0$ will give rise to a solution of $D=7$ maximal gauged supergravity which
can then be uplifted to $D=11$ on $S^4$ \cite{Cvetic:1999xp,Nastase:1999cb}.
In particular, the vacuum AdS$_7$ solution, which has radius 2, uplifts to
the AdS$_7\times S^4$ solution dual to the $d=6$ $\mathcal{N}=(0,2)$ SCFT.

A solution will be supersymmetric provided that there are solutions to the Killing spinor equations    
\cite{Liu:1999ai} 
\begin{align}\label{ksd7eqs}
0&\, = \, \Big[\nabla_{\mu}+\frac12(A^{(1)}_\mu\GGamma^{12}+A^{(2)}_\mu\GGamma^{34})+\frac{1}{4}\ex^{-4\lambda_1-4\lambda_2}\Gamma_{\mu}+\frac12\, \Gamma_{\mu}\,\slashed{\partial}(\lambda_1+\lambda_2)\nn
& \qquad\qquad\qquad\qquad\qquad+\frac{1}{4}\ex^{-2\lambda_1}(F^{(1)}_{\mu\nu}\,\Gamma^{\nu})\GGamma_{12}+\frac{1}{4}\ex^{-2\lambda_2}(F^{(2)}_{\mu\nu}\,\Gamma^{\nu})\,\GGamma_{34}\Big]\,\epsilon
\,,\nn
0&\, = \, \Big[\slashed{\partial}(3\lambda_1+2\lambda_2)-\left(\ex^{2\lambda_1}-\ex^{-4\lambda_1-4\lambda_2}\right)+\frac{1}{4}\ex^{-2\lambda_1}(\Gamma^{\mu\nu}F^{(1)}_{\mu\nu})\GGamma_{12}\Big]\,\epsilon\,,\nn
0&\, = \, \Big[\slashed{\partial}(2\lambda_1+3\lambda_2)-\left(\ex^{2\lambda_2}-\ex^{-4\lambda_1-4\lambda_2}\right)+\frac{1}{4}\ex^{-2\lambda_2}(\Gamma^{\mu\nu}F^{(2)}_{\mu\nu})\GGamma_{34}\Big]\,\epsilon\,.
\end{align}
Here $\epsilon$ is a $Spin(1,6)$ spinor transforming in the ${\bf 4}$ of the $SO(5)$  R-symmetry, and is acted
on by both the $D=7$ gamma matrices $\Gamma_\mu$, $\mu=0,...,6$, and the gamma matrices $\GGamma_A$, $A=1,...,5$,
which are associated with $SO(5)$. We also note that solutions of minimal $D=7$ gauged supergravity require $A^{(1)}=A^{(2)}$
and $\lambda_1=\lambda_2$. Finally,
the R-symmetry gauge field, normalized as in section \ref{sectwists}, is given by\footnote{Note that this has the opposite sign
to that defined in \cite{Ferrero:2021wvk}.}
\begin{align}\label{rsymM5}
A^R\, \equiv \, -(A^{(1)}+A^{(2)})\,.
\end{align}

\subsection{Supergravity solution and Killing spinors}

The local supersymmetric AdS$_5\times \Sigma$ solutions discussed in \cite{Ferrero:2021wvk} are given\footnote{Note that we have slightly changed the notation compared with \cite{Ferrero:2021wvk}, so that it is similar to previous sections, and we are also using a
different gauge. We also note that we should identify the parameters $n_+,n_-$ in \cite{Ferrero:2021wvk} with $n_1,n_2$, respectively, below.} 
by 
\begin{align}\label{M5sugrasol}
\diff s^2_7& \, = \, (yH)^{1/5}\,\left[\diff s^2_{\text{AdS}_5}+\frac{y}{P}\diff y^2+\frac{P}{4H}\diff z^2\right]\,,\nn
A^{(I)}&\, = \, -\frac{y^2}{h_I}\diff z\,,\qquad
X^{(I)}\, = \, \ex^{2\lambda_I}\, = \, \frac{(yH)^{2/5}}{h_I}\,,
\end{align}
where $\diff s^2_{\text{AdS}_5}$ is a unit radius metric, and
\begin{align}
h_I\, =\, y^2+q_I\,,\qquad
H\, = \, h_1\,h_2\,, \qquad
P\, = \, H-4y^3\,.
\end{align}
The solutions are parametrized by two constants $q_I$, $I=1,2$.
Note that we can write the field strength as
\begin{align}
\left(X^{(I)}\right)^{-2} F^{(I)}\, = \, -\frac{4 q_I}{(y H)^{1/2}}\widetilde{\text{vol}}_{\Sigma}\,,
\end{align}
where $\widetilde{\text{vol}}(\Sigma)$ is the volume form associated to the conformally rescaled spindle metric 
$\widetilde{\diff s^2}_{\Sigma}=(y H)^{1/5}\diff s^2_{\Sigma}$.

The conditions that need to be imposed in order to obtain AdS$_5\times \Sigma$ solutions, with $\Sigma$ a compact spindle 
parametrized by $(y,z)$, were described in detail
in \cite{Ferrero:2021wvk}. One requires $y>0$, $h_I>0$, $H>0$. We further require $q_1 q_2<0$ and the quartic polynomial $P$ to have four real 
roots, with the middle two, $y_1,y_2$, both being positive. Choosing $y\in [y_1,y_2]$ then gives AdS$_5\times\Sigma$ solutions
within the twist class, with no solutions in the anti-twist class.
The condition $q_1 q_2<0$ also shows that there are no such solutions lying within minimal 
$D=7$ gauged supergravity, as also pointed out in \cite{Ferrero:2021wvk}. 
Similar to previous cases we can derive the useful result that at the roots $y=y_1, y_2$ we have
\begin{align}\label{mficeiden}
\Big[-\frac{3}{2}+\frac{y_1^2}{4}\sum_{I=1}^2\frac{1}{h_I(y_1)}\Big]\frac{\Delta z}{2\pi}\, = \, \frac{1}{n_1}\,,\qquad
\Big[-\frac{3}{2}+\frac{y_2^2}{4}\sum_{I=1}^2\frac{1}{h_I(y_2)}\Big]\frac{\Delta z}{2\pi}\, = \, -\frac{1}{n_2}\, ,
\end{align}
with the signs on the right hand associated with $P'(y_1)>0$ and $P'(y_2)<0$ and the fact that we only have the twist case.
As in \cite{Ferrero:2021wvk} the gauge field fluxes through the spindle are given by
\begin{align}\label{M52fluxes}
Q_I&\, \equiv\, \frac{1}{2\pi}\int_{\Sigma}F^{(I)}\, = \, -\left(\frac{y^2_2}{h_I(y_2)}-\frac{y^2_1}{h_I(y_1)}\right)\,\frac{\Delta z}{2\pi}\,,\nn
&\, \equiv \, \frac{p_I}{n_1\,n_2}\,,\qquad p_I\in \mathbb{Z}\,.
\end{align}
With the definition \eqref{rsymM5}, the total R-symmetry flux is then calculated to by
\begin{align}\label{m5twistfvg}
Q^R\equiv -(Q_1+Q_2)=-\chi(\Sigma)\,,
\end{align} 
corresponding to the twist case only. 

Here we would like to present explicit expressions for the $D=7$ Killing spinors, which were not given in \cite{Ferrero:2021wvk}.
We choose the following orthonormal frame 
\begin{align}
e^a&\, = \, \left(yH\right)^{1/10}\,\bar{e}^a\,,\qquad
e^5\, = \, \frac{y^{3/5}H^{1/10}}{P^{1/2}}\,\diff y\,, \qquad 
e^6\, = \, \frac{y^{1/10}P^{1/2}}{2H^{2/5}}\,\diff z\,,
\end{align}
with $\bar{e}^a$, $a=0,...,4$ an orthonormal frame for the unit radius metric on $\text{AdS}_5$.
We use explicit $D=7$ gamma matrices given by
\begin{align}
\Gamma_a\, = \, \beta_a\otimes\gamma_3\,,\qquad
\Gamma_5\, = \, 1\otimes \gamma_1\,,\qquad
\Gamma_6\, = \, 1\otimes \gamma_2\,,
\end{align}
where $\beta_a$ are five-dimensional gamma matrices given by
$\beta_0 =  \ii\,\sigma^2\otimes 1$,
$\beta_1 =  \sigma^1\otimes 1$,
$\beta_2 =  \sigma^3\otimes \sigma^1$, 
$\beta_3 =  \sigma^3\otimes \sigma^2$,
$\beta_4  =  \sigma^3\otimes \sigma^3$,
and
$\gamma_1=\sigma^1$, $\gamma_2=\sigma^2$ are again gamma matrices for the spindle directions with
$\gamma_3=\sigma^3$.  We can also choose an explicit set of $SO(5)$ gamma matrices as follows:
$\GGamma_1=-\ii\,\beta_0$,
$\GGamma_a=\beta_{a-1}$, for $a=2,...,5$.

We next introduce the Killing spinor equation for AdS$_5$,  which reads
\begin{align}
\nabla_{\mu} \vartheta\, = \, \frac{\cc}{2}\,\beta_{\mu}\vartheta\,,
\end{align}
with $\cc=\pm 1$. For each value of $\cc$ the solution admits four independent solutions,  
which we label $\vartheta_\cc^{(A)}$,  $A=1,...,4$ (see appendix \ref{AdSspinors}).  
To describe the $D=7$ Killing spinors, we need two more ingredients.  
The first is the common eigenvectors of the $SO(5)$ matrices $\GGamma_{12}$ and $\GGamma_{34}$ that have the same eigenvalue.  There are two of them, which we label $u_{\pm}$, satisfying
\begin{align}
\Gamma_{12}\,u_{\pm}\, = \, \Gamma_{34}\,u_{\pm}\,= \, \pm\ii\,u_{\pm}\,.
\end{align}
The second ingredient is the charge conjugation matrix on the spindle $\Sigma$,  which we take to be $B=\gamma_1$, so that
$B\,(\gamma_{1,2})^*\,B^{-1}=\gamma_{1,2}$,  where $\gamma_{1,2}$ are two-dimensional gamma matrices on $\Sigma$, given above.  
We then define the charge conjugate of a two-dimensional spinor $\zeta$ on the spindle as $\zeta^c=B\,\zeta^*$.

The $\text{AdS}_5\times \Sigma$ solutions preserve 1/2 of the supersymmetry and we can now write the eight Killing spinors
as 
\begin{align}\label{M5spinorsfull}
\epsilon^{(A)}_-\, = \, \vartheta_-^{(A)}\otimes \zeta\otimes u_-\,, \qquad \text{and}\qquad
\epsilon^{(A)}_+\, = \, \vartheta_+^{(A)}\otimes \zeta^c\otimes u_+\,,
\end{align}
where $\zeta$ is the two-dimensional spinor on the spindle given by  
\begin{align}\label{exm5ksesols}
\zeta\, \equiv \, 
\begin{pmatrix}
\zeta_+\\
\zeta_-
\end{pmatrix}\, &  = \, \ex^{-\tfrac{3\ii z}{4}}\,\frac{y^{1/20}}{\sqrt{2}H^{1/5}}\,\begin{pmatrix}
[\sqrt{H}+2y^{3/2}]^{1/2}\\
-[\sqrt{H}-2y^{3/2}]^{1/2}
\end{pmatrix}\nn
 \, & =\, \ex^{-\tfrac{3\ii z}{4}} (y H)^{1/20}\,
\begin{pmatrix}
\cos \frac{\alpha}{2}\\
-\sin \frac{\alpha}{2}
\end{pmatrix}\,,
\end{align}
where the function $\alpha=\alpha(y)$ satisfies
\begin{align}
\sin\alpha\, = \, \frac{P^{1/2}}{H^{1/2}}\,, \qquad \cos \alpha\, = \, \frac{2y^{3/2}}{H^{1/2}}\,.
\end{align}
In particular then the spinors $\epsilon^{(A)}$ given in \eqref{M5spinorsfull} satisfy
\begin{align}
M_-\,\epsilon^{(A)}_-\, = \, -\ii\,\epsilon^{(A)}_-\,, \qquad
M_+\,\epsilon^{(A)}_+\, = \, +\ii\,\epsilon^{(A)}_+\,, 
\end{align}
where 
\begin{align}
M_{\pm}\, = \, \cos\alpha\,\Gamma_{56}\mp \ii\,\sin\alpha\,\Gamma_5\,,
\end{align}
and one can verify that $M^2=-1$.  We thus preserve half of the supersymmetries,  corresponding to a $4d$, $\mathcal{N}=1$ SCFT.  The squared norm of $\zeta$ is
\begin{align}
\zeta^\dagger\zeta \, = \, (yH)^{1/10}\,,
\end{align}
which is proportional to the square root of the warp factor, as expected from the general discussion in section \ref{sec:nowhere}.
We then calculate that on the spindle we have the Killing vector bilinear
\begin{align}
\frac{\ii}{\zeta^\dagger\zeta}(\zeta^\dagger\gamma^m\gamma_3\zeta)\partial_m\, = \, 2\partial_z\, ,
\end{align}
where here $x^m=(y,z)$ are coordinates on the spindle.

As noted above, the supergravity solutions constructed in \cite{Ferrero:2021wvk} all lie within the twist class. In particular $y\in [y_1,y_2]$
with $y_1,y_2>0$ and satisfying $H(y_i)=4y_i^3$. Correspondingly we see from \eqref{exm5ksesols} that at both poles of the spindle
one has $\zeta_-(y_{1,2})=0$,  with $\zeta_+(y_{1,2})\neq 0$, in precise alignment with section \ref{sectwists}.

Note that in the gauge we are using for $A^{(I)}$, the R-symmetry gauge field 
is not regular at either pole located at $y=y_1$ or $y=y_2$.
Using \eqref{mficeiden} we can obtain a smooth gauge field at $y=y_1$ via the gauge transformation: $-A^R\to -A^R_{(0)}=-A^R+\frac{3}{2}\diff z+\frac{1}{n_1} \diff\varphi$. Here we have introduced $\varphi=\frac{2\pi}{\Delta z} z$, with $\Delta\varphi=2\pi$. 
This gauge transformation will act on the spinor, which is only charged under the R-symmetry, and replace the $\ex^{-\tfrac{3\ii z}{4}}$ phase in
\eqref{exm5ksesols}
with $\ex^{\tfrac{\ii \varphi}{2n_1}}$. Noting that the frame we are using is the analogue of the hatted frame in section 2 
(which is the frame that is not regular at $y=y_1$), we see that this behaviour is
in agreement with \eqref{hatnothatspin}. Furthermore, we observe from \eqref{exm5ksesols} that in this frame and the new gauge, we have 
$\zeta_-\sim \ex^{\tfrac{\ii \varphi}{2n_1}}\rho_1$, where $\rho_1$ is the radial coordinate at $y=y_1$. Converting the spinor to the 
locally Euclidean frame we then have $\zeta_-\sim \ex^{\tfrac{\ii \varphi}{n_1}}\rho_1$, and we see that it has a
simple holomorphic zero at the pole $\rho_1=0$. Similar comments apply, \emph{mutatis mutandis}, to the behaviour of the spinors at the other pole located
at $y=y_2$.

We can also define the following one-parameter family of gauge transformations
\begin{align}\label{M5tildeAR}
\tilde{A}^{(I)}\, \equiv\, A^{(I)}+\sa^I\diff z \, = \, \rho_I(y)\,\diff\varphi\, ,
\end{align}
where the $\sa^I$ are constants satisfying the constraint
\begin{align}\label{aIcon}
\sa^1+\sa^2\, = \, +\frac{3}{2}\,.
\end{align}
Here we have introduced the functions
\begin{align}\label{rhoIdefm5}
\rho_I(y)\, \equiv\, \left(-\frac{y^2}{h_I}+\sa^I\right)\frac{\Delta z}{2\pi}\, .
\end{align}
These functions, which play a role in the field theory analysis, generalize the analogous quantities in \cite{Ferrero:2021wvk} (which had
$\sa^1=\sa^2=3/4$), as we remark upon in the next subsection.
In the frame that we are using and in the tilded gauge the spinors are independent of the $\varphi$ coordinate.
From the results of \cite{Ferrero:2021wvk} one can also check that 
\begin{align}\label{m5tildeARpoles}
\left.\tilde A^R\, \right|_{y=y_1}&\, = \, \frac{1}{n_1}\diff\varphi\,, \qquad
\left.\tilde A^R\, \right|_{y=y_2}\, = \, -\frac{1}{n_2}\diff\varphi\,.
\end{align}
This again agrees with the general conditions \eqref{Aatpoles} we derived for the twist
case earlier. 

\subsection{Comparison with field theory}
The AdS$_5\times \Sigma$ solutions were uplifted on an $S^4$ from $D=7$ to $D=11$ supergravity in \cite{Ferrero:2021wvk}.
While the $D=11$ solutions still have orbifold singularities, it is straightforward to carry out flux quantization and
calculate the $a$ central charge for the dual $d=4$ SCFT, and the result \cite{Ferrero:2021wvk} is given by
\begin{align}\label{extrsolutionM5}
a & \, = \, \frac{3\,p_1^2\,p_2^2\,(\mathtt{s}+p_1+p_2)}{8\,n_2n_1(n_2-p_1)(p_2-n_2)\,[\mathtt{s}+2\,(p_1+p_2)]^2}\,N^3\,,
\end{align}
where 
\begin{align}
\mathtt{s}\, = \, \sqrt{(p_1+p_2)^2-12(n_2-p_1)(n_2-p_2)}\,.
\end{align}
The supergravity solution is valid for 
$n_2>n_1$ as well as $p_1<0$ or $p_1>n_1+n_2$ (and hence from
\eqref{m5twistfvg} we have $p_2>0$ or $p_2<0$, respectively), 
which ensures that $a>0$. 

The field theory calculations in \cite{Ferrero:2021wvk} were carried out using the tilded gauge fields in \eqref{M5tildeAR}
with $\sa^1=\sa^2=3/4$, but is straightforward to generalize that calculation to an arbitrary gauge with 
$\sa^1+\sa^2\, = \, +\frac{3}{2}$ (which recall is required to ensure that the spinors are uncharged with respect to $\partial_\varphi$), much as in
section \ref{sec:D3} for the case of D3-branes. Doing so has no effect on the central charge and mixing parameter $\epsilon$, both of which are gauge-invariant quantities, but it does give rise to more general values for the extremized trial R-charge parameters $\Delta^*_1,\Delta^*_2$. It would be of interest to understand why the gauge $\sa^1=\sa^2=3/4$ leads to the simple result of
 \cite{Ferrero:2021wvk} that $\Delta^*_1=\Delta^*_2=1$.

The field theory calculation for the anti-twist can also be carried out in a straightforward manner. One finds
that in this case it is certainly possible to find fluxes on the spindle that give rise to a positive central charge. However,
these cases seem to be obstructed in the sense that there are no dual solutions within our gauged supergravity construction. 

\section{Final Comments}\label{fincoms}

In this paper we have given a general proof that supersymmetric geometries involving a
spindle $\Sigma$ with azimuthal symmetry fall into two classes: the twist and anti-twist, where the quantized magnetic 
flux through the spindle is  necessarily given by \eqref{fluxchi}. Previous 
constructions of AdS$_3\times \Sigma$ D3-brane  and AdS$_2\times \Sigma$ 
M2-brane solutions are all in the anti-twist class, but in this paper we have shown 
in both cases that there are also \emph{twist} solutions. The AdS$_5\times \Sigma$ M5-brane solutions 
in \cite{Ferrero:2021wvk} are also twist solutions, and interestingly in contrast to 
the D3-brane and M2-brane cases
here we do not find anti-twist solutions at all, at least in the gauged supergravity
construction presented.  Notice that a common feature in all cases is that we do not find twist solutions in \emph{minimal} gauged 
supergravity, where there is effectively only a single magnetic charge; although 
for D3-branes and M2-branes there are anti-twist solutions in minimal gauged supergravity
 \cite{Ferrero:2020laf}, \cite{Ferrero:2020twa}, that 
 uplift to $D=10$ and $D=11$ supergravity on general regular Sasaki-Einstein manifolds, respectively. 
 
In all cases the twist solutions have a natural physical interpretation as the IR 
limit of branes wrapped on $\Sigma$ holomorphically embedded in a Calabi-Yau 
$(n+1)$-fold, where $n=2, 3$ or $4$, in the cases of M5-branes, D3-branes or M2-branes, respectively. 
The Calabi-Yau $(n+1)$-fold is the total space of a $\C^n$ bundle over $\Sigma$, given by 
\begin{align}\label{CYn}
\bigoplus_{i=1}^n \mathcal{O}(p_i)\, \rightarrow \, \Sigma\, ,
\end{align}
where $p_i\in\Z$ give the quantized magnetic charges of the $n$ $U(1)$ gauge fields 
in the gauged supergravity constructions.
The first Chern class of the space 
\eqref{CYn} is zero\footnote{Note that a little care is required when identifying the $p_i$ in this discussion with
the $p_i$ in the twist solutions in the bulk of the paper. In section \ref{sec:D3}, for the D3-brane twist solutions they are the same. 
In section \ref{sec:M2}, for the case of the M2-branes they are the same for the twist case A, but for the twist case C the Calabi-Yau is instead
$\bigoplus_{i=1}^n \mathcal{O}(-p_i)\, \rightarrow \, \Sigma$ with $\sum_{i=1}^n \frac{-p_i}{n_1 n_2} + \chi = 0$, 
as it is also for the M5-brane solutions in section \ref{sec:M5}.} if and only if 
\begin{align}\label{pincon}
\sum_{i=1}^n \frac{p_i}{n_1 n_2} + \chi \, = \, 0\, ,
\end{align}
where $\chi=\chi(\Sigma)$ is the Euler number of the spindle, which is precisely the 
twist R-symmetry flux condition in \eqref{fluxchi}. 
One starts with a Minkowski$\times \mathrm{CY}_n$ solution 
to string theory or M-theory, appropriately, and wraps 
$N$ D3-branes, M2-branes or M5-branes over the holomorphically embedded 
$\Sigma\subset \mathrm{CY}_n$. The resulting low-energy theory on
the Minkowski space part of the worldvolume is supersymmetric, 
utilizing a topological twist. 
The AdS twist solutions we have constructed in all cases 
are then naturally interpreted as the near-horizon limits 
of these wrapped brane configurations, which correspondingly 
flow to superconformal field theories in the IR. 
For the D3-brane and M5-brane solutions we have presented dual field theory 
calculations that corroborate this picture. 
{It is also interesting to highlight that for the D3-brane and M5-brane spindle twist solutions, we showed
that the expression for the central charge exactly reproduces the central charge for known supergravity solutions
describing D3-brane and M5-branes wrapping an $S^2$ with a topological twist, upon setting $n_1=n_2=1$. This is despite the
fact that we cannot take a smooth limit of the spindle solutions to get the topologically twisted solutions. 

However, this discussion also raises a number of interesting open 
questions. Firstly, note that \eqref{CYn} is a Calabi-Yau 
for \emph{any} values of $p_i\in\Z$ that satisfy the constraint \eqref{pincon}.  
Since $\chi>0$, certainly the $p_i$ cannot all be positive. 
However, we only find AdS solutions when the $p_i$ satisfy additional inequalities. 
For example, the D3-brane solutions exist only when $n_2>n_1$ and 
two of the three $p_i$ ($i=1,2,3$) are negative, while one is positive. 
Similarly, in the M5-brane case also $n_2>n_1$ and one of $p_1$ or $p_2$ 
must be negative, with the other positive. What happens to the 
field theories on the wrapped branes in the IR when the charges 
do \emph{not} satisfy these conditions? We have no candidate AdS solutions, and it's 
possible the theories do not flow to a superconformal fixed point. It would be interesting
to understand these conditions on the magnetic charges from a purely field theory
point of view. Notice that whenever $p_i>0$ in \eqref{CYn}, for a given $i$, the corresponding normal bundle direction  
has non-trivial holomorphic sections and hence the wrapped brane has corresponding linear deformations. The existence (or not) of this 
moduli space presumably plays some role in the IR behaviour. 

More puzzling is the corresponding interpretation of the anti-twist solutions. 
One can formally still write down a 
complex manifold \eqref{CYn} in this case, but it 
 is \emph{not} Calabi-Yau, as 
the first Chern class is not zero. However, deleting the 
zero section gives a complex cone that does have zero first Chern class; indeed this is a necessary condition
for all GK geometries \cite{Gauntlett:2007ts}, which are relevant for the 
D3-brane AdS$_3\times Y_7$ and M2-brane AdS$_2\times Y_9$ solutions. 
Thus there is a Calabi-Yau cone geometry, but there is no ``blown-up'' Calabi-Yau
geometry with $\Sigma$ a holomorphic curve for the anti-twist. 
The AdS anti-twist solutions strongly suggest an interpretation in terms 
of $N$ D3-branes or $N$ M2-branes wrapping $\Sigma$, and the field theory 
calculation in section \ref{sec:D3anom} assumes this in the D3-brane case and then
correctly reproduces the supergravity central charge. 
But there is not a UV interpretation as $N$ branes wrapping 
a holomorphic curve in a Calabi-Yau, as there is for the twist solutions. 
 On the other hand, the anti-twist solutions have Killing spinors which are sections 
 of non-trivial bundles, and clearly preserve supersymmetry differently 
 than for standard wrapped branes. It remains an 
 interesting open problem to identify a UV geometry for this case.

In this paper, we have also explicitly calculated the Killing spinors for various AdS$\times\Sigma$ solutions of gauged supergravity and, by constructing vector spinor bilinears, elucidated the associated supersymmetry algebra. At the level of the gauged supergravity, the
rotation symmetry of the spindle enters the supersymmetry algebra as an R-symmetry and this is in contrast to the case of the topological twist where it does not. After uplifting to $D=10,11$ supergravity the internal space of the AdS solutions has an R-symmetry Killing vector which appears in the SCFT algebra. In the case of the topological twist, this is just an R-symmetry Killing vector of the space
one uplifts upon, and dual to an R-symmetry of the brane world-volume theory. For the spindle solutions, it is instead a linear combination of an R-symmetry Killing vector of the space
one uplifts upon with the Killing vector of the spindle, and from a dual point of view corresponds 
to the fact that in this case there is a mixing of the R-symmetry in the brane world-volume theory.

Supergravity solutions associated with D4-branes wrapping a spindle were recently constructed in \cite{Faedo:2021nub}. 
Supersymmetric AdS$_4\times\Sigma$ solutions were constructed in a
$D=6$ gauged supergravity theory
and then uplifted on a hemisphere to obtain solutions of massive type IIA supergravity. All of the solutions presented lie within the twist class,
consistent with the general result of this paper, and it appears that there are no solutions within the anti-twist case.
Once again it would be interesting to know why that is. It was also shown that the rotation symmetry of the spindle enters the supersymmetry algebra within the gauged supergravity theory, in the same way that we have shown in this paper for the case of D3-, M2- and M5-brane solutions.

Our general result concerning the existence of just two classes of solutions for the spindle, the twist and the anti-twist, crucially relied
on the assumption that the spindle has an azimuthal symmetry. It would be of much interest
to construct analogous supergravity solutions for spindles without such symmetry, or more generally,
for orbifold Riemann surfaces  of arbitrary genus and with an arbitrary number of orbifold points. In the case of
D3-branes and M2-branes one can aim to make progress using the recent advances in understanding GK geometry
\cite{Couzens:2018wnk,Gauntlett:2018dpc,Hosseini:2019ddy,Gauntlett:2019roi,Kim:2019umc,Gauntlett:2019pqg}.

\section*{Acknowledgments}
\noindent 
We would like to thank Nikolay Bobev, 
Dario Martelli and Nikita Nekrasov for helpful discussions. 
This work was supported in part by STFC grants  ST/T000791/1 and 
ST/T000864/1. 
JPG is supported as a Visiting Fellow at the Perimeter Institute. 

\appendix
\addtocontents{toc}{\protect\setcounter{tocdepth}{1}}

\section{Spheres from spindles }\label{appA}

\subsection{AdS$_3\times S^2$ solutions as a limit}\label{appA_D3}

Here we show how the AdS$_3\times S^2$ solutions of \cite{Benini:2013cda}, associated with the usual topological twist, can be obtained as a limit of the local class of solutions given in \eqref{D3sugrasol}.  

The solutions of \cite{Benini:2013cda} have both the 
warp factor $H$ and the scalar fields constant,  
which clearly cannot be obtained just by tuning the parameters $q_I$. We thus need to take a scaling limit.  We begin by writing
\begin{align}
y\, = \, \alpha +\lambda\,x\,, \quad
q_I\, = \, -\alpha+r_I+s \lambda^2\,,
\end{align}
where $\alpha, r_I,s$ and $\lambda$ are constants.
Taking $\lambda\to 0$ one finds that 
\begin{align}
H\,\longrightarrow\,  r_1r_2r_3\,,
\end{align}
which is constant as desired.  Moreover, since in the scaling limit $\diff y^2\to \lambda^2\, \diff x^2$ we want $P$ to scale with $\lambda^2$ in the limit,  and this will ultimately lead to a round $S^2$ in the compact directions.  This is obtained by setting
\begin{align}
\alpha \, = \,  \sqrt{r_1r_2r_3}\,, \qquad
s \, = \, \frac{1-(r_1+r_2+r_3)}{2\sqrt{r_1r_2 r_3}}\,,
\end{align}
and moreover the parameters $r_I$ must be constrained to satisfy
\begin{align}\label{r3constraint}
\frac{\alpha}{2}\sum_{I=1}^3 r_I^{-1}\, = \, 1\quad\Leftrightarrow\quad   r_1r_2+r_1r_3+r_2r_3\, = \, 
2\sqrt{r_1r_2 r_3}\,.
\end{align}
After redefining the azimuthal coordinate via
\begin{align}
z\, = \, \frac{1}{4\lambda s}\phi\,,
\end{align}
we obtain the metric
\begin{align} \label{AdS3S2metric}
\diff s^2\, = \,   F^{1/3}\left(\diff s^2_{\text{AdS}_3}+G\, \diff s^2_{S^2}\right)\,,
\end{align}
where $\diff s^2_{S^2}$ is a round metric on $S^2$ in spherical coordinates, after setting $x=\cos\theta$.
The functions of the parameters $F$ and $G$ are given by
\begin{align}
F\, = \, \alpha^2\, = \, r_1r_2r_3\,, \qquad 
G\, = \, (8\alpha s)^{-1}\,.
\end{align}

Using the same coordinate transformations,  as well as a gauge transformation to remove terms in the gauge fields that are singular as $\lambda\to 0$,  we obtain the gauge fields and (constant) scalars given by
\begin{align}\label{AdS3S2gaugefields}
A^{(I)}& \, = \,   \frac{2F^{1/2}G}{r_I}\left(1-\frac{F^{1/2}}{r_I}\right)\cos\theta\, \diff\phi \,  \equiv  \, a_I\, \cos\theta \diff \phi\,, \nn
X^{(I)}& \, = \,   \frac{F^{1/3}}{r_I}\,.
\end{align}
Here we have introduced the parameters $a_I$,  which are constrained by \eqref{r3constraint}.  Choosing the correct sign on solving that equation,  one finds
\begin{align}
a_1+a_2+a_3\, = \, -1\,,
\end{align}
or equivalently
\begin{align}
Q^R\, = \, \frac{1}{2\pi}\int_{S^2}\diff\left(A^{(1)}+A^{(2)}+A^{(3)}\right)\, = \, 2\,,
\end{align}
corresponding to a twist case.
It is straightforward to check that this reproduces the AdS$_3\times S^2$ solution of \cite{Benini:2013cda}.  Moreover, from the expression for the Killing spinors given in \eqref{d3kses}, we see that this scaling limit will
give rise to constant, anti-chiral spinors, as expected for the topological twist.

Finally, the form \eqref{AdS3S2metric} of the metric leads to a simple expression for the central charge of the dual two-dimensional SCFT arising from placing $d=4$ $N=4$ SYM theory on $S^2$ with a topological twist. Recall that 
the vacuum solution of our $D=5$ theory \eqref{d3lagoverall} is an AdS$_5$ with unit 
radius, the five-dimensional Newton constant is given by $(G_{(5)})^{-1}=2N^2/\pi$.
Using \eqref{AdS3S2metric} we can then obtain the three-dimensional Newton constant 
for a theory admitting a unit radius AdS$_3$ solution:
$(G_{(3)})^{-1}=8F^{1/2}G N^2$ and hence we deduce the $d=2$ central charge
\begin{align} 
c\, = \, \frac{3}{2G_{(3)}}
\, = \, 12 F^{1/2} G N^2 \, = \, 
\frac{3 \sqrt{r_1r_2r_3}}{1-(r_1+r_2+r_3)}N^2\,.
\end{align}
Using the relation between the parameters $a_I$ and the $r_I$ given implicitly in \eqref{AdS3S2gaugefields},  as well as the constraint \eqref{r3constraint},  it is straightforward to check that this expression agrees with the central charge of the solutions found in \cite{Benini:2013cda} in the case where the internal manifold is a sphere,  which is also recalled in eq. \eqref{c_BB}.

\subsection{AdS$_2\times S^2$ solutions as a limit}\label{appA_M2}

Here we show how the AdS$_2\times S^2$ solutions of \cite{Cacciatori:2009iz},  associated with the usual topological twist and arising as the near-horizon limit of an asymptotically AdS$_4$ black hole,  can be obtained as a limit of the local class of solutions given in \eqref{4chargemagneticAdS22}. 

As in section \ref{appA_D3},  both the 
warp factor $H$ and the scalar fields should be constant, so we need to take a scaling limit.  We write
\begin{align}
y\, = \, \alpha +\lambda\,x\,, \qquad
q_I\, = \, -\alpha+r_I+s \lambda^2\,,
\end{align}
where $\alpha, r_I,s$ and $\lambda$ are constants. Taking
$\lambda\to 0$ one finds that square root removed
\begin{align}
H\,\longrightarrow\,  {r_1r_2r_3r_4}\,,
\end{align}
which is constant as desired.  Moreover,  since in the scaling limit $\diff y^2\to \lambda^2\, \diff x^2$ we want $P$ to scale with $\lambda^2$ in the limit,  and this will ultimately lead to a round $S^2$ in the compact directions.  This is obtained by setting 
\begin{align}
\alpha\, = \,  \frac{\sqrt{r_1r_2r_3r_4}}{2}\,, \qquad
s \, = \, \frac{4-(r_1r_2+r_1r_3+r_1r_4+r_2r_3+r_2r_4+r_3r_4)}{4 \sqrt{r_1r_2r_3r_4}}\,,
\end{align}
and moreover the parameters $r_I$ must be constrained to satisfy
\begin{align}\label{constraintM2_sphere}
\frac{\alpha}{2}\sum_{I=1}^4 r_I^{-1}\, = \, 1\quad\Leftrightarrow\quad r_1r_2r_3+r_1r_2r_4+r_1r_3r_4+r_2r_3r_4 \, = \,  4\sqrt{r_1r_2r_3r_4}\,.
\end{align}
After redefining the azimuthal coordinate via
\begin{align}
z\, = \, \frac{1}{2\lambda s}\phi\,,
\end{align}
we obtain the metric
\begin{align}\label{s2metricapp}
\diff s^2 \, = \,  F^{1/2}\left(\frac{1}{4}\diff s^2_{\text{AdS}_2}+G\, \diff s^2_{S^2}\right)\,,
\end{align}
where $\diff s^2_{S^2}$ is a round metric on $S^2$ in spherical coordinates, after setting $x=\cos\theta$.
\begin{align}
F\, = \, (2\alpha)^2 \, = \, r_1r_2r_3r_4\,, \qquad
G\, = \, (8\alpha s)^{-1}
\end{align}
Using the same coordinate transformations,  as well as a gauge transformation to remove terms in the gauge fields that are singular as $\lambda\to 0$,  we obtain the gauge fields and (constant) scalars given by
\begin{align}\label{gfsm2app}
A^{(I)}&\, = \, \frac{F^{1/2}G}{r_I}\left(2-\frac{F^{1/2}}{r_I}\right)\cos\theta\, \diff \phi\,, \nn
X^{(I)}&\, = \, \frac{F^{1/4}}{r_I}\,.
\end{align}
One can check that, the constraint \eqref{constraintM2_sphere} constrains the total magnetic flux to be
\begin{align}
Q^R\, = \, \frac{1}{2\pi}\int_{S^2}\frac{1}{2}\diff\left(A^{(1)}+A^{(2)}+A^{(3)}+A^{(4)}\right)\, = \, 2\,,
\end{align}
which corresponds to a twist case.  Moreover,  after the scaling limit and the gauge transformation,  the Killing spinors \eqref{M2spinors1} reduce to constant,  chiral spinors on the sphere,  which corresponds to a standard topological twist.  Although we have not matched the parameters explicitly,  this should reproduce the near-horizon limit of the supersymmetric black holes given in \cite{Cacciatori:2009iz}.  In terms of the parameters $r_I$ used here, from \eqref{s2metricapp} we deduce that the area of the horizon is $A_{\text{horizon}}=4\pi F^{1/2}G$ and
hence the black hole entropy reads
\begin{align}
S_{BH}
\, = \, \frac{\pi}{G_{(4)}}\,\frac{\sqrt{r_1r_2r_3r_4}}{4-(r_1r_2+r_1r_3+r_1r_4+r_2r_3+r_2r_4+r_3r_4)}\,,
\end{align}
where we remind the reader that the parameters $r_I$ are subject to the constraint \eqref{constraintM2_sphere}. This can
be expressed in terms of magnetic charges via \eqref{gfsm2app}.
Also, recalling that the vacuum solution of our $D=4$ theory \eqref{m2lagoverall} is an AdS$_4$ with unit radius, which uplifts to AdS$_4\times S^7$ dual to ABJM theory, the four-dimensional Newton constant is given by $(G_{(4)})^{-1}=2^{3/2}N^{3/2}/(3\pi)$.

\subsection{AdS$_5\times S^2$ solutions as a limit}\label{appA_M5}

Here we show how the AdS$_5\times S^2$ solutions of \cite{Bah:2012dg}, associated with the usual topological twist, can be obtained as a limit of the local class of solutions given in \eqref{M5sugrasol}.  

As in appendix \ref{appA_D3},  the solutions of \cite{Bah:2012dg} have both the 
warp factor $H$ and the scalar fields constant, so we take a scaling limit.  We set
\begin{align}
y\, = \, \alpha +\lambda\,x\,, \qquad
q_I\, = \, -\alpha^2+r_I+s \lambda^2\,,
\end{align}
where $\alpha, r_I,s$ and $\lambda$ are constants. Taking
taking $\lambda\to 0$ one finds that
\begin{align}
H\,\longrightarrow\, \,  r_1r_2\,,
\end{align}
which is constant as desired.  Moreover,  since in the scaling limit $\diff y^2\to \lambda^2\, \diff x^2$ we want $P$ to scale with $\lambda^2$ in the limit,  and this will ultimately lead to a round $S^2$ in the compact directions.  This is obtained by setting
\begin{align}
\alpha \, = \,  \left(\frac{r_1r_2}{4}\right)^{1/3}\,, 
\qquad
s\, = \, 1-\frac{r_1+r_2}{9}\,,
\end{align}
and moreover the parameters $r_I$ must be constrained to satisfy
\begin{align}\label{constraintM5_sphere}
\frac{2\alpha^2}{3}\sum_{I=1}^2 r_I^{-1}\, = \, 1\quad\Leftrightarrow\quad r_1+r_2 \, = \, 3(2r_1 r_2)^{1/3}\,.
\end{align}
After setting
\begin{align}
z \, = \, -\frac{2 \alpha}{3\lambda s}\phi\,,
\end{align}
we obtain the metric
\begin{align}\label{AdS5S2metric}
\diff s^2\, = \,  F^{1/5}\left(\diff s^2_{\text{AdS}_5}+G\, \diff s^2_{S^2}\right)\,,
\end{align}
where $\diff s^2_{S^2}$ is a round metric on $S^2$ in spherical coordinates, after setting $x=\cos\theta$.
The functions of the parameters $F$ and $G$ are given by
\begin{align}
F\, = \, 4\alpha^4\,, \quad
G\, = \, (6 s)^{-1}\,.
\end{align}
Using the same coordinate transformations,  as well as a gauge transformation to remove terms in the gauge fields that are singular as $\lambda\to 0$,  we obtain the gauge fields and (constant) scalars given by
\begin{align}
A^{(I)}&\, = \, \frac{2 F^{1/2}G}{r_I}\left(2-\frac{F^{1/2}}{r_I}\right)\cos\theta\, \diff \phi\,, \nn
X^{(I)}&\, = \, \frac{F^{2/5}}{r_I}\,.
\end{align}
One can check that, the constraint \eqref{constraintM5_sphere} constrains the total magnetic flux to be
\begin{align}
Q^R\, =\,  - \frac{1}{2\pi}\int_{S^2}\diff\left(A^{(1)}+A^{(2)}\right)\, = \, -2\,,
\end{align}
which corresponds to a twist case.  Moreover,  after the scaling limit and the gauge transformation,  the Killing spinors \eqref{exm5ksesols} reduce to constant,  chiral spinors on the sphere,  which corresponds to a standard topological twist.  The local solution matches exactly that of \cite{Bah:2012dg} after setting their parameter $z$ to be\footnote{Note that in \cite{Bah:2012dg} the authors set the gauge coupling $m=2$,  while our conventions correspond to $m=1$.  Thus,  with this choice of $z$ we reproduce their solution modulo adjusting the value of the gauge coupling.}
\begin{align}\label{AdS5S2_zpar}
z\, = \, -1-\frac{2G}{3F^{1/2}}(2r_1-r_2)r_2\,.
\end{align}

Finally, the form \eqref{AdS5S2metric} of the metric leads to a simple expression for the central charge of the dual four-dimensional SCFT arising from placing the 
$d=6$, $\mathcal{N}=(0,2)$ theory on $S^2$ with a topological twist. Recalling that the vacuum solution of our $D=7$ theory \eqref{dsevenlag} is an AdS$_7$ with radius 2, the seven-dimensional Newton constant, is given by $(G_{(7)})^{-1}=N^3/(6\pi^2)$.
Using \eqref{AdS5S2metric} we can then obtain the five-dimensional Newton constant 
for a theory admitting a unit radius AdS$_5$ solution:
$(G_{(5)})^{-1}=2F^{1/2}GN^3/(3\pi)$ and hence the $d=4$ central charge
\begin{align} 
a\, =\, \frac{\pi}{8G_{(5)}}
\, =\, 
\frac{F^{1/2} G}{12}N^3
\, = \, \frac{2(r_1+r_2)^2}{9[9-(r_1+r_2)]}N^3\,.
\end{align}
Using the relation \eqref{AdS5S2_zpar},  as well as the constraint \eqref{constraintM5_sphere},  it is straightforward to check that this expression agrees with the central charge of the solutions found in eq.  (2.22) of\cite{Bah:2012dg} in the case where the internal manifold is a sphere,  which we recall here for the reader's convenience:
\begin{align}
a\, = \, \frac{1-9z^2+(1+3z^2)^{3/2}}{48 z^2}N^3\,.
\end{align}
This central charge also arises in the limit $n_{1,2}\to 1$ of the central charge found in \cite{Ferrero:2021wvk} (see the discussion at the end of section 4) for the AdS$_5$ spindle solutions discussed in section \ref{sec:M5}.

\section{Killing spinors in AdS}\label{AdSspinors}

Here we include a few comments on  AdS Killing spinors that we use in the main text (see also \cite{Lu:1996rhb,Lu:1998nu}).
Using Poincar\'e coordinates $\bar{x}^a=(t,\,x_1,\dots x_{d-1},\,r)$, we can write the metric on a unit radius AdS$_{d+1}$ in the form
\begin{align}
\diff s^2_{\text{AdS}_{d+1}}\, = \, \frac{1}{r^2}\Big({-\diff t^2+\sum_{i=1}^{d-1}\diff x_i^2+\diff r^2}\Big)\,,
\end{align}
with an associated orthonormal frame given by
\begin{align}
\bar{e}^0\, = \, \frac{\diff t}{r}\,, \quad
\bar{e}^1\, = \, \frac{\diff x_1}{r}\,, \quad \dots\,, \quad
\bar{e}^{d-1}\, = \, \frac{\diff x_{d-1}}{r}\,, \quad
\bar{e}^d \, = \, \frac{\diff r}{r}\,.
\end{align}
We introduce arbitrary gamma matrices $\beta_a$ satisfying $\{\beta_a\,,\,\beta_b\}=2\eta_{ab}$
with signature $(-,+,...,+)$. We want to construct solutions to the Killing spinor equation
\begin{align}\label{AdSKSEall}
\nabla_a\vartheta\,=\,\frac{\cc}{2}\,\beta_a\vartheta\,,
\end{align}
where $\cc=\pm 1$.  
For each choice of $\cc$ we can find $\mathtt{d}=2^{\lfloor (d+1)/2\rfloor}$ solutions; half of them correspond to 
Poincar\'e supercharges $\mathcal{Q}_\cc^{(\alpha)}$,  and their bilinears give rise to translations in the conformal algebra,  while the remaining half corresponds to conformal supercharges $\mathcal{S}_\cc^{(\alpha)}$,  and their bilinears give rise to special conformal transformations.   
We can express them in terms of the eigenvectors of $\beta_{d}$,  associated with the $\diff r/r$ component of the frame.  Since
$(\beta_{d})^2=1$, the eigenvectors of $\beta_{d}$ come in two sets $v_{\pm}^{(\alpha)}$,  satisfying
\begin{align}
\beta_d\,v_{\pm}^{(\alpha)}\, = \, \pm\,v_{\pm}^{(\alpha)}\,, \qquad \alpha\, = \, 1,...,\tfrac{\mathtt{d}}{2}\,.
\end{align}
The solutions to the Killing spinor equation \eqref{AdSKSEall} can then be written as
\begin{align}
\mathcal{Q}^{(\alpha)}_{\pm}\, = \, \frac{1}{\sqrt{r}}\,v_{\mp}^{(\alpha)}\,,\qquad
\mathcal{S}^{(\alpha)}_{\pm}\, = \, \frac{1}{\sqrt{r}}\,\beta_a\,\bar{x}^a\,v_{\pm}^{(\alpha)}\,.
\end{align}
We shall refer to the $\mathtt{d}$ solutions of \eqref{AdSKSEall}, for a given $\cc$, as $\vartheta_\cc^{(A)}$, with
\begin{align}
\vartheta_\cc^{(A)}\, = \, (\mathcal{Q}^{(\alpha)}_\cc\,,\,\,\mathcal{S}^{(\alpha)}_\cc)\,, \qquad
A=1,\dots,\mathtt{d}\,.
\end{align}

\section{Killing spinors for the D3-brane solutions}\label{app:c}
Here we provide some details of the derivation of the solution to the Killing spinor equations for the D3-brane solutions\footnote{The 
derivation of the Killing spinors for the case of M2-brane and M5-brane solutions runs along similar lines.}
given
in \eqref{D3sugrasol}, \eqref{D3functions}.
As in the text we choose the orthonormal frame for the metric given by
\begin{align}
\ex^a&\, = \, H^{1/6}\,\bar \ex^a,\quad\text{$a=0,1,2$}\, ,\qquad \ex^3\, =\, \frac{H^{1/6}}{2P^{1/2}}\,\diff y\,, \qquad
\ex^4\, = \, \frac{P^{1/2}}{H^{1/3}}\,\diff z\,,
\end{align}
where $\bar\ex^a$ is an orthonormal frame for the unit radius metric on
 AdS$_3$. 
We use five-dimensional gamma matrices
\begin{align}\label{gammas5dapp}
\Gamma_a&\, = \, \beta_a\otimes \gamma_3\,, \qquad 
\Gamma_3 \, = \, 1\otimes \gamma_1\,, \qquad
\Gamma_4\, = \, 1\otimes \gamma_2\,,
\end{align}
where $\beta_0=\ii \sigma^2$, $\beta_1=\sigma^1$, $\beta_2=\sigma^3$
are three-dimensional gamma matrices for AdS$_3$ directions,  while $\gamma_1=\sigma^1$,  $\gamma_2=\sigma^2$ are two-dimensional gamma matrices for the spindle directions,  with chirality matrix $\gamma_3=-\ii\,\gamma_1\gamma_2=\sigma^3$.  
We recall that the functions appearing in the solution, given in \eqref{D3functions}, allows us to introduce 
an angle $\alpha$ defined via
\begin{align}
\sin\alpha \, = \,  \frac{P^{1/2}}{H^{1/2}}\, , \qquad \cos\alpha \, = \,  \frac{y}{H^{1/2}}\, .
\end{align}

We first consider the gaugino Killing spinor equation given in \eqref{vargrav}:
\begin{align}\label{appgaugino}
\Big[\slashed{\partial}\varphi_i-2\sum_{I=1}^3\partial_{\varphi_i} X^{(I)}+\frac{\ii}{2}\sum_{I=1}^3\partial_{\varphi_i} \left(X^{(I)}\right)^{-1}\,{F}^{(I)}_{\mu\nu}\Gamma^{\mu\nu}\Big]\epsilon\, = \, 0\,.
\end{align}
For the solution, and utilizing $X^{(1)}X^{(2)}X^{(3)}=1$, we can rewrite
\begin{align}
\slashed{\partial}\varphi_i\, = \, -2\Big(\sum_{I=1}^3\partial_{\varphi_i} X^{(I)}\Big)\sin\alpha\,  \Gamma_3\,.
\end{align}
Next, using the condition \eqref{fluxD3branes} for the flux we have
\begin{align}\label{fluxD3branesnew2}
\left(X^{(I)}\right)^{-1} F^{(I)}_{34}
\, = \, \frac{2 X^{(I)}\q_I}{H^{1/2}}
\, = \, -2\Big(X^{(I)}\cos\alpha-\frac{1}{H^{1/6}}\Big)\,,
\end{align}
and thus we can also rewrite
\begin{align}\label{D3FslashpieceXI}
\frac{\ii}{2}\partial_{\varphi_i} \Big(X^{(I)}\Big)^{-1}\,{F}^{(I)}_{\mu\nu}\Gamma^{\mu\nu}\, = \, 2\ii \partial_{\varphi_i} X^{(I)} \Big[ 1-\Big( X^{(I)}\Big)^{-1}\frac{H^{1/3}}{y}\Big]\cos\alpha\, \Gamma_{34}\,.
\end{align}
Taking the sum of over $I$ we deduce that
\begin{align}
\frac{\ii}{2}\sum_{I=1}^3\partial_{\varphi_i} \Big(X^{(I)}\Big)^{-1}\,{F}^{(I)}_{\mu\nu}\Gamma^{\mu\nu}\, = \, 2\ii\Big(\sum_{I=1}^3\partial_{\varphi_i} X^{(I)}\Big)\cos\alpha\,  \Gamma_{34}\,,
\end{align}
where we used the fact that the sum for the second term in the brackets in \eqref{D3FslashpieceXI} vanishes,  due to
\begin{align}
\sum_{I=1}^3\left( X^{(I)}\right)^{-1}\partial_{\varphi_i} X^{(I)}\, = \, \sum_{I=1}^3 \partial_{\varphi_i} \log X^{(I)}\, = \, \partial_{\varphi_i} \log \left[X^{(1)}X^{(2)}X^{(3)}\right]\, = \, 0\,,
\end{align}
since $X^{(1)}X^{(2)}X^{(3)}=1$. 
Putting this together, for the solution we can thus rewrite 
\eqref{appgaugino} in the form
\begin{align}
2\ii \Big(\sum_{I=1}^3\partial_{\varphi_i} X^{(I)}\Big)\left(M+\ii\right)\epsilon\, = \, 0\,,
\end{align}
where the matrix $M$ is given by
\begin{align}\label{emmdef}
M\, = \, \cos\alpha\, \Gamma_{34}+\ii \sin\alpha\, \Gamma_3\,,
\end{align}
introduced in \eqref{spinorprojectionD3},  satisfying $M^2=-1$.  We thus see that the vanishing of the gaugino variation, for both $i=1,2$,  reduces to the condition that the Killing spinors $\epsilon$ satisfy
\begin{align}\label{projectionD3}
M\epsilon\, =\, -\ii\epsilon\,.
\end{align} 

We now turn to the gravitino equation in \eqref{vargrav} given by
\begin{align}
\Big[\nabla_{\mu}-\frac{\ii}{2}\sum_{I=1}^3A^{(I)}_{\mu}+\frac{1}{6}\sum_{I=1}^3 X^{(I)}\,\Gamma_{\mu}+\frac{\ii}{24}\sum_{I=1}^3\,\left(X^{(I)}\right)^{-1}\,\left(\Gamma_{\mu}^{\,\,\,\nu\rho}-4\,\delta_{\mu}^{\nu}\,\Gamma^{\rho}\right)\,F^{(I)}_{\nu\rho}\Big]\epsilon\, = \, 0\,.
\end{align}
We make an ansatz for the spinor of the form
\begin{align}
\epsilon\, = \, \vartheta\otimes \zeta\,,
\end{align}
where $\vartheta$ an AdS$_3$ Killing spinor satisfying
\begin{align}\label{AdS3KSE_D3}
\bar\nabla_a\vartheta\, = \, \frac{1}{2}\beta_a\vartheta\,,
\end{align}
where here $\bar\nabla_a$ is the covariant derivative on the unit radius AdS$_3$ (as discussed in appendix \ref{AdSspinors})
and $\zeta$ is a two-dimensional spinor on the spindle,  which we can write as
\begin{align}
\zeta\, = \, 
\begin{pmatrix}
\zeta_+(y,z)\\
\zeta_-(y,z)
\end{pmatrix}\,.
\end{align}
The condition \eqref{projectionD3} that we derived above implies that $\zeta_\pm$ are related via
\begin{align}\label{constraintzetapm}
\cos\frac{\alpha}{2}\, \zeta_++\sin\frac{\alpha}{2}\, \zeta_-\, = \, 0\,.
\end{align}

We first consider the components of the Killing spinor equation that are along the directions tangent to
AdS$_3$.  Using the fact that
\begin{align}\label{exxachrel}
\sum_{I=1}^3X^{(I)}\, = \, \frac{H'}{H^{2/3}}\,,
\end{align}
as well as \eqref{AdS3KSE_D3} we obtain
\begin{align}
\nabla_a\epsilon\, = \, \Big[-\frac{\ii}{2H^{1/6}}\Gamma_{34}-\frac{1}{6}\Big(\sum_{I=1}^3X^{(I)}\Big)\sin\alpha \Gamma_3\Big]\Gamma_a\epsilon\,.
\end{align}
Then using \eqref{fluxD3branesnew2} 
we deduce that for these components we must have
\begin{align}
-\frac{\ii}{6}\Big(\sum_{I=1}^3X^{(I)}\Big)\Gamma_a\left(M+\ii\right)\epsilon\, = \, 0\,,
\end{align}
which vanishes by virtue of the projection \eqref{projectionD3}.

We next consider the components tangent to the spindle $\Sigma$. The spin connection associated 
to $\widetilde{\diff s^2}_{\Sigma}$ is 
given by
$\widetilde{\omega}^{12}\, = \, \widetilde{\omega}^{12}_4\ex^4$, where 
\begin{align}
\widetilde{\omega}^{12}_4\, = \, \frac{2PH'-3HP'}{3H^{7/6}P^{1/2}}\,.
\end{align}
We now consider the components of the gravitino variation along the $\ex^4$ direction.  
We have
\begin{align}
\nabla_4\epsilon\, = \, \frac{1}{H^{1/6}\sin\alpha}\partial_z \epsilon+\frac{1}{2}\widetilde{\omega}^{12}_4 \Gamma_{34} \epsilon\,.
\end{align}
The gauge fields can be rewritten as
\begin{align}
A^{(I)}_4\, = \, \frac{\cos\alpha}{\sin\alpha}\, X^{(I)}\,,
\end{align}
and using \eqref{fluxD3branesnew2} 
we obtain the condition
\begin{align}
\Big[\frac{1}{H^{1/6}\sin\alpha}\partial_z+\frac{1}{2}\widetilde{\omega}^{12}_4\Gamma_{34}-\Big(\sum_{I=1}^3X^{(I)}\Big)\Big(\frac{\ii}{2}\frac{\cos\alpha}{\sin\alpha}-\frac{1}{6}\Gamma_4+\frac{\ii}{3}\cos\alpha\Gamma_3\Big)+\frac{\ii}{H^{1/6}}\Gamma_3\Big]\epsilon\, = \, 0\,.
\end{align}
This requires some massaging.  First,  we use \eqref{emmdef} to rewrite
\begin{align}
\frac{\ii}{2}\frac{\cos\alpha}{\sin\alpha}\, = \, -\frac{1}{2}\left(\frac{\cos^2\alpha}{\sin\alpha}\Gamma_{34}+\ii\cos\alpha\Gamma_3\right)+\frac{1}{2}\frac{\cos\alpha}{\sin\alpha}(M+\ii)\,.
\end{align}
Combining this with
\begin{align}
\frac{1}{2}\widetilde{\omega}^{12}_4+\frac{1}{2}\frac{\cos^2\alpha}{\sin\alpha}\Big(\sum_{I=1}^3X^{(I)}\Big)\, = \, \frac{\cos\alpha}{H^{1/6}\sin\alpha}-\frac{1}{6}\sin\alpha\Big(\sum_{I=1}^3X^{(I)}\Big)\,,
\end{align}
we obtain
\begin{align}
\frac{1}{H^{1/6}\sin\alpha}\Big[\partial_z+M-H^{1/6}\Big(\sum_{I=1}^3X^{(I)}\Big)\Big(\frac{\ii}{6}\sin\alpha\Gamma_4+\frac{1}{2}\cos\alpha\Big)\left(M+\ii\right)\Big]\epsilon\, =\, 0\,.
\end{align}
Using \eqref{projectionD3} this finally reduces to
$\left(\partial_z-\ii\right)\epsilon=0$ so that the spinor $\zeta$ on the spindle must satisfy
\begin{align}
\partial_z\zeta\, = \, \ii\zeta\,.
\end{align}
This can be solved by setting
\begin{align}
\zeta\, = \, \ex^{\ii z}\bar{\zeta}\, = \, 
\begin{pmatrix}
\bar{\zeta}_+(y)\\
\bar{\zeta}_-(y)
\end{pmatrix}\,,
\end{align}
with $\bar{\zeta}_{\pm}$ subject to the same constraint \eqref{constraintzetapm} as $\zeta_{\pm}$.

Finally, let us turn to the components along $\ex^3$.  Here we again use \eqref{fluxD3branesnew2}
to obtain the condition
\begin{align}
\Big[2H^{1/3}\sin\alpha\partial_y+\frac{1}{6}\Big(\sum_{I=1}^3X^{(I)}\Big)\left(\Gamma_3+2\ii\cos\alpha \Gamma_4\right)-\frac{\ii}{H^{1/6}}\Gamma_4\Big]\epsilon\, = \, 0\,.
\end{align}
Writing $\epsilon=\ex^{\ii z}\vartheta\otimes \bar{\zeta}$, and using the explicit choice of gamma matrices given in and below
\eqref{gammas5dapp}, as well as \eqref{exxachrel}
we obtain the ODE
\begin{align}\label{finalODE}
\partial_y\bar{\zeta}\, = \, \Big[-\frac{H'}{12 H^{1/2}P^{1/2}}\left(\sigma^1+2\ii\sigma^2\right)+\frac{\ii}{2P^{1/2}}\sigma^2\Big]\bar{\zeta}\,.
\end{align}
Now, using the relation between $\bar\zeta_\pm$ given in \eqref{constraintzetapm}, we find that the first component of this ODE is
\begin{align}
\partial_y\bar{\zeta}_+\, = \, \frac{-6H+H'(2y+H^{1/2})}{12 H P^{1/2}}\cot\frac{\alpha}{2}\bar{\zeta}_+\,,
\end{align}
which is solved by
\begin{align}
\bar \zeta_+\, = \, H^{1/12}\sin\frac{\alpha}{2}\,,\qquad\Rightarrow \qquad
\bar \zeta_-\, = \, -H^{1/12}\cos\frac{\alpha}{2}\,.
\end{align}
One can check that this solution also solves the other component of the ODE \eqref{finalODE}. We have thus
derived the Killing spinor solution given in \eqref{d3kses}.

\newpage

\providecommand{\href}[2]{#2}\begingroup\raggedright\endgroup

\end{document}